\documentclass[prd,showpcs,amsmath,amssymb,nofootinbib,longbibliography,twocolumn,showpacs,notitlepage]{revtex4-1}
\usepackage{multirow}
\usepackage{epsfig}
\usepackage{amsmath}
\usepackage{bm}
\usepackage{times}
\usepackage{graphicx}
\usepackage{color}
\usepackage{slashed}
\usepackage{graphicx}
\usepackage{amsmath} 
\usepackage[latin1]{inputenc}
\usepackage{hyperref}
\usepackage{soul}
\usepackage{multirow}
\usepackage{epsfig}
\usepackage{amsmath}
\usepackage{bm}  
\usepackage{times}
\usepackage{graphicx} 
\usepackage{color}
\usepackage{slashed}
\usepackage{graphicx}
\usepackage{amsmath}

\def\bea{\begin{eqnarray}}
\def\eea{\end{eqnarray}}
\def\bean{\begin{eqnarray*}}
\def\eean{\end{eqnarray*}} 
\def\nn{\nonumber}
\def\beaal{\begin{align}}
\def\eeaal{\end{align}}

\begin{document}  
 
\title{Probing the Neutrino Seesaw Scale with Gravitational Waves}

\author{Bartosz~Fornal}
\affiliation{Department of Chemistry and Physics, Barry University, Miami Shores, Florida 33161, USA\vspace{1mm}}
\author{Dyori~Polynice\vspace{1mm}}
\affiliation{Department of Chemistry and Physics, Barry University, Miami Shores, Florida 33161, USA\vspace{1mm}}
\author{Luka~Thompson\vspace{1mm}}
\affiliation{Department of Chemistry and Physics, Barry University, Miami Shores, Florida 33161, USA\vspace{1mm}}

\date{\today}

\begin{abstract}
Neutrinos are the most elusive particles of the Standard Model. The physics behind their masses remains unknown and requires introducing new particles and interactions. 
An elegant solution to this problem is provided by the seesaw mechanism. 
Typically considered at a high scale, it is potentially testable in gravitational wave experiments by searching for a spectrum from  cosmic strings, which offers a rather generic signature across many high-scale seesaw models. Here we consider the possibility of a low-scale seesaw mechanism at the PeV scale, generating neutrino masses within the framework of a model with gauged U(1) lepton number. In this case, the gravitational wave signal at high frequencies arises from a first order phase transition in the early Universe, whereas at low frequencies it is generated by domain wall annihilation, leading to a double-peaked structure in the gravitational wave spectrum. The signals discussed here can be searched for in upcoming experiments, including gravitational wave interferometers, pulsar timing arrays, and astrometry observations.
\end{abstract}

\maketitle

\section{Introduction}
\label{beginning}

The Standard Model \cite{Glashow:1961tr,Higgs:1964pj,Englert:1964et,Weinberg:1967tq,Salam:1968rm,Fritzsch:1973pi,Gross:1973id,Politzer:1973fx} is a triumph of theoretical and experimental elementary particle physics in describing the world at the subatomic level. It also enables us to dive back in time and understand what happened in the  Universe starting from roughly one-trillionth of a second after the Big Bang. Despite those undeniable successes, a few key observations  still escape theoretical understanding, with the most pressing ones, on the particle physics end, considering questions about the nature of dark matter, the origin of the matter-antimatter asymmetry, and the mechanism behind neutrino masses. While dark matter \cite{1970ApJ...159..379R,Boomerang:2000efg,Gavazzi:2007vw} may be entirely decoupled from the Standard Model  and direct detection experiments may never see it, the other two problems require additional ingredients and interactions with the Standard Model particle content, confirming that the currently established elementary particle picture is not yet complete. In this work we focus on the neutrino mass puzzle.

Several frameworks for generating small neutrino masses have been proposed. The simplest one is the type I seesaw mechanism \cite{Minkowski:1977sc,seesaw,seesaw2,Mohapatra:1979ia}, which requires introducing only right-handed neutrinos with a large mass term,
 \bea
-\mathcal{L}_I \supset y_\nu \, \bar{l}_L \widetilde{H} \nu_R +  M \overline{\nu_R^c} \nu_R  + {\rm h.c.}\ ,
\eea
leading to the left-handed neutrino masses $m_\nu \sim (y_\nu v)^2/M$, where $v$ is the Higgs vacuum expectation value (vev).
 The other tree-level mechanisms include type II seesaw \cite{Konetschny:1977bn,Schechter:1980gr,PhysRevD.22.2860,Mohapatra:1980yp} involving a new ${\rm SU}(2)_L$ triplet scalar $\Delta$,
  \bea
-\mathcal{L}_{I\!I} \supset y_\nu \, \overline{{l}_L^c} i \tau_2\Delta\, l_L +  \mu\, H^T i\tau_2  \Delta^\dagger H\   + {\rm h.c.},
\eea
resulting in neutrino masses $m_\nu \sim y_\nu \mu \,v_0^2 /M_\Delta^2$, where $v_0$ is the vev of the neutral component of $\Delta$,
 and  type III seesaw \cite{Foot:1988aq} with new  heavy ${\rm SU}(2)_L$ triplet fermions $\rho$,
   \bea
-\mathcal{L}_{I\!I\!I} \supset y_\nu \, {l}_L i \tau_2 \rho_L H + {\rm Tr}\!\left(\rho \,M_\rho \rho\right)  + {\rm h.c.},
\eea
yielding neutrino masses  $m_\nu \sim (y_\nu v)^2 /M_\rho$.
 The  commonly considered neutrino mass generation mechanism at the loop level is the Zee model \cite{Zee:1980ai} requiring a new  electroweak singlet and an electroweak doublet scalar.

 Given that the Standard Model is a gauge theory with baryon and lepton number as accidental global symmetries, it is natural to consider a framework in which one of them or both   are promoted to be gauge symmetries. The efforts to do this started fifty years ago \cite{Pais:1973mi}, with only a few other attempts  \cite{Rajpoot:1987yg, Foot:1989ts, Carone:1995pu, Georgi:1996ei} until fairly recently, when modern phenomenologically consistent models of this type were constructed  \cite{FileviezPerez:2010gw,Duerr:2013dza,Perez:2014qfa}.
 This idea was further expanded to supersymmetric theories  \cite{Arnold:2013qja}, unified models \cite{Fornal:2015boa,Fornal:2015one}, and generalized to non-Abelian gauged lepton number  \cite{Fornal:2017owa}. Detailed analyses of dark matter candidates in models with gauged baryon/lepton number  were conducted in \cite{Duerr:2014wra,Ohmer:2015lxa,FileviezPerez:2018jmr}, and solutions to the matter-antimatter asymmetry puzzle  were considered  through a new sphaleron process
 \cite{Fornal:2017owa} and   high-scale leptogenesis \cite{FileviezPerez:2021hbc}.

The nontrivial  symmetry breaking pattern in theories with gauged baryon and lepton number presents an opportunity to search for signatures of those models in gravitational wave experiments. Indeed, the 2016 first direct detection of gravitational waves by the Laser Interferometer Gravitational Wave
Observatory (LIGO) and Virgo collaboration \cite{LIGOScientific:2016aoc} 
brought new hope for particle physics by searching for a stochastic gravitational wave background from the early Universe arising from  first order phase transitions \cite{Kosowsky:1991ua}, dynamics of cosmic strings \cite{Vachaspati:1984gt,Sakellariadou:1990ne}, domain wall annihilation \cite{Hiramatsu:2010yz}, and inflation \cite{Turner:1996ck}. Among those, the processes resulting in signatures that  most heavily depend on the particle physics details  are first order phase transitions and domain wall annihilation.

First order phase transitions occur when a new true vacuum is formed with an energy density lower than that of the high temperature false vacuum, and both are  separated by a potential barrier. When the Universe undergoes a transition from the false to the true vacuum, this triggers the nucleation of bubbles, which expand and fill up the Universe. During this process, gravitational waves are generated from sound  waves, bubble wall collisions, and turbulence. This has been  studied in the context of numerous particle physics models \cite{Grojean:2006bp,Schwaller:2015tja,Vaskonen:2016yiu,Dorsch:2016nrg,Baldes:2017rcu,Bernon:2017jgv,Chala:2018ari,Angelescu:2018dkk,Brdar:2018num,Okada:2018xdh,Croon:2018kqn,Alves:2018jsw,Breitbach:2018ddu,Croon:2018erz,Hall:2019ank,Ellis:2019oqb,Dev:2019njv,Hasegawa:2019amx,VonHarling:2019rgb,DelleRose:2019pgi,Lewicki:2019gmv,Greljo:2019xan,Huang:2020bbe,Okada:2020vvb,Lewicki:2020jiv,Ellis:2020nnr,Fornal:2020esl,Han:2020ekm,Fornal:2020ngq,Craig:2020jfv,Fornal:2021ovz,DiBari:2021dri,Azatov:2021ifm,Zhou:2022mlz,Benincasa:2022elt,Kawana:2022fum,Costa:2022oaa,Costa:2022lpy,Fornal:2022qim,Kierkla:2022odc,Azatov:2022tii,Fornal:2022qim,Fornal:2023hri,Bosch:2023spa,Bunji:2024ovg} (for a 
 review see \cite{Caldwell:2022qsj,Athron:2023xlk} and for constraints from LIGO/Virgo data see \cite{LIGO_FOPT}). 
Among those works, the first order phase transition gravitational wave signatures of various models with separately gauged baryon and lepton number symmetries were studied in \cite{Fornal:2020esl,Fornal:2022qim,Fornal:2023hri,Bosch:2023spa}.

If there existed more than one  vacuum the Universe could transition to, topological defects such as domain walls would be produced. Those are  two-dimensional field configurations formed at the boundaries of  regions of different vacua. To avoid overclosing the Universe, domain walls need to undergo annihilation, which is possible if there exists a slight energy density mismatch (potential bias) between the vacua. The resulting stochastic gravitational wave background depends both on the scale of the symmetry breaking and the potential bias, determined by the details of the scalar potential of the model. Such signatures have been considered in many particle physics models \cite{Kadota:2015dza,Eto:2018hhg,Eto:2018tnk,Chen:2020soj,Battye:2020jeu,Craig:2020bnv,Dunsky:2021tih,Blasi:2022ayo,Barman:2022yos,Borah:2022wdy,Fornal:2023hri,King:2023cgv,Bosch:2023spa} (for a review see
 \cite{Saikawa:2017hiv} and for the bounds from LIGO/Virgo data see \cite{Jiang:2022svq}).

Since the theories for neutrino masses usually studied in the literature are high-scale seesaw models (allowing for an $\mathcal{O}(1)$ neutrino Yukawa coupling), the commonly considered gravitational wave signatures  are those arising from cosmic strings, which are topological structures forming when a ${\rm U}(1)$ symmetry is broken and correspond to one-dimensional field configurations along the direction of the unbroken symmetry. The reason is that gravitational radiation from the dynamics of cosmic strings provides signals from models with high symmetry breaking scales which are within the reach of current and upcoming gravitational wave experiments \cite{Blanco-Pillado:2017oxo,Ringeval:2017eww,Cui:2017ufi,Cui:2018rwi,Guedes:2018afo,Dror:2019syi,Gouttenoire:2019rtn,Buchmuller:2019gfy,King:2020hyd,Fornal:2020esl} (for a review see \cite{Gouttenoire:2019kij} and for the  limits from LIGO/Virgo data see \cite{LIGOScientific:2021nrg}). Nevertheless, cosmic string signatures are generic and depend only on the symmetry breaking scale, thus they are not useful in probing particle physics details.

In this paper  we consider the  seesaw mechanism at the PeV scale within the framework of the model with gauged lepton number proposed in \cite{Debnath:2023akj}. The  theory predicts   a first order phase transition in the early Universe and subsequent formation of  domain walls which undergo annihilation. This results in a very unique scenario when a double-peaked gravitational wave signature is expected from just a single ${\rm U}(1)$ symmetry breaking. The peak in the spectrum arising from the first order phase transition appears at high frequencies and is testable in future  experiments such as Cosmic Explorer (CE) \cite{Reitze:2019iox}, Einstein Telescope (ET) \cite{Punturo:2010zz}, DECIGO \cite{Kawamura:2011zz}, and Big Bang Observer (BBO) \cite{Crowder:2005nr}. Additionally, the peak from domain wall annihilation is within the reach of the Laser Interferometer Space Antenna (LISA) \cite{Audley:2017drz} and the other upcoming space-based interferometer  $\mu$ARES \cite{Sesana:2019vho}, pulsar timing arrays  NANOGrav \cite{NANOGRAV:2018hou} and SKA \cite{Weltman:2018zrl}, and the astrometry experiments THEIA \cite{Garcia-Bellido:2021zgu} and GAIA \cite{Gaia1,Moore:2017ity}.

We begin by describing  the model in Section \ref{gaugL}, including the particle content of the theory, symmetry breaking pattern, the resulting seesaw mechanism and the dark matter particle. In Section \ref{PPTT} we determine the shape of the effective potential and provide formulas for deriving the parameters of the first order phase transition, followed by a discussion of the resulting gravitational wave spectrum in Section \ref{PTspec}. The following Sections \ref{DWw} and \ref{DWw2} are devoted to describing the formation of domain walls and the gravitational wave signal from their subsequent annihilation. The final gravitational wave signal of the model is discussed in Section \ref{signat}, followed by a brief summary in Section \ref{sum}.

\section{The model}\label{gaugL}

A natural gauge extension of the Standard Model which can accommodate the type I seesaw mechanism for the neutrinos is based on the symmetry \cite{Debnath:2023akj}
\bea\label{groupN}
{\rm SU}(3)_c \times {\rm SU}(2)_L \times {\rm U}(1)_Y \times {\rm U}(1)_\ell  \ ,
\eea
where ${\rm U}(1)_\ell$ is gauged lepton number. 

\subsection{Extra fermionic representations}

To cancel the gauge anomalies, the following new fermions are added to the Standard Model,
\bea
&&\Psi_L = \left(1,2,-\tfrac12, \ell_1\right)  , \ \Psi_R = \left(1,2,-\tfrac12, \ell_2\right)  ,\nn\\
&&\eta_L= \left(1,1,-1, \ell_2\right)  ,  \ \ \eta_R= \left(1,1,-1, \ell_1\right)  ,\\
&&\chi_L= \left(1,1,0, \ell_2\right)  , \  \ \chi_R= \left(1,1,0, \ell_1\right) , \ \ \nu_{Ri}= \left(1,1,0, 1\right)  , \ \ \nn
\eea
where $\ell_1 - \ell_2 = -3$. We also assume, following \cite{Debnath:2023akj}, that $\ell_1 \ne - \ell_2$ and  $\ell_1, \ell_2 \ne \pm 1$. The corresponding Lagrangian fermionic kinetic terms are,
\bea
\mathcal{L}_{\rm kin}^f &=& i \,\overline{\Psi}_L \slashed{D}\,{\Psi}_L +  i \,\overline{\Psi}_R \slashed{D}\,{\Psi}_R +  i \,\overline{\eta}_L \slashed{D}\,{\eta}_L +  i \,\overline{\eta}_R \slashed{D}\,{\eta}_R\nn\\
&+&i \,\overline{l}_L \slashed{D}\,{l}_L +  i \,\overline{e}_R \slashed{D}\,{e}_R +  i \,\overline{\nu}_R \slashed{D}\,{\nu}_R  \ ,
\eea
where the covariant derivative $D_\mu = \partial_\mu + i g_\ell L_{\mu} L$, and after symmetry breaking $L_\mu$ gives rise to the new gauge boson $Z_\ell$.

\subsection{Symmetry breaking and type I seesaw}

The gauged lepton number symmetry ${\rm U}(1)_\ell$ is broken when the two complex scalar fields,
\bea
S = (1,1,0,3) \ , \ \ \ \ \phi = (1,1,0,-2) 
\eea
develop vevs, $\langle S \rangle = v_S/\sqrt2$ and   $\langle \phi \rangle = v_\phi/\sqrt2$, respectively. 
This provides masses to the new fermions through the terms
\bea
-\mathcal{L}_Y^f &=& y_\Psi \overline{\Psi}_R S \Psi_L + y_\eta \overline{\eta}_L S\eta_R + y_\chi \overline{\chi}_L S\chi_R + y_1 \overline{\Psi}_R H \eta_L \nn\\
&+& y_2 \overline{\Psi}_L H \eta_R + y_3 \overline{\Psi}_R \widetilde{H} \chi_L  + y_4 \overline{\Psi}_L \widetilde{H} \chi_R \nn\\
&+& y_\nu \,\overline{l}_L \widetilde{H} \nu_R + Y_\nu \,\overline{\nu_R^c} \,\phi\, \nu_R     + {\rm h.c.}\ , \ \ \ 
\eea
where the last line introduces the type I seesaw mechanism, with the resulting neutrino mass matrix
\bea
\mathcal{M}_\nu = \frac{v^2}{2\sqrt2 \,v_\phi} \, y_\nu \,Y_\nu^{-1} y_\nu^T \ .
\eea
In our analysis we consider the ${\rm U}(1)_\ell$ breaking  scale
\bea\label{vevv}
v_\ell \equiv \sqrt{v_S^2 + v_\phi^2} \,\sim\, 1 \  \rm PeV \ . 
\eea
Assuming that 
the elements of the $Y_\nu$ matrix are $\mathcal{O}(1)$, the neutrino Yukawa matrix entries required to reproduce the measured neutrino mass splittings
are
\bea
y_\nu \sim 10^{-5} \ ,
\eea
which is larger than the Standard Model electron Yukawa. 

\subsection{Scalar potential and boson masses}

The scalar potential of the theory can be written as
\bea\label{dim5}
V(S,\phi) &=& -  \,\mu_S^2 |S|^2  +  \lambda_{S} |S|^4  -\mu_\phi^2 |\phi|^2  + \lambda_{\phi} |\phi|^4\nn\\
&+& \lambda_{S\phi} |S|^2 |\phi|^2 + \left(\frac{\lambda_M}{\Lambda}S^2 \phi^3 +{\rm h.c.}\right)\ ,
\eea
where we included a  dimension-five term, as in \cite{Debnath:2023akj}, permitted by the quantum numbers of $S$ and $\phi$,  small compared to the other terms, ${\lambda_M}/{\Lambda} \ll 1/v_S,1/v_\phi$. We also assumed that the terms coupling the heavy scalars to the Higgs are small. 

\vspace{1mm}

Upon symmetry breaking, the mass of $Z_\ell$ is given by
\bea
m_{Z_\ell} = g_\ell \sqrt{9v_S^2 + 4 v_\phi^2} \ .
\eea
The squared masses for the two $CP$-even scalars are
\bea
m_{\rm even}^2 &=& \lambda_S v_S^2 + \lambda_\phi v_\phi^2 \nn\\
&\pm& \sqrt{(\lambda_S v_S^2 - \lambda_\phi v_\phi^2)^2 + (\lambda_{S\phi} v_S v_\phi)^2} \ .
\eea
There are also two $CP$-odd scalars -- one is the Goldstone boson that becomes the longitudinal component of $Z_\ell$, and the other  is the Majoron with mass
\bea
m_J = \sqrt{\frac{\lambda_M v_\phi}{2\sqrt2\,\Lambda}\left(9v_S^2 + 4 v_\phi^2\right)} \ .
\eea
Constraints on the Majoron mass and couplings to neutrinos were studied in \cite{Heeck:2019guh} and for our choice of symmetry breaking scale they are all satisfied.

\subsection{Dark matter}

After the ${\rm U}(1)_\ell$ gauge symmetry is spontaneously broken, the model exhibits an accidental global ${\rm U}(1)$
 symmetry under which the new fields transform according to
 \bea
 &&\Psi_L \to e^{i\theta} \Psi_L \ , \ \  \Psi_R \to e^{i\theta} \Psi_R \ , \ \  \eta_L \to e^{i\theta} \eta_L \ , \nn\\
 && \eta_R \to e^{i\theta} \eta_R \ , \ \  \chi_L \to e^{i\theta} \chi_L \ , \ \  \chi_R \to e^{i\theta} \chi_R \ .
 \eea
 Because of this residual symmetry, the lightest field among the new fermions remains stable. If it is also a Standard Model singlet, such as $\chi$, it becomes a good dark matter candidate. 
  
As  argued in \cite{Debnath:2023akj},  to remain consistent with the observed dark matter relic abundance of $h^2 \Omega_{\rm DM} \approx 0.12$ \cite{Planck:2018vyg}, for most of the dark matter annihilation channels an upper bound on the ${\rm U}(1)_\ell$ breaking scale arises. For instance, assuming the $s$-channel annihilation,
 \bea
 \bar\chi\,\chi \to Z_\ell^* \to \bar{l} \,l \ ,
 \eea
perturbativity of  couplings requires that $m_{Z_\ell} \lesssim 30 \ {\rm TeV}$.\break Nevertheless, there are other annihilation channels that are not velocity suppressed, e.g., the $t$-channel annihilation via
   \bea
\bar \chi\,\chi \to \phi^* \phi \ ,
 \eea
with a sufficiently large cross section even when the symmetry is broken at the scale $\sim \rm PeV$. Of course one can always assume nonthermal dark matter production, in which case the upper bound on the ${\rm U}(1)_\ell$ breaking scale  does not apply.

\section{First order phase transition}
\label{PPTT}

To determine the range of parameters for which a first order phase transition occurs in the model, one first needs to find  the shape of the effective potential, including  its dependence on temperature. In our case there are two scalar fields participating in the breaking of ${\rm U}(1)_\ell$. Denoting by $\varphi_S$ and $\varphi_\phi$ the background fields corresponding to $S$ and $\phi$, respectively, the effective potential takes the form,
\bea\label{efp}
&&V_{\rm eff}(\varphi_S, \varphi_\phi, T) = -\frac12 \left(\lambda_S v_S^2  +\frac12\lambda_{S\phi}v_\phi^2\right)\varphi_S^2 + \frac14\lambda_S \varphi_S^4 \nn\\
&&- \ \frac12 \left(\lambda_\phi v_\phi^2  +\frac12\lambda_{S\phi}v_S^2\right)\varphi_\phi^2 + \frac14\lambda_\phi \varphi_\phi^4  + \frac14 \lambda_{S\phi} \varphi_S^2 \varphi_\phi^2\nn\\
&&+ \  \sum_i \frac{n_i }{64\pi^2}\,m_i^4(\varphi_S,\varphi_\phi)\left[\log \left(\frac{m_i^2(\varphi_S,\varphi_\phi)}{\Lambda_\ell^2}\right) - c_i \right] \\
&& +\ \frac{T^4}{2\pi^2}\sum_i n_i \int_0^\infty dx\,x^2 \log\left(1\mp e^{-\sqrt{{x^2 + m_i^2(\varphi_S,\varphi_\phi)}/{T^2}}}\right)\nn\\
&& + \ \frac{T}{12\pi} \sum_j n'_j \left\{m_j^3(\varphi_S,\varphi_\phi)- \left[m^2(\varphi_S,\varphi_\phi) + \Pi(T)\right]_j^{\frac32}\right\} \ .\nn
\eea
The first two lines of Eq.\,(\ref{efp}) correspond to the tree level part of the potential determined using the minimization conditions. The third line reflects the one-loop Coleman-Weinberg zero temperature contribution in the $\overline{\rm MS}$ renormalization scheme \cite{Quiros:1999jp}; the sum is over all particles charged under ${\rm U}(1)_\ell$, with $m_i(\varphi_S,\varphi_\phi)$ being their field-dependent masses, $n_i$ denoting the corresponding number of degrees of freedom (with a negative sign for fermions), $c_i = 3/2$ for scalar and fermions, $c_i = 5/6$ for gauge bosons, and $\Lambda_\ell$ being the renormalization scale which we take to be at the ${\rm U}(1)_\ell$ breaking scale. The last two lines correspond to the finite temperature contribution, with the plus sign corresponding to bosons and the minus sign to fermions, the sum over $i$ including all particles charges under  ${\rm U}(1)_\ell$ and the sum over $j$ involving only bosons, $n'_j$ denoting all degrees of freedom for scalars and only longitudinal ones for vector bosons, $\Pi(T)$ being the thermal mass matrix, and the subscript $j$ for $\left[m^2(\varphi_S,\varphi_\phi) + \Pi(T)\right]$ in the last line denoting the eigenvalues.

Because of choosing a high symmetry breaking scale, we can assume that the Yukawa couplings $y_\Psi$, $y_\eta$, and $y_\chi$ are small, consistent with the experimental constraints on the new fermions, so that their contribution does not affect the shape of the effective potential considerably. The field-dependent squared mass for the gauge boson $Z_\ell$ is,
\bea
m_{Z_\ell}^2(\varphi_S, \varphi_\phi) = g_\ell^2 \left(9\varphi_S^2 + 4 \varphi_\phi^2\right) \ ,
\eea
and the corresponding numbers of degrees of freedom are $n_{Z_\ell}=3$ and $n'_{Z_\ell}=1$. 
In the case of the $CP$-even scalars, their field-dependent squared mass matrix is given by
\bea
m^2_{\rm even}(\varphi_S, \varphi_\phi)  =
\begin{pmatrix}
2 \lambda_S \varphi_S^2 &  \lambda_{S\phi}  \varphi_S  \varphi_\phi\\
\lambda_{S\phi}  \varphi_S  \varphi_\phi & 2\lambda_{\phi}  \varphi_\phi^2 
\end{pmatrix}  ,
\eea
whereas for the $CP$-odd scalar $J$ and the Goldstone  it is
\bea
m^2_{J, GB}(\varphi_S, \varphi_\phi)  =\frac{\lambda_M\varphi_\phi}{2\sqrt2 \Lambda}
\begin{pmatrix}
9 \, \varphi_S^2 & 6 \,\varphi_S  \varphi_\phi \\
6 \,\varphi_S  \varphi_\phi  &  4\, \varphi_\phi^2 
\end{pmatrix}  .
\eea

Assuming  the quartic couplings are much smaller than the gauge coupling, which is precisely the parameter space region of interest for first order phase transitions, and choosing as a benchmark the new fermion ${\rm U}(1)_\ell$ charges to be $\ell_1 = -1/2$ and $\ell_2 = 5/2$, the  thermal mass for the gauge boson $Z_\ell$ is
\bea\label{thm}
\Pi_{Z_\ell}(T) &=& \frac{13}{2} g_\ell^2 T^2 \ .
\eea
For the $CP$-even scalars, as well as the $CP$-odd scalar and the Goldstone boson, the thermal mass matrices are the same  and given by
\bea
 \Pi_{\rm even}(T) &=& \Pi_{{J, GB}} (T) =  \frac14
 \begin{pmatrix}
9 & 0\\
 0 & 4
 \end{pmatrix} g_\ell^2 T^2 \ .
\eea

We find that for a wide range of quartic couplings and gauge couplings  the effective potential develops vacua at nonzero field values which have a lower energy density than the high temperature vacuum at $(\varphi_S, \varphi_\phi) = (0,0)$, separated from each other by a potential bump. The new minima appear around the field values $(v_S,v_\phi)$, $(v_S, -v_\phi)$,  $(-v_S,v_\phi)$, and $(-v_S,-v_\phi)$, however, two pairs of them are physically equivalent (for a related discussion in the more general case of a two Higgs doublet model, see \cite{Ginzburg:2004vp,Battye:2011jj}). In particular, since the effective potential is invariant under the $\mathcal{Z}_2$ transformation $\varphi_S \to -\varphi_S$, $\varphi_\phi \to \varphi_\phi$, just the two vacua around $(v_S,v_\phi)$ and  $(v_S,-v_\phi)$ are physically distinct. 
Their energy densities differ only slightly due to the dimension five term in Eq.\,(\ref{dim5}). As will be discussed in Section \ref{DWw}, this leads to the formation of domain walls, since the phase transition will populate both of those vacua.

\begin{figure}[t!]
\includegraphics[trim={0cm 0.0cm 0cm 0cm},clip,width=8.8cm]{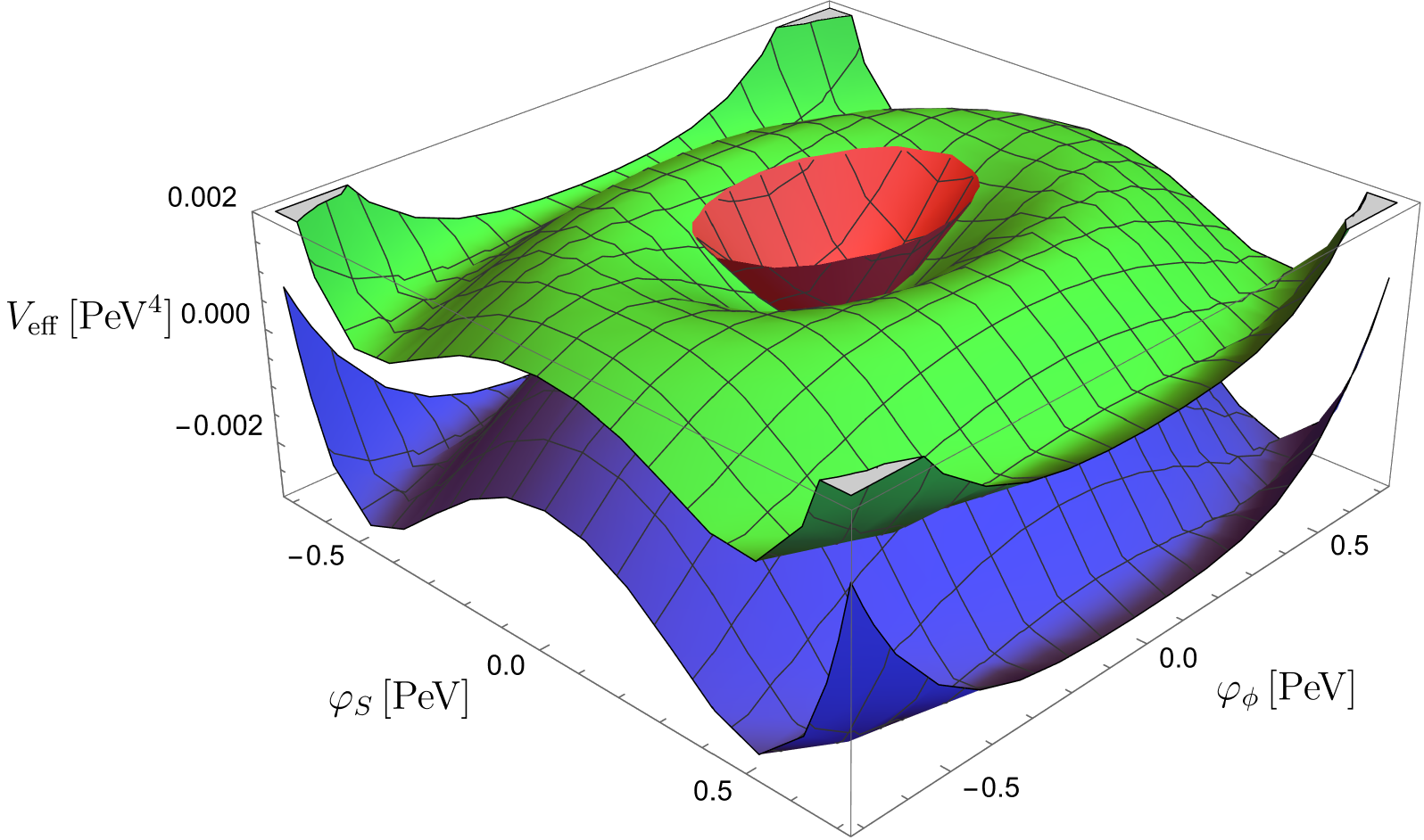} \vspace{-6mm}
\caption{Effective potential of the model, $V_{\rm eff}(\varphi_S, \varphi_\phi, T)$, plotted for $v_S = v_\phi = 1 \ \rm PeV$, $g_\ell = 0.6$, $\lambda_S = \lambda_\phi = 0.001$, $\lambda_{S\phi}=0$, and the temperatures: $450 \ \rm TeV$ (blue), $550 \ \rm TeV$ (green), $650 \ \rm TeV$ (red).}\label{fig:one}
\end{figure}

Figure \ref{fig:one} shows a plot of the effective  potential (prepared with {\fontfamily{cmtt}\selectfont Mathematica} \cite{Wolfram}) assuming the parameter values: $v_S = v_\phi = 1 \ \rm PeV$, $g_\ell = 0.6$, $\lambda_S = \lambda_\phi = 0.001$, and for several temperature values: $450 \ \rm TeV$ (blue), $550 \ \rm TeV$ (green), and $650 \ \rm TeV$ (red).
For concreteness, we focus on a phase transition from the false vacuum $(0,0)$ to the true vacuum around $(v_S,v_\phi)$. Since the two vacua are separated by a potential bump, the transition will be first order, resulting in the formation of true vacuum bubbles  in various patches of the Universe. If the conditions are right,  those bubbles expand and eventually fill out the entire Universe. 

The nucleation process can be initiated if the bubble nucleation rate becomes comparable to the Hubble expansion rate. This is described by the equation \cite{LINDE1983421}
 \bea\label{nucli}
 \bigg(\frac{S(T_*)}{2\pi T_*}\bigg)^{\!\frac32}T_*^4 \,\exp\left({-\frac{S(T_*)}{T_*}} \right) \approx H(T_*)^4 \ ,
\eea
where $T_*$ is the nucleation temperature at which this happens  and $S(T)$ is the Euclidean action
\bea
S(T)= \int d^3 r \bigg[\,\frac12 \bigg(\frac{d \varphi_{\rm bubble}}{dr}\bigg)^2+V_{\rm eff}(\varphi_{\rm bubble}, T)\bigg] . \ \ \ \ \ \ \ \ 
\eea
Here  $\varphi_{\rm bubble}$ is  the solution of the equation,
\bea
\frac{d^2 \varphi}{dr^2}+\frac{2}{r}\frac{d\varphi}{dr}-\frac{d}{d\varphi} V_{\rm eff}(\varphi,T) = 0 \ ,
\eea
constrained by the two conditions:
\bea
\frac{d\varphi}{dr}\bigg|_{r=0} = 0 \ \ \   \ \ {\rm and } \ \ \ \ \ \varphi(\infty) = \varphi_{\rm false} \ .
\eea
Writing the Hubble constant in terms of the temperature, the Planck mass, and the number of active degrees of freedom,
\bea
H(T) \approx \frac{T^2}{ M_{P}}\sqrt{\frac{4\pi^3 g_*}{45}} \ ,
\eea
Eq.\,(\ref{nucli}) takes the following form,
\bea\label{STi444}
\frac{S(T_*)}{T_*}   \approx  \log\!\left(\frac{M_{P}^4}{T_*^4}\right)  \!-\! \log\left[\left(\frac{4\pi^3g_*}{45}\right)^{\!\!2}\!\left(\frac{2\pi \,T_*}{S(T_*)}\right)^{\!\!\frac32}\right]\!. \  \ \ \ 
\eea

Upon finding the nucleation temperature from Eq.\,(\ref{STi444}), one can calculate the other two model-dependent parameters describing the first order phase transition: the strength of the phase transition $\alpha$ and its duration $1/\tilde\beta$. The parameter $\alpha$ is given by  the difference between the energy densities of the true and false vacuum divided by the energy density of radiation at the nucleation temperature,
\bea\label{aaa}
\alpha = \frac{\Delta \rho_{v}(T_*)}{ \frac{\pi^2}{30} g_* T_*^4} \ ,
\eea
where the numerator is given by
\bea
\Delta \rho_{v}(T) &=& V_{\rm eff}(\varphi_{\rm false},T) -  V_{\rm eff}(\varphi_{\rm true},T)\nn\\
&-&T \frac{\partial}{\partial T} {\left[ V_{\rm eff}(\varphi_{\rm false},T) -  V_{\rm eff}(\varphi_{\rm true},T)\right]} \ . \ \ \
\eea
The parameter $\tilde\beta$, being the inverse of the duration of the phase transition, is determined from
\bea\label{bta}
\tilde{\beta} = T_* \frac{d}{dT} \bigg(\frac{S(T)}{T}\bigg)\bigg|_{T=T_*} \ .
\eea
The dynamics of a first order phase transition also depends on the bubble wall velocity  $v_w$. Here we assume that it is close to the speed of light, i.e.,  $v_w = c$. For the rationale behind other choices  see  \cite{Espinosa:2010hh,Caprini:2015zlo}.

\section{Gravitational waves from\\ \ \ phase transition}\label{PTspec}

As described in Section \ref{beginning}, collisions between expanding bubbles, sound shock waves propagating through plasma, and magnetohydrodynamic turbulence  generate a stochastic gravitational wave background permeating the Universe since then. The expectation regarding the shape of each contribution to the spectrum was derived from numerical simulations and the empirical formulas have been determined.\vspace{2mm}

The  contribution from bubbles collisions is \cite{Kosowsky:1991ua,Huber:2008hg,Caprini:2015zlo,Lewicki:2020azd}
\bea\label{eq1}
h^2 \Omega_{\rm coll}(\nu) \,&\approx&\, \big({4.9\times 10^{-6}}\big)\frac{1}{\tilde\beta^2}\left(\frac{\kappa_c \, \alpha}{\alpha+1}\right)^2\left(\frac{100}{g_*}\right)^{\frac13}\nn\\
&\times&\frac{\big(\frac{\nu}{\nu_c}\big)^{2.8}}{1+2.8\big(\frac{\nu}{\nu_c}\big)^{3.8}} \ , \ \ \ \ \ \ 
\eea
where $\kappa_c$ is the fraction of the latent heat deposited into the bubble front \cite{Kamionkowski:1993fg} and $\nu_c$ is the peak frequency,
\bea
\kappa_c &=& \frac{\frac{4}{27}\sqrt{\frac32\alpha}+0.72\,\alpha}{1+0.72 \,\alpha} \ ,\nn\\
\nu_c &=& (0.037 \ {\rm Hz} ) \ \tilde\beta\left(\frac{g_{*}}{100}\right)^\frac16 \! \left(\frac{T_*}{1 \ {\rm PeV}}\right) \ .
\eea
The sound wave contribution   is given by \cite{Hindmarsh:2013xza,Caprini:2015zlo}
\bea\label{sound}
h^2 \Omega_{\rm sw}(\nu) \,&\approx&\, \big(1.9\times 10^{-5}\big)\frac1{\tilde\beta}\left(\frac{\kappa_s\,\alpha}{\alpha+1}\right)^2\left(\frac{100}{g_*}\right)^{\frac13} \Upsilon\nn\\
&\times&\frac{\big(\frac{\nu}{\nu_s}\big)^3}{\big[1+0.75 \big(\frac{\nu}{\nu_s}\big)^2\big]^{\frac72}}  \ ,
\eea
where $\kappa_s$ is the  fraction of the latent heat going into the bulk motion of the plasma \cite{Espinosa:2010hh},  $\nu_s$ is the peak frequency, and $\Upsilon$ is a suppression factor \cite{Ellis:2020awk,Guo:2020grp},
\bea\label{sp}
\kappa_s &=& \frac{\alpha}{0.73+0.083\sqrt\alpha + \alpha} \ ,\nn\\
\nu_s &=& (0.19 \ {\rm Hz} )  \ \tilde\beta\left(\frac{g_*}{100}\right)^\frac16 \!\left(\frac{T_*}{1 \ {\rm PeV}}\right) \ ,\nn\\
\Upsilon\, &=& 1- {\bigg[1+\frac{8\sqrt[3]\pi}{\tilde{\beta}}\sqrt{\frac{{\alpha+1}}{3\kappa_s\alpha}} \ \bigg]^{-\frac12}} \ .
\eea
Magnetohydrodynamic turbulence results in \cite{Caprini:2006jb,Caprini:2009yp}
\bea\label{tur}
h^2 \Omega_{\rm turb}(\nu) \,&\approx&\, \big({3.4\times 10^{-4}}\big) \frac1{\tilde\beta}\left(\frac{\epsilon\,\kappa_s \, \alpha}{\alpha+1}\right)^{\frac32}\left(\frac{100}{g_*}\right)^{\frac13} \nn\\
&\times&\frac{ \big(\frac{\nu}{\nu_t}\big)^{3}}{\big(1+\frac{8\pi \nu}{h_*}\big)\big(1+\frac{\nu}{\nu_t}\big)^{\frac{11}{3}}}\ ,
\eea
where we assumed $\epsilon=0.05$ \cite{Caprini:2015zlo} and
\bea
\nu_t &=& (0.27 \ {\rm Hz} ) \ \tilde\beta\left(\frac{g_*}{100}\right)^\frac16\!\left(\frac{T_*}{1 \ {\rm PeV}}\right) \ ,\nn\\
h_* &=& (0.17 \ {\rm Hz})\left(\frac{g_*}{100}\right)^\frac16\!\left(\frac{T_*}{1 \ {\rm PeV}}\right) \ .
\eea

\begin{figure}[t!]
\includegraphics[trim={1.9cm 0.4cm 0.8cm 0cm},clip,width=9.5cm]{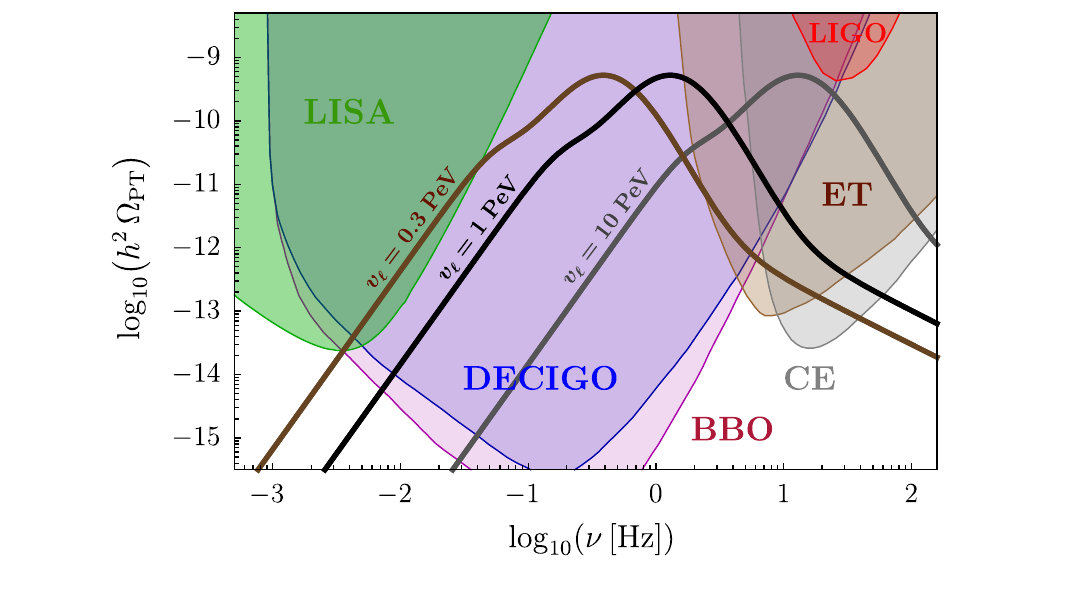} \vspace{-7mm}
\caption{First order phase transition-generated stochastic gravitational wave background from ${\rm U}(1)_\ell$ breaking for the model parameters: $g_\ell = 0.22$, $\lambda_S = \lambda_\phi \equiv \lambda = 10^{-4}$,  and for the three symmetry breaking scales: $0.3 \ \rm PeV$ (brown), $1 \ \rm PeV$ (black), $10 \ \rm PeV$ (gray). Sensitivities of future gravitational wave detectors: LISA, DECIGO, BBO, ET, and CE, as well as the reach of LIGO's O5 observing run  are shown as shaded regions.\\}\label{fig:two}
\end{figure}

\begin{figure}[t!]
\includegraphics[trim={2cm 0.8cm 1cm 0cm},clip,width=9.0cm]{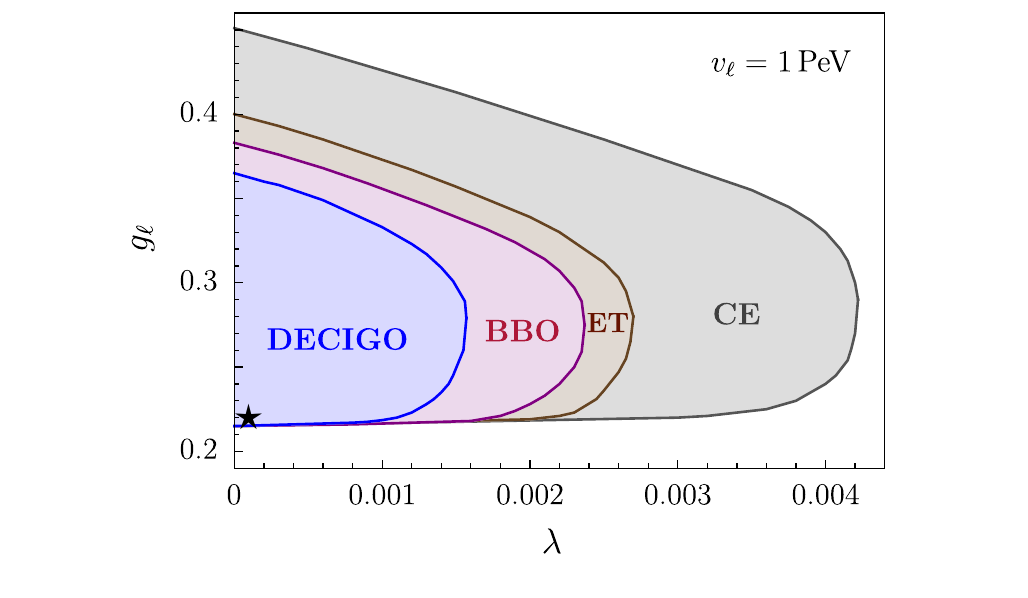} \vspace{-5mm}
\caption{Parameter space $(\lambda, g_\ell)$ for which a gravitational wave signal originating from a first order phase transition at the scale $1 \ {\rm PeV}$   is within the reach of upcoming gravitational wave experiments (DECIGO, BBO, ET, CE) with a signal-to-noise ratio greater than five. Colors for the various experiments are the same as in Figure \ref{fig:two}, and the  star corresponds to the $1 \ \rm PeV$  curve.  }\label{fig:three}
\end{figure}

The total gravitational wave signal is
\bea\label{eq2}
h^2 \Omega_{\rm PT}(\nu) = h^2\Omega_{\rm sw}(\nu) + h^2\Omega_{\rm coll}(\nu) + h^2\Omega_{\rm turb}(\nu) \ . \ \ \ \ 
\eea

\vspace{1mm}

Figure \ref{fig:two} shows the gravitational wave spectrum of the model from a first order phase transition happening at three different symmetry breaking scales:   $v_\ell = 0.3 \ \rm PeV$ (brown curve), $1 \ {\rm PeV}$ (black curve) and $10 \ \rm PeV$ (gray curve). The signal was plotted  using Eqs.\,(\ref{eq1})-(\ref{eq2}), with the phase transition  parameters calculated using Eqs.\,(\ref{nucli})-(\ref{bta}), for the gauge coupling $g_\ell = 0.22$, quartic couplings $\lambda_S = \lambda_\phi \equiv \lambda = 0.0001$, and assuming that the mixed quartic coupling $\lambda_{S\phi}$ and the new fermion Yukawas are negligibly small. 

The strength of the phase transition and the inverse of its duration  corresponding to the signals in Figure \ref{fig:two} are: $\alpha \approx 4$ and $\tilde{\beta} \approx 80$, whereas the nucleation temperatures are: $T_* \approx  21 \ \rm TeV$ for the brown curve, $T_* \approx 70 \ \rm TeV$ for the black curve, and $T_* \approx 700 \ \rm TeV$ for the gray curve. For a given symmetry breaking scale, longer phase phase transitions (characterized by a smaller $\tilde\beta$) lead to stronger gravitational wave signals. The shape of the spectrum is dominated by the sound wave contribution in the central part, and  by the bubble collision part at lower and at higher frequencies. The signal would be up to two orders of magnitude stronger if not for the sound wave suppression factor $\Upsilon$ in Eq.\,(\ref{sound}).

The shaded regions in Figure \ref{fig:two} indicate the  reach of several upcoming gravitational wave interferometers, including LISA  \cite{Audley:2017drz} (green), DECIGO  \cite{Kawamura:2011zz} (dark blue), Big Bang Observer \cite{Crowder:2005nr} (magenta), Einstein Telescope \cite{Punturo:2010zz} (brown), and Cosmic Explorer  \cite{Reitze:2019iox} (gray). The expected sensitivity of LIGO's O5 observing run \cite{LIGOScientific:2014pky} (red) is also included, but its reach is limited to theories with supercooled phase transitions, which is not the case for the model under consideration. 
For a symmetry breaking scale $\sim \rm PeV$ the signal is peaked around frequencies $\sim 1\  \rm Hz$, and is within the reach of most of those detectors. 
As the scale of symmetry breaking decreases, the nucleation temperature $T_*$ also decreases and the peak of the spectrum moves to lower frequencies, eventually becoming searchable also in LISA when $v_\ell \lesssim 0.5 \ \rm PeV$.

Figure \ref{fig:three} shows the region of parameter space of the quartic coupling $\lambda$ versus the gauge coupling $g_\ell$ (assuming  a symmetry breaking scale of $v_\ell = 1 \ \rm {PeV}$)  for which the model can be probed in future gravitational wave detectors. The colors of the shaded regions correspond to those in Figure \ref{fig:two} for the various experiments. Not all parameters for which a first order phase transition occurs result in a measurable signal. The upper bound of the shaded regions corresponds to the signal being too weak to be detectable in any of the planned experiments (this includes the first order phase transition example shown in Figure \ref{fig:one}). On the other hand, parameter points below the shaded regions do not result in a first order phase transition at all, since the bubble nucleation rate never becomes comparable to the Hubble expansion rate.

\section{Domain walls}
\label{DWw}

Given that the energy density splitting between the vacua  (potential bias) located in the proximity of the field values $(v_S, v_\phi)$ and $(v_S, -v_\phi)$,
\bea\label{split}
\Delta \rho = \frac{\lambda_M}{2\sqrt2 \, \Lambda} v_S^2 v_\phi^3  \  \ ,
\eea
is small due to our assumption  ${\lambda_M}/{\Lambda} \ll 1/v_S,1/v_\phi$, the first order phase transition discussed in Section \ref{PPTT}  populates both of them at similar rates. This creates domain walls, i.e., two-dimensional topological structures existing on the boundaries between the two vacua regions. Because of the nonzero energy density difference in Eq.\,(\ref{split}), domain walls are unstable, and eventually  undergo annihilation, avoiding the production of 
large density fluctuations in the Universe \cite{Saikawa:2017hiv}.

We denote the domain wall profile by $\vec\phi_{DW}(z)$, where $z$ is the axis perpendicular to the domain wall surface. It can be found as the solution of the following equation  \cite{Chen:2020soj},
\bea
\frac{d^2 \vec\varphi_{DW}(z)}{d z^2} - \vec\nabla_{\!\phi}V_{\rm eff}\big[\vec\varphi_{DW}(z)\big]= 0 \ ,
\eea
with the boundary conditions,
 \bea 
 \vec\varphi_{DW}(-\infty) =(v_S, v_\phi) \ , \ \ \ \  \ \vec\varphi_{DW}(\infty) =(v_S, -v_\phi) \ .\ \ \ 
 \eea 
 Apart from the potential bias $\Delta \rho$, domain walls are described also by the tension parameter $\sigma$ defined as,
\bea
\sigma = \int_{-\infty}^\infty dz \bigg[ \frac12 \bigg(\frac{d\vec\varphi_{DW}(z)}{d z}\bigg)^2 + V_{\rm eff}\big(\vec\varphi_{DW}(z)\big)\bigg].\ \ \ 
\eea
To a good  approximation one can write
\bea
\sigma \,\sim \,  v_\ell^3\ .
\eea
Domain walls undergo efficient annihilation if \cite{Saikawa:2017hiv}
\bea\label{condd11}
\sqrt{\Delta \rho} \ \gtrsim \frac{\sigma}{M_{P}} \ . 
\eea
In our case $v_S \sim v_\phi \sim 1 \ \rm PeV$, and the above inequality is  satisfied  even for $\lambda_M$ as small as $10^{-13}$ (since $\Lambda < M_P$). 
Another constraint on $\Delta \rho$ comes from the requirement of the domain walls annihilating before Big Bang nucleosynthesis, not to alter the ratios of the produced elements; however, that bound is weaker than the one  in Eq.\,(\ref{condd11}).

\section{Gravitational waves from domain walls}\label{DWw2}

\vspace{-1mm}

Numerical simulations yield the following spectrum of the
 stochastic gravitational wave background generated by annihilating domain walls \cite{Kadota:2015dza,Saikawa:2017hiv},
 \vspace{-2mm}
\bea\label{dme}
h^2 \Omega_{\rm DW}(\nu)&\approx& 7.1\times 10^{-33}\left(\frac{\sigma}{\rm PeV^3}\right)^4\left(\frac{\rm TeV^4}{\Delta \rho}\right)^2\left(\frac{100}{g_*}\right)^{\frac13}\nn\\
&\times& \!\!\bigg[\!\left(\frac{\nu}{\nu_d}\right)^3\!\theta(\nu_d-\nu) + \left(\frac{\nu_d}{\nu}\right)\theta(\nu-\nu_d)\bigg]. \ \ \ \ \ \ \ 
\eea
Here $\theta(\nu)$ is the Heaviside step function, $\mathcal{A} = 0.8$ was taken for the area parameter and $\tilde\epsilon_{\rm gw} = 0.7$ for the efficiency parameter \cite{Hiramatsu:2013qaa}, and $\nu_d$ is the signal's peak frequency given by
\bea
\nu_d \approx (0.14 \ {\rm Hz}) \,  \sqrt{\bigg(\frac{\rm PeV^3}{\sigma}\bigg)\bigg(\frac{\Delta \rho}{\rm TeV^4}\bigg)} \ .
\eea

There is a constraint on the strength of the gravitational wave signal arising from cosmic microwave background measurements which requires   $h^2 \Omega(\nu) < 2.9 \times 10^{-7}$ \cite{Clarke:2020bil}. This translates to the condition
\bea\label{cmb}
\sqrt{\Delta\rho} \  \gtrsim \frac{\sigma}{2.5\times 10^{12} \ \rm PeV} \ ,
\eea
which is slightly  stronger than the bound imposed by Big Bang nucleosynthesis in Eq.\,(\ref{condd11}).

To investigate how this bound translates into constraints on the model parameters, in particular the coefficient of the $\mathcal{Z}_2$ breaking  term in Eq.\,(\ref{dim5}), we rewrite Eq.\,(\ref{split}) as
\bea
\Delta \rho = \frac{\lambda_M}{2\sqrt2 \, \Lambda} v_\ell^5 \sin^3(\xi) \cos^2(\xi)
\eea
where we parameterized $v_S \equiv v_\ell \cos(\xi)$ and $v_\phi \equiv v_\ell \sin(\xi)$, consistent with Eq.\,(\ref{vevv}).
Taking $v_\phi \ll v_S$, one obtains
\bea
\frac{\lambda_M}{\Lambda} \sin^3(\xi) \, \gtrsim \frac{v_\ell}{2.2\times 10^{24} \ \rm PeV^2} \ .
\eea

\begin{figure}[t!]
\includegraphics[trim={1.8cm 0.4cm 0.8cm 0cm},clip,width=9.5cm]{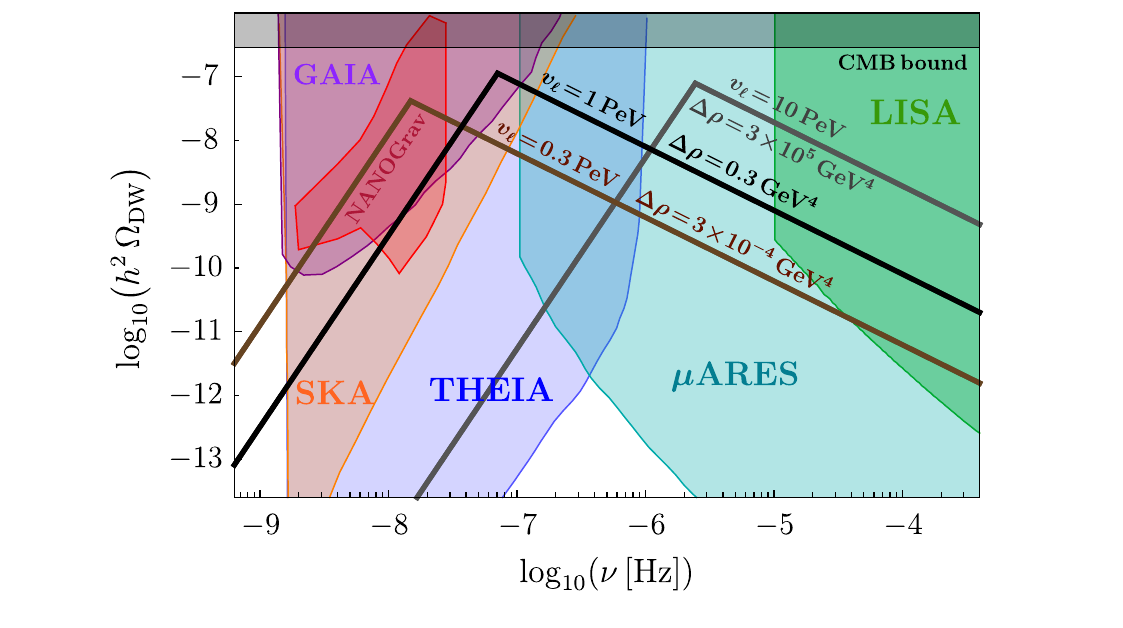} \vspace{-7mm}
\caption{Domain wall annihilation-induced stochastic gravitational wave background  for several  symmetry breaking scales: $0.3 \ \rm PeV$ (brown), $1 \ \rm PeV$ (black), $10 \ \rm PeV$ (gray), and the potential biases as described on the plot. Shaded regions depict the sensitivity of future gravitational wave and astrometry experiments.}\label{fig:four}
\end{figure}

In Figure \ref{fig:four} we included three representative signal curves for the stochastic gravitational wave background arising from domain wall annihilation, assuming the same choice of ${\rm U}(1)_\ell$ breaking scales as for the phase transition case in Figure \ref{fig:two}, i.e., $v_\ell = 0.3 \ \rm PeV$ (brown curve), $1 \ {\rm PeV}$ (black curve) and $10 \ \rm PeV$ (gray curve). The adopted values for the potential bias, as indicated in the figure, are: $\Delta\rho = 3\times 10^{-4} \ \rm GeV^4$, $0.3 \ \rm GeV^4$, and $3\times 10^{5} \ \rm GeV^4$, respectively. As implied by the form of Eq.\,(\ref{dme}), at low frequencies the slope grows $\sim \nu^3$, whereas  for high frequencies it falls $\sim 1/\nu$. 

Overplotted in Figure  \ref{fig:four} as the shaded regions are the sensitivities of several upcoming  experiments relevant for low-frequency gravitational wave search. This includes the planned space-based interferometers LISA \cite{Audley:2017drz} (green) and  $\mu$ARES \cite{Sesana:2019vho} (cyan), the pulsar timing array  SKA \cite{Weltman:2018zrl} (orange), and the astrometry experiments THEIA \cite{Garcia-Bellido:2021zgu} (light blue) and GAIA \cite{Gaia1,Moore:2017ity} (purple). We also indicated the region in which a stochastic gravitational wave signal has been observed by the 
 NANOGrav experiment \cite{NANOGRAV:2018hou}, and we find that the  brown  curve corresponding to  $v_\ell = 0.3 \ \rm PeV$ and $\Delta\rho = 3\times 10^{-4} \ \rm GeV^4$ has the greatest overlap 
with the measured signal region.

\begin{figure}[t!]
\includegraphics[trim={2cm 0.8cm 1cm 0cm},clip,width=9.0cm]{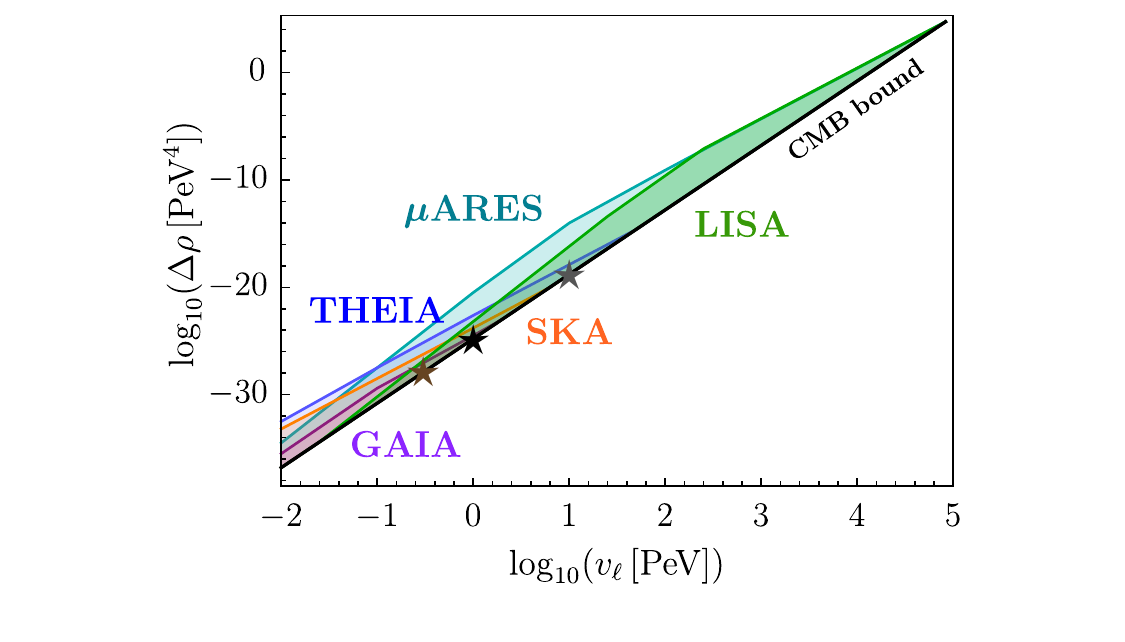} \vspace{-5mm}
\caption{Regions of parameter space $(v_\ell,\Delta\rho)$ with a signal-to-noise ratio of the gravitational wave signature generated by domain wall annihilation greater than five in various experiments. The choice of colors matches that in Figure \ref{fig:four}, and the stars correspond to the three curves for $v_\ell = 0.3 \ \rm PeV$, $1 \ \rm PeV$, $10 \ \rm PeV$.}\label{fig:five}
\end{figure}

Similarly as in the previous case, we performed a scan over $(v_\ell, \Delta\rho)$ and determined the parameter space regions which yield measurable signals in the experiments listed above -- the results are shown in Figure \ref{fig:five}. The lower bound depicted by the black line corresponds to the cosmic microwave background constraint given by Eq.\,(\ref{cmb}). The colors chosen for various experiments correspond to the ones used in Figure \ref{fig:four}. 

We emphasize that the results shown in Figure \ref{fig:five} are model independent and can be applied to any theory with a domain wall gravitational wave signature. Those results are therefore complementary to the model-independent constraints on the domain wall parameters $(v, \Delta\rho)$ derived in \cite{Bosch:2023spa} (see Figure 4 in that reference) derived for higher symmetry breaking scales relevant for DECIGO, BBO, ET, and CE. 

Finally, those constraints can be translated into bounds on the coefficient of the $\mathcal{Z}_2$ breaking dimension five term. For example, assuming $v_S \sim 10 \ \rm PeV$ and $v_\phi \sim 1 \ \rm TeV$ (for the seesaw mechanism this requires $y_\nu \sim 10^{-6}$), the $\mu$ARES experiment will be able to test the parameter region
\bea
{4\times 10^{6} \ \rm PeV }  \lesssim \frac{\Lambda}{\lambda_M}  \lesssim {2\times 10^{11} \ \rm PeV }  \ ,
\eea
providing the opportunity to probe the scale of new physics responsible for the generation of the higher-dimensional term  in Eq.\,(\ref{dim5}) using gravitational waves.

\begin{figure*}[t!]
\includegraphics[trim={0cm 0.0cm 0 0},clip,width=18cm]{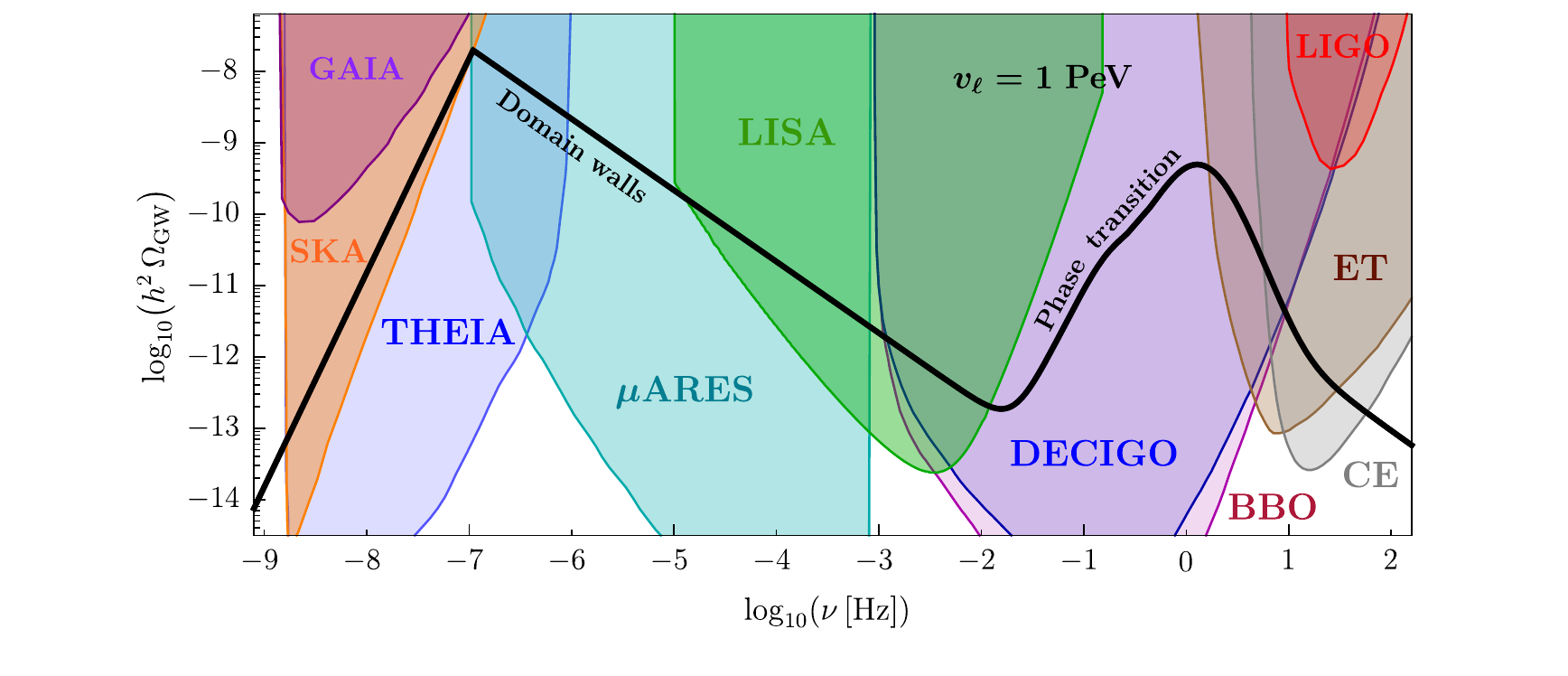} \vspace{-10mm}
\caption{Gravitational wave signature of the model with gauged lepton number broken at the scale $v_\ell = 1 \ \rm PeV$. The domain wall contribution at low frequencies is plotted assuming a potential bias of $\Delta\rho = 0.6 \ \rm GeV^4$, whereas the high frequency first order phase transition part corresponds to the choice of the gauge coupling $g_\ell = 0.22$ and the quartic couplings $\lambda_S = \lambda_\phi = 0.0001$. The sensitivity reach of future  experiments is shown as the shaded regions, which include: GAIA \cite{Gaia1,Moore:2017ity} (purple), SKA \cite{Weltman:2018zrl} (orange),  THEIA \cite{Garcia-Bellido:2021zgu} (light blue), $\mu$ARES \cite{Sesana:2019vho} (cyan), LISA \cite{Audley:2017drz} (green), DECIGO  \cite{Kawamura:2011zz}(dark blue), Big Bang Observer \cite{Crowder:2005nr} (magenta), Einstein Telescope \cite{Punturo:2010zz} (brown), Cosmic Explorer \cite{Reitze:2019iox} (gray), and LIGO after the O5 observing run \cite{LIGOScientific:2014pky} (red).\vspace{5mm}}\label{fig:six}
\end{figure*} 


\section{Gravitational wave spectrum\\ of the model}
\label{signat}

Since the gauged lepton number breaking in the model leads to a first order phase transition with  the formation of domain walls, which subsequently annihilate, this results in  a unique gravitational wave signature involving a domain wall peak at lower frequencies and a phase transition bump at higher frequencies. If  the ${\rm U}(1)_\ell$ breaking occurs  at $\sim  \rm PeV$, this signature can be searched for in most of the upcoming experiments sensitive to gravitational waves. 
This is demonstrated in Figure \ref{fig:six}, which  shows the signal expected from the model if the gauged lepton number symmetry is broken at the scale  $v_\ell = 1 \ \rm PeV$, and assuming the gauge coupling $g_\ell = 0.22$, the quartic couplings $\lambda_S = \lambda_\phi = 10^{-4}$, and the potential bias $\Delta\rho = 0.6 \ \rm GeV^4$. 

As visible in the plot, this signal can be searched for at low frequencies in the upcoming pulsar timing array SKA \cite{Weltman:2018zrl} (orange region), astrometry experiment THEIA \cite{Garcia-Bellido:2021zgu} (light blue region), space-based interferometer $\mu$ARES \cite{Sesana:2019vho} (cyan), throughout intermediate frequencies at gravitational wave  interferometers LISA \cite{Audley:2017drz} (green), DECIGO  \cite{Kawamura:2011zz} (dark blue) and Big Bang Observer \cite{Crowder:2005nr} (magenta), up to high frequencies at the Einstein Telescope \cite{Punturo:2010zz} (brown) and Cosmic Explorer \cite{Reitze:2019iox} (gray). We note that for lower symmetry breaking scales the domain wall signal may be searchable at the astrometry experiment GAIA \cite{Gaia1,Moore:2017ity} (purple), but the phase transition contribution is too weak to be detectable by LIGO \cite{LIGOScientific:2014pky} (red) even  after its O5 observing run.

Given the difference in slope of the domain wall and phase transition contributions, the two can be distinguished. The domain wall spectrum dependence on frequency is $\sim \nu^3$ to the left of the peak and $\sim 1/\nu$  to the right. For the phase transition spectrum the dependence  is $\sim \nu^{2.8}$ at low frequencies where the bubble collision contribution dominates, 
$\sim \nu^3$ just to the left and $\sim 1/\nu^{4}$ just to  the right of the peak from the sound wave contribution, turning into $\sim 1/\nu$ at higher frequencies again from the bubble collision contribution. There is also a nontrivial dependence on frequency  where the two peaks meet, which can be investigated by DECIGO and BBO.

It is quite remarkable that a single symmetry breaking leads to a signature that is searchable both at pulsar timing arrays and at gravitational wave interferometers. In case it is seen in one of the detectors, this scenario will foster collaboration between the different experimental groups, since only their combined efforts would lead to determining the full spectrum of the model.

\section{Summary}\label{sum}

Despite the huge experimental effort to discover beyond-Standard-Model physics at the Large Hadron Collider, in dark matter direct detection experiments, and through indirect detection observations, so far no indisputable evidence of new physics has been found. This prompted particle physics to look for synergies with other areas of physics and expand its search strategies, especially since the guidance from theory is clear -- new particles and phenomena are needed to explain the outstanding open questions, such as the nature of dark matter, the origin of the matter-antimatter asymmetry, or the mechanism behind neutrino masses. One such promising synergy has arisen after the first direct detection of gravitational waves by LIGO. A potential  discovery of a primordial gravitational wave background from the early Universe would certainly  initiate a new golden age for particle physics.

Among processes producing stochastic gravitational wave signals are first order phase transitions and domain wall annihilation in the early Universe. From a theoretical perspective, they arise when one or more symmetries of the theory are spontaneously broken and the vacuum state abruptly changes. The expected shape of the gravitational wave spectrum was determined via simulations and is relatively well understood. Thus far, in the models considered in the literature each gauge  symmetry breaking led to just a single feature in the gravitational wave spectrum which would be within the reach of upcoming experiments. The cases where several measurable features were present in the spectrum required the breaking of two or more gauge groups. 

\break

The uniqueness of the model considered here arises from its prediction of two types of gravitational wave contributions from just a single gauge symmetry breaking. This leads to gravitational wave signatures with a double-peaked spectrum, consisting of a domain wall-induced peak at low frequencies and a peak from phase transitions at high frequencies, with the former searchable mainly in pulsar timing arrays and through astrometry measurements, while  the latter  in interferometers. Since both peaks originate from the breaking of the same gauge group, the domain wall contribution is necessarily shifted to lower frequencies compared to the phase transition one. As mentioned above, this feature differentiates the model from other theories in which, although both such peaks also exist in the spectrum  \cite{Fornal:2023hri,Bosch:2023spa},  two separate gauge symmetries have to be broken for the signature to take this shape. An additional difference is that in these  other models, in contrast to the predictions of the model considered in this paper,  the domain wall peak appears at higher frequencies than the first order phase transition peak.

A natural extension of the model would be to introduce a gauged baryon number symmetry ${\rm U}(1)_B$ broken at a higher energy scale. This would make the model even more attractive, not only accommodating a low-scale seesaw for neutrino masses and a dark matter candidate, but also explaining the matter-antimatter asymmetry through high-scale baryogenesis. This could result in the appearance of an additional feature in the gravitational wave spectrum, such as a cosmic string contribution or an extra domain wall peak at higher frequencies. Apart from that, it would also be intriguing to build an explicit UV completion for the $\mathcal{Z}_2$ breaking dimension-five term, offering additional insight into the particle physics origins  of the domain wall signal.

\subsection*{Acknowledgments}

This research was supported by the National Science Foundation under Grant No. PHY-2213144.

\bibliography{bibliography}

\begin{thebibliography}{146}%
\makeatletter
\providecommand \@ifxundefined [1]{%
 \@ifx{#1\undefined}
}%
\providecommand \@ifnum [1]{%
 \ifnum #1\expandafter \@firstoftwo
 \else \expandafter \@secondoftwo
 \fi
}%
\providecommand \@ifx [1]{%
 \ifx #1\expandafter \@firstoftwo
 \else \expandafter \@secondoftwo
 \fi
}%
\providecommand \natexlab [1]{#1}%
\providecommand \enquote  [1]{#1}%
\providecommand \bibnamefont  [1]{#1}%
\providecommand \bibfnamefont [1]{#1}%
\providecommand \citenamefont [1]{#1}%
\providecommand \href@noop [0]{\@secondoftwo}%
\providecommand \href [0]{\begingroup \@sanitize@url \@href}%
\providecommand \@href[1]{\@@startlink{#1}\@@href}%
\providecommand \@@href[1]{\endgroup#1\@@endlink}%
\providecommand \@sanitize@url [0]{\catcode `\\12\catcode `\$12\catcode
  `\&12\catcode `\#12\catcode `\^12\catcode `\_12\catcode `\%12\relax}%
\providecommand \@@startlink[1]{}%
\providecommand \@@endlink[0]{}%
\providecommand \url  [0]{\begingroup\@sanitize@url \@url }%
\providecommand \@url [1]{\endgroup\@href {#1}{\urlprefix }}%
\providecommand \urlprefix  [0]{URL }%
\providecommand \Eprint [0]{\href }%
\providecommand \doibase [0]{http://dx.doi.org/}%
\providecommand \selectlanguage [0]{\@gobble}%
\providecommand \bibinfo  [0]{\@secondoftwo}%
\providecommand \bibfield  [0]{\@secondoftwo}%
\providecommand \translation [1]{[#1]}%
\providecommand \BibitemOpen [0]{}%
\providecommand \bibitemStop [0]{}%
\providecommand \bibitemNoStop [0]{.\EOS\space}%
\providecommand \EOS [0]{\spacefactor3000\relax}%
\providecommand \BibitemShut  [1]{\csname bibitem#1\endcsname}%
\let\auto@bib@innerbib\@empty
\bibitem [{\citenamefont {Glashow}(1961)}]{Glashow:1961tr}%
  \BibitemOpen
  \bibfield  {author} {\bibinfo {author} {\bibfnamefont {S.~L.}\ \bibnamefont
  {Glashow}},\ }\bibfield  {title} {\enquote {\bibinfo {title} {\emph{Partial
  Symmetries of Weak Interactions}},}\ }\href {\doibase
  10.1016/0029-5582(61)90469-2} {\bibfield  {journal} {\bibinfo  {journal}
  {Nucl. Phys.}\ }\textbf {\bibinfo {volume} {22}},\ \bibinfo {pages}
  {579--588} (\bibinfo {year} {1961})}\BibitemShut {NoStop}%
\bibitem [{\citenamefont {Higgs}(1964)}]{Higgs:1964pj}%
  \BibitemOpen
  \bibfield  {author} {\bibinfo {author} {\bibfnamefont {P.~W.}\ \bibnamefont
  {Higgs}},\ }\bibfield  {title} {\enquote {\bibinfo {title} {\emph{Broken
  Symmetries and the Masses of Gauge Bosons}},}\ }\href {\doibase
  10.1103/PhysRevLett.13.508} {\bibfield  {journal} {\bibinfo  {journal} {Phys.
  Rev. Lett.}\ }\textbf {\bibinfo {volume} {13}},\ \bibinfo {pages} {508--509}
  (\bibinfo {year} {1964})}\BibitemShut {NoStop}%
\bibitem [{\citenamefont {Englert}\ and\ \citenamefont
  {Brout}(1964)}]{Englert:1964et}%
  \BibitemOpen
  \bibfield  {author} {\bibinfo {author} {\bibfnamefont {F.}~\bibnamefont
  {Englert}}\ and\ \bibinfo {author} {\bibfnamefont {R.}~\bibnamefont
  {Brout}},\ }\bibfield  {title} {\enquote {\bibinfo {title} {\emph{Broken
  Symmetry and the Mass of Gauge Vector Mesons}},}\ }\href {\doibase
  10.1103/PhysRevLett.13.321} {\bibfield  {journal} {\bibinfo  {journal} {Phys.
  Rev. Lett.}\ }\textbf {\bibinfo {volume} {13}},\ \bibinfo {pages} {321--323}
  (\bibinfo {year} {1964})}\BibitemShut {NoStop}%
\bibitem [{\citenamefont {Weinberg}(1967)}]{Weinberg:1967tq}%
  \BibitemOpen
  \bibfield  {author} {\bibinfo {author} {\bibfnamefont {S.}~\bibnamefont
  {Weinberg}},\ }\bibfield  {title} {\enquote {\bibinfo {title} {\emph{A Model
  of Leptons}},}\ }\href {\doibase 10.1103/PhysRevLett.19.1264} {\bibfield
  {journal} {\bibinfo  {journal} {Phys. Rev. Lett.}\ }\textbf {\bibinfo
  {volume} {19}},\ \bibinfo {pages} {1264--1266} (\bibinfo {year}
  {1967})}\BibitemShut {NoStop}%
\bibitem [{\citenamefont {Salam}(1968)}]{Salam:1968rm}%
  \BibitemOpen
  \bibfield  {author} {\bibinfo {author} {\bibfnamefont {A.}~\bibnamefont
  {Salam}},\ }\bibfield  {title} {\enquote {\bibinfo {title} {\emph{Weak and
  Electromagnetic Interactions}},}\ }\bibfield  {booktitle} {\emph {\bibinfo
  {booktitle} {{\rm {8th Nobel Symposium Lerum, Sweden, May 19-25, 1968}}}},\
  }\href {\doibase 10.1142/9789812795915_0034} {\bibfield  {journal} {\bibinfo
  {journal} {Conf. Proc. C}\ }\textbf {\bibinfo {volume} {680519}},\ \bibinfo
  {pages} {367--377} (\bibinfo {year} {1968})}\BibitemShut {NoStop}%
\bibitem [{\citenamefont {Fritzsch}\ \emph {et~al.}(1973)\citenamefont
  {Fritzsch}, \citenamefont {Gell-Mann},\ and\ \citenamefont
  {Leutwyler}}]{Fritzsch:1973pi}%
  \BibitemOpen
  \bibfield  {author} {\bibinfo {author} {\bibfnamefont {H.}~\bibnamefont
  {Fritzsch}}, \bibinfo {author} {\bibfnamefont {M.}~\bibnamefont {Gell-Mann}},
  \ and\ \bibinfo {author} {\bibfnamefont {H.}~\bibnamefont {Leutwyler}},\
  }\bibfield  {title} {\enquote {\bibinfo {title} {\emph{Advantages of the
  Color Octet Gluon Picture}},}\ }\href {\doibase 10.1016/0370-2693(73)90625-4}
  {\bibfield  {journal} {\bibinfo  {journal} {Phys. Lett. B}\ }\textbf
  {\bibinfo {volume} {47}},\ \bibinfo {pages} {365--368} (\bibinfo {year}
  {1973})}\BibitemShut {NoStop}%
\bibitem [{\citenamefont {Gross}\ and\ \citenamefont
  {Wilczek}(1973)}]{Gross:1973id}%
  \BibitemOpen
  \bibfield  {author} {\bibinfo {author} {\bibfnamefont {D.~J.}\ \bibnamefont
  {Gross}}\ and\ \bibinfo {author} {\bibfnamefont {F.}~\bibnamefont
  {Wilczek}},\ }\bibfield  {title} {\enquote {\bibinfo {title}
  {\emph{Ultraviolet Behavior of Nonabelian Gauge Theories}},}\ }\href
  {\doibase 10.1103/PhysRevLett.30.1343} {\bibfield  {journal} {\bibinfo
  {journal} {Phys. Rev. Lett.}\ }\textbf {\bibinfo {volume} {30}},\ \bibinfo
  {pages} {1343--1346} (\bibinfo {year} {1973})}\BibitemShut {NoStop}%
\bibitem [{\citenamefont {Politzer}(1973)}]{Politzer:1973fx}%
  \BibitemOpen
  \bibfield  {author} {\bibinfo {author} {\bibfnamefont {H.~D.}\ \bibnamefont
  {Politzer}},\ }\bibfield  {title} {\enquote {\bibinfo {title} {\emph{Reliable
  Perturbative Results for Strong Interactions?}}}\ }\href {\doibase
  10.1103/PhysRevLett.30.1346} {\bibfield  {journal} {\bibinfo  {journal}
  {Phys. Rev. Lett.}\ }\textbf {\bibinfo {volume} {30}},\ \bibinfo {pages}
  {1346--1349} (\bibinfo {year} {1973})}\BibitemShut {NoStop}%
\bibitem [{\citenamefont {{Rubin}}\ and\ \citenamefont
  {{Ford}}(1970)}]{1970ApJ...159..379R}%
  \BibitemOpen
  \bibfield  {author} {\bibinfo {author} {\bibfnamefont {V.~C.}\ \bibnamefont
  {{Rubin}}}\ and\ \bibinfo {author} {\bibfnamefont {Jr.}\ \bibnamefont
  {{Ford}}, \bibfnamefont {W.~K.}},\ }\bibfield  {title} {\enquote {\bibinfo
  {title} {\emph{Rotation of the Andromeda Nebula from a Spectroscopic Survey
  of Emission Regions}},}\ }\href {\doibase 10.1086/150317} {\bibfield
  {journal} {\bibinfo  {journal} {\apj}\ }\textbf {\bibinfo {volume} {159}},\
  \bibinfo {pages} {379} (\bibinfo {year} {1970})}\BibitemShut {NoStop}%
\bibitem [{\citenamefont {de~Bernardis}\ \emph {et~al.}(2000)\citenamefont
  {de~Bernardis} \emph {et~al.}}]{Boomerang:2000efg}%
  \BibitemOpen
  \bibfield  {author} {\bibinfo {author} {\bibfnamefont {P.}~\bibnamefont
  {de~Bernardis}} \emph {et~al.} (\bibinfo {collaboration} {Boomerang}),\
  }\bibfield  {title} {\enquote {\bibinfo {title} {\emph{A Flat Universe from
  High Resolution Maps of the Cosmic Microwave Background Radiation}},}\ }\href
  {\doibase 10.1038/35010035} {\bibfield  {journal} {\bibinfo  {journal}
  {Nature}\ }\textbf {\bibinfo {volume} {404}},\ \bibinfo {pages} {955--959}
  (\bibinfo {year} {2000})},\ \Eprint {http://arxiv.org/abs/astro-ph/0004404}
  {arXiv:astro-ph/0004404} \BibitemShut {NoStop}%
\bibitem [{\citenamefont {Gavazzi}\ \emph {et~al.}(2007)\citenamefont
  {Gavazzi}, \citenamefont {Treu}, \citenamefont {Rhodes}, \citenamefont
  {Koopmans}, \citenamefont {Bolton}, \citenamefont {Burles}, \citenamefont
  {Massey},\ and\ \citenamefont {Moustakas}}]{Gavazzi:2007vw}%
  \BibitemOpen
  \bibfield  {author} {\bibinfo {author} {\bibfnamefont {R.}~\bibnamefont
  {Gavazzi}}, \bibinfo {author} {\bibfnamefont {T.}~\bibnamefont {Treu}},
  \bibinfo {author} {\bibfnamefont {J.~D.}\ \bibnamefont {Rhodes}}, \bibinfo
  {author} {\bibfnamefont {L.~V.}\ \bibnamefont {Koopmans}}, \bibinfo {author}
  {\bibfnamefont {A.~S.}\ \bibnamefont {Bolton}}, \bibinfo {author}
  {\bibfnamefont {S.}~\bibnamefont {Burles}}, \bibinfo {author} {\bibfnamefont
  {R.}~\bibnamefont {Massey}}, \ and\ \bibinfo {author} {\bibfnamefont {L.~A.}\
  \bibnamefont {Moustakas}},\ }\bibfield  {title} {\enquote {\bibinfo {title}
  {\emph{The Sloan Lens ACS Survey. 4. The Mass Density Profile of Early-Type
  Galaxies out to 100 Effective Radii}},}\ }\href {\doibase 10.1086/519237}
  {\bibfield  {journal} {\bibinfo  {journal} {Astrophys. J.}\ }\textbf
  {\bibinfo {volume} {667}},\ \bibinfo {pages} {176--190} (\bibinfo {year}
  {2007})},\ \Eprint {http://arxiv.org/abs/astro-ph/0701589}
  {arXiv:astro-ph/0701589} \BibitemShut {NoStop}%
\bibitem [{\citenamefont {Minkowski}(1977)}]{Minkowski:1977sc}%
  \BibitemOpen
  \bibfield  {author} {\bibinfo {author} {\bibfnamefont {P.}~\bibnamefont
  {Minkowski}},\ }\bibfield  {title} {\enquote {\bibinfo {title} {\emph{$\mu
  \to e\gamma$ at a Rate of One Out of $10^{9}$ Muon Decays?}}}\ }\href
  {\doibase 10.1016/0370-2693(77)90435-X} {\bibfield  {journal} {\bibinfo
  {journal} {Phys. Lett. B}\ }\textbf {\bibinfo {volume} {67}},\ \bibinfo
  {pages} {421--428} (\bibinfo {year} {1977})}\BibitemShut {NoStop}%
\bibitem [{\citenamefont {Gell-Mann}\ \emph {et~al.}()\citenamefont
  {Gell-Mann}, \citenamefont {Ramond},\ and\ \citenamefont {Slansky}}]{seesaw}%
  \BibitemOpen
  \bibfield  {author} {\bibinfo {author} {\bibfnamefont {M.}~\bibnamefont
  {Gell-Mann}}, \bibinfo {author} {\bibfnamefont {P.}~\bibnamefont {Ramond}}, \
  and\ \bibinfo {author} {\bibfnamefont {R.}~\bibnamefont {Slansky}},\
  }\href@noop {} {\bibinfo  {journal} {in \emph{Supergravity}, eds. P. van
  Nieuwenhuizen and D. Freedman, Amsterdam, NL: North Holland. pp. 315-321
  (1979)}\ }\BibitemShut {NoStop}%
\bibitem [{\citenamefont {Yanagida}()}]{seesaw2}%
  \BibitemOpen
\bibfield  {journal} {  }\bibfield  {author} {\bibinfo {author} {\bibfnamefont
  {T.}~\bibnamefont {Yanagida}},\ }\bibfield  {title} {\enquote {\bibinfo
  {title} {\emph{Horizontal gauge symmetry and masses of neutrinos}},}\
  }\href@noop {} {\bibinfo  {journal} {Proceedings: Workshop on the Unified
  Theories and the Baryon Number in the Universe: published in KEK Japan, Conf.
  Proc. C7902131, pp. 95-99 (1979)}\ }\BibitemShut {NoStop}%
\bibitem [{\citenamefont {Mohapatra}\ and\ \citenamefont
  {Senjanovic}(1980)}]{Mohapatra:1979ia}%
  \BibitemOpen
\bibfield  {journal} {  }\bibfield  {author} {\bibinfo {author} {\bibfnamefont
  {R.~N.}\ \bibnamefont {Mohapatra}}\ and\ \bibinfo {author} {\bibfnamefont
  {G.}~\bibnamefont {Senjanovic}},\ }\bibfield  {title} {\enquote {\bibinfo
  {title} {\emph{Neutrino Mass and Spontaneous Parity Nonconservation}},}\
  }\href {\doibase 10.1103/PhysRevLett.44.912} {\bibfield  {journal} {\bibinfo
  {journal} {Phys. Rev. Lett.}\ }\textbf {\bibinfo {volume} {44}},\ \bibinfo
  {pages} {912} (\bibinfo {year} {1980})}\BibitemShut {NoStop}%
\bibitem [{\citenamefont {Konetschny}\ and\ \citenamefont
  {Kummer}(1977)}]{Konetschny:1977bn}%
  \BibitemOpen
  \bibfield  {author} {\bibinfo {author} {\bibfnamefont {W.}~\bibnamefont
  {Konetschny}}\ and\ \bibinfo {author} {\bibfnamefont {W.}~\bibnamefont
  {Kummer}},\ }\bibfield  {title} {\enquote {\bibinfo {title}
  {\emph{Nonconservation of Total Lepton Number with Scalar Bosons}},}\ }\href
  {\doibase 10.1016/0370-2693(77)90407-5} {\bibfield  {journal} {\bibinfo
  {journal} {Phys. Lett. B}\ }\textbf {\bibinfo {volume} {70}},\ \bibinfo
  {pages} {433--435} (\bibinfo {year} {1977})}\BibitemShut {NoStop}%
\bibitem [{\citenamefont {Schechter}\ and\ \citenamefont
  {Valle}(1980)}]{Schechter:1980gr}%
  \BibitemOpen
  \bibfield  {author} {\bibinfo {author} {\bibfnamefont {J.}~\bibnamefont
  {Schechter}}\ and\ \bibinfo {author} {\bibfnamefont {J.~W.~F.}\ \bibnamefont
  {Valle}},\ }\bibfield  {title} {\enquote {\bibinfo {title} {\emph{Neutrino
  Masses in SU(2) $\times$ U(1) Theories}},}\ }\href {\doibase
  10.1103/PhysRevD.22.2227} {\bibfield  {journal} {\bibinfo  {journal} {Phys.
  Rev. D}\ }\textbf {\bibinfo {volume} {22}},\ \bibinfo {pages} {2227}
  (\bibinfo {year} {1980})}\BibitemShut {NoStop}%
\bibitem [{\citenamefont {{Cheng, T. P. and Li,
  Ling-Fong}}({1980})}]{PhysRevD.22.2860}%
  \BibitemOpen
  \bibfield  {author} {\bibinfo {author} {\bibnamefont {{Cheng, T. P. and Li,
  Ling-Fong}}},\ }\bibfield  {title} {\enquote {\bibinfo {title}
  {\emph{Neutrino Masses, Mixings, and Oscillations in SU(2) $\times$ U(1)
  Models of Electroweak Interactions}},}\ }\href {\doibase
  10.1103/PhysRevD.22.2860} {\bibfield  {journal} {\bibinfo  {journal} {{Phys.
  Rev. D}}\ }\textbf {\bibinfo {volume} {{22}}},\ \bibinfo {pages}
  {{2860--2868}} (\bibinfo {year} {{1980}})}\BibitemShut {NoStop}%
\bibitem [{\citenamefont {Mohapatra}\ and\ \citenamefont
  {Senjanovic}(1981)}]{Mohapatra:1980yp}%
  \BibitemOpen
  \bibfield  {author} {\bibinfo {author} {\bibfnamefont {R.~N.}\ \bibnamefont
  {Mohapatra}}\ and\ \bibinfo {author} {\bibfnamefont {G.}~\bibnamefont
  {Senjanovic}},\ }\bibfield  {title} {\enquote {\bibinfo {title}
  {\emph{Neutrino Masses and Mixings in Gauge Models with Spontaneous Parity
  Violation}},}\ }\href {\doibase 10.1103/PhysRevD.23.165} {\bibfield
  {journal} {\bibinfo  {journal} {Phys. Rev. D}\ }\textbf {\bibinfo {volume}
  {23}},\ \bibinfo {pages} {165} (\bibinfo {year} {1981})}\BibitemShut
  {NoStop}%
\bibitem [{\citenamefont {Foot}\ \emph
  {et~al.}(1989{\natexlab{a}})\citenamefont {Foot}, \citenamefont {Lew},
  \citenamefont {He},\ and\ \citenamefont {Joshi}}]{Foot:1988aq}%
  \BibitemOpen
  \bibfield  {author} {\bibinfo {author} {\bibfnamefont {R.}~\bibnamefont
  {Foot}}, \bibinfo {author} {\bibfnamefont {H.}~\bibnamefont {Lew}}, \bibinfo
  {author} {\bibfnamefont {X.~G.}\ \bibnamefont {He}}, \ and\ \bibinfo {author}
  {\bibfnamefont {G.~C.}\ \bibnamefont {Joshi}},\ }\bibfield  {title} {\enquote
  {\bibinfo {title} {\emph{Seesaw Neutrino Masses Induced by a Triplet of
  Leptons}},}\ }\href {\doibase 10.1007/BF01415558} {\bibfield  {journal}
  {\bibinfo  {journal} {Z. Phys. C}\ }\textbf {\bibinfo {volume} {44}},\
  \bibinfo {pages} {441} (\bibinfo {year} {1989}{\natexlab{a}})}\BibitemShut
  {NoStop}%
\bibitem [{\citenamefont {Zee}(1980)}]{Zee:1980ai}%
  \BibitemOpen
  \bibfield  {author} {\bibinfo {author} {\bibfnamefont {A.}~\bibnamefont
  {Zee}},\ }\bibfield  {title} {\enquote {\bibinfo {title} {\emph{A Theory of
  Lepton Number Violation, Neutrino Majorana Mass, and Oscillation}},}\ }\href
  {\doibase 10.1016/0370-2693(80)90349-4} {\bibfield  {journal} {\bibinfo
  {journal} {Phys. Lett. B}\ }\textbf {\bibinfo {volume} {93}},\ \bibinfo
  {pages} {389} (\bibinfo {year} {1980})},\ \bibinfo {note} {[Erratum:
  Phys.Lett.B 95, 461 (1980)]}\BibitemShut {NoStop}%
\bibitem [{\citenamefont {Pais}(1973)}]{Pais:1973mi}%
  \BibitemOpen
  \bibfield  {author} {\bibinfo {author} {\bibfnamefont {A.}~\bibnamefont
  {Pais}},\ }\bibfield  {title} {\enquote {\bibinfo {title} {\emph{Remark on
  Baryon Conservation}},}\ }\href {\doibase 10.1103/PhysRevD.8.1844} {\bibfield
   {journal} {\bibinfo  {journal} {Phys. Rev. D}\ }\textbf {\bibinfo {volume}
  {8}},\ \bibinfo {pages} {1844--1846} (\bibinfo {year} {1973})}\BibitemShut
  {NoStop}%
\bibitem [{\citenamefont {Rajpoot}(1988)}]{Rajpoot:1987yg}%
  \BibitemOpen
  \bibfield  {author} {\bibinfo {author} {\bibfnamefont {S.}~\bibnamefont
  {Rajpoot}},\ }\bibfield  {title} {\enquote {\bibinfo {title} {\emph{Gauge
  Symmetries of Electroweak Interactions}},}\ }\href {\doibase
  10.1007/BF00669312} {\bibfield  {journal} {\bibinfo  {journal} {Int. J.
  Theor. Phys.}\ }\textbf {\bibinfo {volume} {27}},\ \bibinfo {pages} {689}
  (\bibinfo {year} {1988})}\BibitemShut {NoStop}%
\bibitem [{\citenamefont {Foot}\ \emph
  {et~al.}(1989{\natexlab{b}})\citenamefont {Foot}, \citenamefont {Joshi},\
  and\ \citenamefont {Lew}}]{Foot:1989ts}%
  \BibitemOpen
  \bibfield  {author} {\bibinfo {author} {\bibfnamefont {R.}~\bibnamefont
  {Foot}}, \bibinfo {author} {\bibfnamefont {G.~C.}\ \bibnamefont {Joshi}}, \
  and\ \bibinfo {author} {\bibfnamefont {H.}~\bibnamefont {Lew}},\ }\bibfield
  {title} {\enquote {\bibinfo {title} {\emph{Gauged Baryon and Lepton
  Numbers}},}\ }\href {\doibase 10.1103/PhysRevD.40.2487} {\bibfield  {journal}
  {\bibinfo  {journal} {Phys. Rev. D}\ }\textbf {\bibinfo {volume} {40}},\
  \bibinfo {pages} {2487--2489} (\bibinfo {year}
  {1989}{\natexlab{b}})}\BibitemShut {NoStop}%
\bibitem [{\citenamefont {Carone}\ and\ \citenamefont
  {Murayama}(1995)}]{Carone:1995pu}%
  \BibitemOpen
  \bibfield  {author} {\bibinfo {author} {\bibfnamefont {C.~D.}\ \bibnamefont
  {Carone}}\ and\ \bibinfo {author} {\bibfnamefont {H.}~\bibnamefont
  {Murayama}},\ }\bibfield  {title} {\enquote {\bibinfo {title}
  {\emph{Realistic Models with a Light U(1) Gauge Boson Coupled to Baryon
  Number}},}\ }\href {\doibase 10.1103/PhysRevD.52.484} {\bibfield  {journal}
  {\bibinfo  {journal} {Phys. Rev. D}\ }\textbf {\bibinfo {volume} {52}},\
  \bibinfo {pages} {484--493} (\bibinfo {year} {1995})},\ \Eprint
  {http://arxiv.org/abs/hep-ph/9501220} {arXiv:hep-ph/9501220} \BibitemShut
  {NoStop}%
\bibitem [{\citenamefont {Georgi}\ and\ \citenamefont
  {Glashow}(1996)}]{Georgi:1996ei}%
  \BibitemOpen
  \bibfield  {author} {\bibinfo {author} {\bibfnamefont {H.}~\bibnamefont
  {Georgi}}\ and\ \bibinfo {author} {\bibfnamefont {S.~L.}\ \bibnamefont
  {Glashow}},\ }\bibfield  {title} {\enquote {\bibinfo {title} {\emph{Decays of
  a Leptophobic Gauge Boson}},}\ }\href {\doibase 10.1016/0370-2693(96)00997-5}
  {\bibfield  {journal} {\bibinfo  {journal} {Phys. Lett. B}\ }\textbf
  {\bibinfo {volume} {387}},\ \bibinfo {pages} {341--345} (\bibinfo {year}
  {1996})},\ \Eprint {http://arxiv.org/abs/hep-ph/9607202}
  {arXiv:hep-ph/9607202} \BibitemShut {NoStop}%
\bibitem [{\citenamefont {Fileviez~Perez}\ and\ \citenamefont
  {Wise}(2010)}]{FileviezPerez:2010gw}%
  \BibitemOpen
  \bibfield  {author} {\bibinfo {author} {\bibfnamefont {P.}~\bibnamefont
  {Fileviez~Perez}}\ and\ \bibinfo {author} {\bibfnamefont {M.~B.}\
  \bibnamefont {Wise}},\ }\bibfield  {title} {\enquote {\bibinfo {title}
  {\emph{Baryon and Lepton Number as Local Gauge Symmetries}},}\ }\href
  {\doibase 10.1103/PhysRevD.82.079901} {\bibfield  {journal} {\bibinfo
  {journal} {Phys. Rev. D}\ }\textbf {\bibinfo {volume} {82}},\ \bibinfo
  {pages} {011901} (\bibinfo {year} {2010})},\ \bibinfo {note} {[Erratum:
  Phys.Rev.D 82, 079901 (2010)]},\ \Eprint {http://arxiv.org/abs/1002.1754}
  {arXiv:1002.1754 [hep-ph]} \BibitemShut {NoStop}%
\bibitem [{\citenamefont {Duerr}\ \emph {et~al.}(2013)\citenamefont {Duerr},
  \citenamefont {Fileviez~Perez},\ and\ \citenamefont {Wise}}]{Duerr:2013dza}%
  \BibitemOpen
  \bibfield  {author} {\bibinfo {author} {\bibfnamefont {M.}~\bibnamefont
  {Duerr}}, \bibinfo {author} {\bibfnamefont {P.}~\bibnamefont
  {Fileviez~Perez}}, \ and\ \bibinfo {author} {\bibfnamefont {M.~B.}\
  \bibnamefont {Wise}},\ }\bibfield  {title} {\enquote {\bibinfo {title}
  {\emph{Gauge Theory for Baryon and Lepton Numbers with Leptoquarks}},}\
  }\href {\doibase 10.1103/PhysRevLett.110.231801} {\bibfield  {journal}
  {\bibinfo  {journal} {Phys. Rev. Lett.}\ }\textbf {\bibinfo {volume} {110}},\
  \bibinfo {pages} {231801} (\bibinfo {year} {2013})},\ \Eprint
  {http://arxiv.org/abs/1304.0576} {arXiv:1304.0576 [hep-ph]} \BibitemShut
  {NoStop}%
\bibitem [{\citenamefont {Fileviez~Perez}\ \emph {et~al.}(2014)\citenamefont
  {Fileviez~Perez}, \citenamefont {Ohmer},\ and\ \citenamefont
  {Patel}}]{Perez:2014qfa}%
  \BibitemOpen
  \bibfield  {author} {\bibinfo {author} {\bibfnamefont {P.}~\bibnamefont
  {Fileviez~Perez}}, \bibinfo {author} {\bibfnamefont {S.}~\bibnamefont
  {Ohmer}}, \ and\ \bibinfo {author} {\bibfnamefont {H.~H.}\ \bibnamefont
  {Patel}},\ }\bibfield  {title} {\enquote {\bibinfo {title} {\emph{Minimal
  Theory for Lepto-Baryons}},}\ }\href {\doibase
  10.1016/j.physletb.2014.06.057} {\bibfield  {journal} {\bibinfo  {journal}
  {Phys. Lett. B}\ }\textbf {\bibinfo {volume} {735}},\ \bibinfo {pages}
  {283--287} (\bibinfo {year} {2014})},\ \Eprint
  {http://arxiv.org/abs/1403.8029} {arXiv:1403.8029 [hep-ph]} \BibitemShut
  {NoStop}%
\bibitem [{\citenamefont {Arnold}\ \emph {et~al.}(2013)\citenamefont {Arnold},
  \citenamefont {Fileviez~Perez}, \citenamefont {Fornal},\ and\ \citenamefont
  {Spinner}}]{Arnold:2013qja}%
  \BibitemOpen
  \bibfield  {author} {\bibinfo {author} {\bibfnamefont {J.~M.}\ \bibnamefont
  {Arnold}}, \bibinfo {author} {\bibfnamefont {P.}~\bibnamefont
  {Fileviez~Perez}}, \bibinfo {author} {\bibfnamefont {B.}~\bibnamefont
  {Fornal}}, \ and\ \bibinfo {author} {\bibfnamefont {S.}~\bibnamefont
  {Spinner}},\ }\bibfield  {title} {\enquote {\bibinfo {title} {\emph{B and L
  at the Supersymmetry Scale, Dark Matter, and R-Parity Violation}},}\ }\href
  {\doibase 10.1103/PhysRevD.88.115009} {\bibfield  {journal} {\bibinfo
  {journal} {Phys. Rev. D}\ }\textbf {\bibinfo {volume} {88}},\ \bibinfo
  {pages} {115009} (\bibinfo {year} {2013})},\ \Eprint
  {http://arxiv.org/abs/1310.7052} {arXiv:1310.7052 [hep-ph]} \BibitemShut
  {NoStop}%
\bibitem [{\citenamefont {Fornal}\ \emph {et~al.}(2015)\citenamefont {Fornal},
  \citenamefont {Rajaraman},\ and\ \citenamefont {Tait}}]{Fornal:2015boa}%
  \BibitemOpen
  \bibfield  {author} {\bibinfo {author} {\bibfnamefont {B.}~\bibnamefont
  {Fornal}}, \bibinfo {author} {\bibfnamefont {A.}~\bibnamefont {Rajaraman}}, \
  and\ \bibinfo {author} {\bibfnamefont {T.~M.~P.}\ \bibnamefont {Tait}},\
  }\bibfield  {title} {\enquote {\bibinfo {title} {\emph{Baryon Number as the
  Fourth Color}},}\ }\href {\doibase 10.1103/PhysRevD.92.055022} {\bibfield
  {journal} {\bibinfo  {journal} {Phys. Rev. D}\ }\textbf {\bibinfo {volume}
  {92}},\ \bibinfo {pages} {055022} (\bibinfo {year} {2015})},\ \Eprint
  {http://arxiv.org/abs/1506.06131} {arXiv:1506.06131 [hep-ph]} \BibitemShut
  {NoStop}%
\bibitem [{\citenamefont {Fornal}\ and\ \citenamefont
  {Tait}(2016)}]{Fornal:2015one}%
  \BibitemOpen
  \bibfield  {author} {\bibinfo {author} {\bibfnamefont {B.}~\bibnamefont
  {Fornal}}\ and\ \bibinfo {author} {\bibfnamefont {T.~M.~P.}\ \bibnamefont
  {Tait}},\ }\bibfield  {title} {\enquote {\bibinfo {title} {\emph{Dark Matter
  from Unification of Color and Baryon Number}},}\ }\href {\doibase
  10.1103/PhysRevD.93.075010} {\bibfield  {journal} {\bibinfo  {journal} {Phys.
  Rev. D}\ }\textbf {\bibinfo {volume} {93}},\ \bibinfo {pages} {075010}
  (\bibinfo {year} {2016})},\ \Eprint {http://arxiv.org/abs/1511.07380}
  {arXiv:1511.07380 [hep-ph]} \BibitemShut {NoStop}%
\bibitem [{\citenamefont {Fornal}\ \emph {et~al.}(2017)\citenamefont {Fornal},
  \citenamefont {Shirman}, \citenamefont {Tait},\ and\ \citenamefont
  {West}}]{Fornal:2017owa}%
  \BibitemOpen
  \bibfield  {author} {\bibinfo {author} {\bibfnamefont {B.}~\bibnamefont
  {Fornal}}, \bibinfo {author} {\bibfnamefont {Y.}~\bibnamefont {Shirman}},
  \bibinfo {author} {\bibfnamefont {T.~M.~P.}\ \bibnamefont {Tait}}, \ and\
  \bibinfo {author} {\bibfnamefont {J.~Rittenhouse}\ \bibnamefont {West}},\
  }\bibfield  {title} {\enquote {\bibinfo {title} {\emph{Asymmetric Dark Matter
  and Baryogenesis from $SU(2)_{\ell}$}},}\ }\href {\doibase
  10.1103/PhysRevD.96.035001} {\bibfield  {journal} {\bibinfo  {journal} {Phys.
  Rev. D}\ }\textbf {\bibinfo {volume} {96}},\ \bibinfo {pages} {035001}
  (\bibinfo {year} {2017})},\ \Eprint {http://arxiv.org/abs/1703.00199}
  {arXiv:1703.00199 [hep-ph]} \BibitemShut {NoStop}%
\bibitem [{\citenamefont {Duerr}\ and\ \citenamefont
  {Fileviez~Perez}(2015)}]{Duerr:2014wra}%
  \BibitemOpen
  \bibfield  {author} {\bibinfo {author} {\bibfnamefont {M.}~\bibnamefont
  {Duerr}}\ and\ \bibinfo {author} {\bibfnamefont {P.}~\bibnamefont
  {Fileviez~Perez}},\ }\bibfield  {title} {\enquote {\bibinfo {title}
  {\emph{Theory for Baryon Number and Dark Matter at the LHC}},}\ }\href
  {\doibase 10.1103/PhysRevD.91.095001} {\bibfield  {journal} {\bibinfo
  {journal} {Phys. Rev. D}\ }\textbf {\bibinfo {volume} {91}},\ \bibinfo
  {pages} {095001} (\bibinfo {year} {2015})},\ \Eprint
  {http://arxiv.org/abs/1409.8165} {arXiv:1409.8165 [hep-ph]} \BibitemShut
  {NoStop}%
\bibitem [{\citenamefont {Ohmer}\ and\ \citenamefont
  {Patel}(2015)}]{Ohmer:2015lxa}%
  \BibitemOpen
  \bibfield  {author} {\bibinfo {author} {\bibfnamefont {S.}~\bibnamefont
  {Ohmer}}\ and\ \bibinfo {author} {\bibfnamefont {H.~H.}\ \bibnamefont
  {Patel}},\ }\bibfield  {title} {\enquote {\bibinfo {title}
  {\emph{Leptobaryons as Majorana Dark Matter}},}\ }\href {\doibase
  10.1103/PhysRevD.92.055020} {\bibfield  {journal} {\bibinfo  {journal} {Phys.
  Rev. D}\ }\textbf {\bibinfo {volume} {92}},\ \bibinfo {pages} {055020}
  (\bibinfo {year} {2015})},\ \Eprint {http://arxiv.org/abs/1506.00954}
  {arXiv:1506.00954 [hep-ph]} \BibitemShut {NoStop}%
\bibitem [{\citenamefont {Fileviez~Perez}\ \emph {et~al.}(2019)\citenamefont
  {Fileviez~Perez}, \citenamefont {Golias}, \citenamefont {Li},\ and\
  \citenamefont {Murgui}}]{FileviezPerez:2018jmr}%
  \BibitemOpen
  \bibfield  {author} {\bibinfo {author} {\bibfnamefont {P.}~\bibnamefont
  {Fileviez~Perez}}, \bibinfo {author} {\bibfnamefont {E.}~\bibnamefont
  {Golias}}, \bibinfo {author} {\bibfnamefont {R.-H.}\ \bibnamefont {Li}}, \
  and\ \bibinfo {author} {\bibfnamefont {C.}~\bibnamefont {Murgui}},\
  }\bibfield  {title} {\enquote {\bibinfo {title} {\emph{Leptophobic Dark
  Matter and the Baryon Number Violation Scale}},}\ }\href {\doibase
  10.1103/PhysRevD.99.035009} {\bibfield  {journal} {\bibinfo  {journal} {Phys.
  Rev. D}\ }\textbf {\bibinfo {volume} {99}},\ \bibinfo {pages} {035009}
  (\bibinfo {year} {2019})},\ \Eprint {http://arxiv.org/abs/1810.06646}
  {arXiv:1810.06646 [hep-ph]} \BibitemShut {NoStop}%
\bibitem [{\citenamefont {Fileviez~Perez}\ \emph {et~al.}(2021)\citenamefont
  {Fileviez~Perez}, \citenamefont {Murgui},\ and\ \citenamefont
  {Plascencia}}]{FileviezPerez:2021hbc}%
  \BibitemOpen
  \bibfield  {author} {\bibinfo {author} {\bibfnamefont {P.}~\bibnamefont
  {Fileviez~Perez}}, \bibinfo {author} {\bibfnamefont {C.}~\bibnamefont
  {Murgui}}, \ and\ \bibinfo {author} {\bibfnamefont {A.~D.}\ \bibnamefont
  {Plascencia}},\ }\bibfield  {title} {\enquote {\bibinfo {title}
  {\emph{Baryogenesis via Leptogenesis: Spontaneous B and L Violation}},}\
  }\href {\doibase 10.1103/PhysRevD.104.055007} {\bibfield  {journal} {\bibinfo
   {journal} {Phys. Rev. D}\ }\textbf {\bibinfo {volume} {104}},\ \bibinfo
  {pages} {055007} (\bibinfo {year} {2021})},\ \Eprint
  {http://arxiv.org/abs/2103.13397} {arXiv:2103.13397 [hep-ph]} \BibitemShut
  {NoStop}%
\bibitem [{\citenamefont {Abbott}\ \emph {et~al.}(2016)\citenamefont {Abbott}
  \emph {et~al.}}]{LIGOScientific:2016aoc}%
  \BibitemOpen
  \bibfield  {author} {\bibinfo {author} {\bibfnamefont {B.~P.}\ \bibnamefont
  {Abbott}} \emph {et~al.} (\bibinfo {collaboration} {LIGO Scientific,
  Virgo}),\ }\bibfield  {title} {\enquote {\bibinfo {title} {\emph{Observation
  of Gravitational Waves from a Binary Black Hole Merger}},}\ }\href {\doibase
  10.1103/PhysRevLett.116.061102} {\bibfield  {journal} {\bibinfo  {journal}
  {Phys. Rev. Lett.}\ }\textbf {\bibinfo {volume} {116}},\ \bibinfo {pages}
  {061102} (\bibinfo {year} {2016})},\ \Eprint
  {http://arxiv.org/abs/1602.03837} {arXiv:1602.03837 [gr-qc]} \BibitemShut
  {NoStop}%
\bibitem [{\citenamefont {Kosowsky}\ \emph {et~al.}(1992)\citenamefont
  {Kosowsky}, \citenamefont {Turner},\ and\ \citenamefont
  {Watkins}}]{Kosowsky:1991ua}%
  \BibitemOpen
  \bibfield  {author} {\bibinfo {author} {\bibfnamefont {A.}~\bibnamefont
  {Kosowsky}}, \bibinfo {author} {\bibfnamefont {M.~S.}\ \bibnamefont
  {Turner}}, \ and\ \bibinfo {author} {\bibfnamefont {R.}~\bibnamefont
  {Watkins}},\ }\bibfield  {title} {\enquote {\bibinfo {title}
  {\emph{Gravitational Radiation from Colliding Vacuum Bubbles}},}\ }\href
  {\doibase 10.1103/PhysRevD.45.4514} {\bibfield  {journal} {\bibinfo
  {journal} {Phys. Rev. D}\ }\textbf {\bibinfo {volume} {45}},\ \bibinfo
  {pages} {4514--4535} (\bibinfo {year} {1992})}\BibitemShut {NoStop}%
\bibitem [{\citenamefont {Vachaspati}\ and\ \citenamefont
  {Vilenkin}(1985)}]{Vachaspati:1984gt}%
  \BibitemOpen
  \bibfield  {author} {\bibinfo {author} {\bibfnamefont {T.}~\bibnamefont
  {Vachaspati}}\ and\ \bibinfo {author} {\bibfnamefont {A.}~\bibnamefont
  {Vilenkin}},\ }\bibfield  {title} {\enquote {\bibinfo {title}
  {\emph{Gravitational Radiation from Cosmic Strings}},}\ }\href {\doibase
  10.1103/PhysRevD.31.3052} {\bibfield  {journal} {\bibinfo  {journal} {Phys.
  Rev. D}\ }\textbf {\bibinfo {volume} {31}},\ \bibinfo {pages} {3052}
  (\bibinfo {year} {1985})}\BibitemShut {NoStop}%
\bibitem [{\citenamefont {Sakellariadou}(1990)}]{Sakellariadou:1990ne}%
  \BibitemOpen
  \bibfield  {author} {\bibinfo {author} {\bibfnamefont {M.}~\bibnamefont
  {Sakellariadou}},\ }\bibfield  {title} {\enquote {\bibinfo {title}
  {\emph{Gravitational Waves Emitted from Infinite Strings}},}\ }\href
  {\doibase 10.1103/PhysRevD.42.354} {\bibfield  {journal} {\bibinfo  {journal}
  {Phys. Rev. D}\ }\textbf {\bibinfo {volume} {42}},\ \bibinfo {pages}
  {354--360} (\bibinfo {year} {1990})},\ \bibinfo {note} {[Erratum: Phys. Rev.
  D 43, 4150 (1991)]}\BibitemShut {NoStop}%
\bibitem [{\citenamefont {Hiramatsu}\ \emph {et~al.}(2010)\citenamefont
  {Hiramatsu}, \citenamefont {Kawasaki},\ and\ \citenamefont
  {Saikawa}}]{Hiramatsu:2010yz}%
  \BibitemOpen
  \bibfield  {author} {\bibinfo {author} {\bibfnamefont {T.}~\bibnamefont
  {Hiramatsu}}, \bibinfo {author} {\bibfnamefont {M.}~\bibnamefont {Kawasaki}},
  \ and\ \bibinfo {author} {\bibfnamefont {K.}~\bibnamefont {Saikawa}},\
  }\bibfield  {title} {\enquote {\bibinfo {title} {\emph{Gravitational Waves
  from Collapsing Domain Walls}},}\ }\href {\doibase
  10.1088/1475-7516/2010/05/032} {\bibfield  {journal} {\bibinfo  {journal}
  {JCAP}\ }\textbf {\bibinfo {volume} {05}},\ \bibinfo {pages} {032} (\bibinfo
  {year} {2010})},\ \Eprint {http://arxiv.org/abs/1002.1555} {arXiv:1002.1555
  [astro-ph.CO]} \BibitemShut {NoStop}%
\bibitem [{\citenamefont {Turner}(1997)}]{Turner:1996ck}%
  \BibitemOpen
  \bibfield  {author} {\bibinfo {author} {\bibfnamefont {M.~S.}\ \bibnamefont
  {Turner}},\ }\bibfield  {title} {\enquote {\bibinfo {title}
  {\emph{Detectability of Inflation Produced Gravitational Waves}},}\ }\href
  {\doibase 10.1103/PhysRevD.55.R435} {\bibfield  {journal} {\bibinfo
  {journal} {Phys. Rev. D}\ }\textbf {\bibinfo {volume} {55}},\ \bibinfo
  {pages} {R435--R439} (\bibinfo {year} {1997})},\ \Eprint
  {http://arxiv.org/abs/astro-ph/9607066} {arXiv:astro-ph/9607066} \BibitemShut
  {NoStop}%
\bibitem [{\citenamefont {Grojean}\ and\ \citenamefont
  {Servant}(2007)}]{Grojean:2006bp}%
  \BibitemOpen
  \bibfield  {author} {\bibinfo {author} {\bibfnamefont {C.}~\bibnamefont
  {Grojean}}\ and\ \bibinfo {author} {\bibfnamefont {G.}~\bibnamefont
  {Servant}},\ }\bibfield  {title} {\enquote {\bibinfo {title}
  {\emph{Gravitational Waves from Phase Transitions at the Electroweak Scale
  and Beyond}},}\ }\href {\doibase 10.1103/PhysRevD.75.043507} {\bibfield
  {journal} {\bibinfo  {journal} {Phys. Rev. D}\ }\textbf {\bibinfo {volume}
  {75}},\ \bibinfo {pages} {043507} (\bibinfo {year} {2007})},\ \Eprint
  {http://arxiv.org/abs/hep-ph/0607107} {arXiv:hep-ph/0607107} \BibitemShut
  {NoStop}%
\bibitem [{\citenamefont {Schwaller}(2015)}]{Schwaller:2015tja}%
  \BibitemOpen
  \bibfield  {author} {\bibinfo {author} {\bibfnamefont {P.}~\bibnamefont
  {Schwaller}},\ }\bibfield  {title} {\enquote {\bibinfo {title}
  {\emph{Gravitational Waves from a Dark Phase Transition}},}\ }\href {\doibase
  10.1103/PhysRevLett.115.181101} {\bibfield  {journal} {\bibinfo  {journal}
  {Phys. Rev. Lett.}\ }\textbf {\bibinfo {volume} {115}},\ \bibinfo {pages}
  {181101} (\bibinfo {year} {2015})},\ \Eprint
  {http://arxiv.org/abs/1504.07263} {arXiv:1504.07263 [hep-ph]} \BibitemShut
  {NoStop}%
\bibitem [{\citenamefont {Vaskonen}(2017)}]{Vaskonen:2016yiu}%
  \BibitemOpen
  \bibfield  {author} {\bibinfo {author} {\bibfnamefont {V.}~\bibnamefont
  {Vaskonen}},\ }\bibfield  {title} {\enquote {\bibinfo {title}
  {\emph{Electroweak Baryogenesis and Gravitational Waves from a Real Scalar
  Singlet}},}\ }\href {\doibase 10.1103/PhysRevD.95.123515} {\bibfield
  {journal} {\bibinfo  {journal} {Phys. Rev. D}\ }\textbf {\bibinfo {volume}
  {95}},\ \bibinfo {pages} {123515} (\bibinfo {year} {2017})},\ \Eprint
  {http://arxiv.org/abs/1611.02073} {arXiv:1611.02073 [hep-ph]} \BibitemShut
  {NoStop}%
\bibitem [{\citenamefont {Dorsch}\ \emph {et~al.}(2017)\citenamefont {Dorsch},
  \citenamefont {Huber}, \citenamefont {Konstandin},\ and\ \citenamefont
  {No}}]{Dorsch:2016nrg}%
  \BibitemOpen
  \bibfield  {author} {\bibinfo {author} {\bibfnamefont {G.~C.}\ \bibnamefont
  {Dorsch}}, \bibinfo {author} {\bibfnamefont {S.~J.}\ \bibnamefont {Huber}},
  \bibinfo {author} {\bibfnamefont {T.}~\bibnamefont {Konstandin}}, \ and\
  \bibinfo {author} {\bibfnamefont {J.~M.}\ \bibnamefont {No}},\ }\bibfield
  {title} {\enquote {\bibinfo {title} {\emph{A Second Higgs Doublet in the
  Early Universe: Baryogenesis and Gravitational Waves}},}\ }\href {\doibase
  10.1088/1475-7516/2017/05/052} {\bibfield  {journal} {\bibinfo  {journal}
  {JCAP}\ }\textbf {\bibinfo {volume} {05}},\ \bibinfo {pages} {052} (\bibinfo
  {year} {2017})},\ \Eprint {http://arxiv.org/abs/1611.05874} {arXiv:1611.05874
  [hep-ph]} \BibitemShut {NoStop}%
\bibitem [{\citenamefont {Baldes}(2017)}]{Baldes:2017rcu}%
  \BibitemOpen
  \bibfield  {author} {\bibinfo {author} {\bibfnamefont {I.}~\bibnamefont
  {Baldes}},\ }\bibfield  {title} {\enquote {\bibinfo {title}
  {\emph{Gravitational Waves from the Asymmetric-Dark-Matter Generating Phase
  Transition}},}\ }\href {\doibase 10.1088/1475-7516/2017/05/028} {\bibfield
  {journal} {\bibinfo  {journal} {JCAP}\ }\textbf {\bibinfo {volume} {05}},\
  \bibinfo {pages} {028} (\bibinfo {year} {2017})},\ \Eprint
  {http://arxiv.org/abs/1702.02117} {arXiv:1702.02117 [hep-ph]} \BibitemShut
  {NoStop}%
\bibitem [{\citenamefont {Bernon}\ \emph {et~al.}(2018)\citenamefont {Bernon},
  \citenamefont {Bian},\ and\ \citenamefont {Jiang}}]{Bernon:2017jgv}%
  \BibitemOpen
  \bibfield  {author} {\bibinfo {author} {\bibfnamefont {J.}~\bibnamefont
  {Bernon}}, \bibinfo {author} {\bibfnamefont {L.}~\bibnamefont {Bian}}, \ and\
  \bibinfo {author} {\bibfnamefont {Y.}~\bibnamefont {Jiang}},\ }\bibfield
  {title} {\enquote {\bibinfo {title} {\emph{A New Insight into the Phase
  Transition in the Early Universe with Two Higgs Doublets}},}\ }\href
  {\doibase 10.1007/JHEP05(2018)151} {\bibfield  {journal} {\bibinfo  {journal}
  {JHEP}\ }\textbf {\bibinfo {volume} {05}},\ \bibinfo {pages} {151} (\bibinfo
  {year} {2018})},\ \Eprint {http://arxiv.org/abs/1712.08430} {arXiv:1712.08430
  [hep-ph]} \BibitemShut {NoStop}%
\bibitem [{\citenamefont {Chala}\ \emph {et~al.}(2018)\citenamefont {Chala},
  \citenamefont {Krause}, \citenamefont {Nardini}}]{Chala:2018ari}%
  \BibitemOpen
  \bibfield  {author} {\bibinfo {author} {\bibfnamefont {M.}~\bibnamefont
  {Chala}}, \bibinfo {author} {\bibfnamefont {C.}~\bibnamefont {Krause}}, \bibinfo {author} {\bibfnamefont {G.}~\bibnamefont {Nardini}},\
  }\bibfield  {title} {\enquote {\bibinfo {title} {\emph{Signals of the
  Electroweak Phase Transition at Colliders and Gravitational Wave
  Observatories}},}\ }\href {\doibase 10.1007/JHEP07(2018)062} {\bibfield
  {journal} {\bibinfo  {journal} {JHEP}\ }\textbf {\bibinfo {volume} {07}},\
  \bibinfo {pages} {062} (\bibinfo {year} {2018})},\ \Eprint
  {http://arxiv.org/abs/1802.02168} {arXiv:1802.02168 [hep-ph]} \BibitemShut
  {NoStop}%
\bibitem [{\citenamefont {Angelescu}\ and\ \citenamefont
  {Huang}(2019)}]{Angelescu:2018dkk}%
  \BibitemOpen
  \bibfield  {author} {\bibinfo {author} {\bibfnamefont {A.}~\bibnamefont
  {Angelescu}}\ and\ \bibinfo {author} {\bibfnamefont {P.}~\bibnamefont
  {Huang}},\ }\bibfield  {title} {\enquote {\bibinfo {title} {\emph{Multistep
  Strongly First Order Phase Transitions from New Fermions at the TeV
  Scale}},}\ }\href {\doibase 10.1103/PhysRevD.99.055023} {\bibfield  {journal}
  {\bibinfo  {journal} {Phys. Rev. D}\ }\textbf {\bibinfo {volume} {99}},\
  \bibinfo {pages} {055023} (\bibinfo {year} {2019})},\ \Eprint
  {http://arxiv.org/abs/1812.08293} {arXiv:1812.08293 [hep-ph]} \BibitemShut
  {NoStop}%
\bibitem [{\citenamefont {Brdar}\ \emph {et~al.}(2019)\citenamefont {Brdar},
  \citenamefont {Helmboldt},\ and\ \citenamefont {Kubo}}]{Brdar:2018num}%
  \BibitemOpen
  \bibfield  {author} {\bibinfo {author} {\bibfnamefont {V.}~\bibnamefont
  {Brdar}}, \bibinfo {author} {\bibfnamefont {A.~J.}\ \bibnamefont
  {Helmboldt}}, \ and\ \bibinfo {author} {\bibfnamefont {J.}~\bibnamefont
  {Kubo}},\ }\bibfield  {title} {\enquote {\bibinfo {title}
  {\emph{Gravitational Waves from First-Order Phase Transitions: LIGO as a
  Window to Unexplored Seesaw Scales}},}\ }\href {\doibase
  10.1088/1475-7516/2019/02/021} {\bibfield  {journal} {\bibinfo  {journal}
  {JCAP}\ }\textbf {\bibinfo {volume} {02}},\ \bibinfo {pages} {021} (\bibinfo
  {year} {2019})},\ \Eprint {http://arxiv.org/abs/1810.12306} {arXiv:1810.12306
  [hep-ph]} \BibitemShut {NoStop}%
\bibitem [{\citenamefont {Okada}\ and\ \citenamefont
  {Seto}(2018)}]{Okada:2018xdh}%
  \BibitemOpen
  \bibfield  {author} {\bibinfo {author} {\bibfnamefont {N.}~\bibnamefont
  {Okada}}\ and\ \bibinfo {author} {\bibfnamefont {O.}~\bibnamefont {Seto}},\
  }\bibfield  {title} {\enquote {\bibinfo {title} {\emph{Probing the Seesaw
  Scale with Gravitational Waves}},}\ }\href {\doibase
  10.1103/PhysRevD.98.063532} {\bibfield  {journal} {\bibinfo  {journal} {Phys.
  Rev. D}\ }\textbf {\bibinfo {volume} {98}},\ \bibinfo {pages} {063532}
  (\bibinfo {year} {2018})},\ \Eprint {http://arxiv.org/abs/1807.00336}
  {arXiv:1807.00336 [hep-ph]} \BibitemShut {NoStop}%
\bibitem [{\citenamefont {Croon}\ \emph {et~al.}(2019)\citenamefont {Croon},
  \citenamefont {Gonzalo},\ and\ \citenamefont {White}}]{Croon:2018kqn}%
  \BibitemOpen
  \bibfield  {author} {\bibinfo {author} {\bibfnamefont {D.}~\bibnamefont
  {Croon}}, \bibinfo {author} {\bibfnamefont {T.~E.}\ \bibnamefont {Gonzalo}},
  \ and\ \bibinfo {author} {\bibfnamefont {G.}~\bibnamefont {White}},\
  }\bibfield  {title} {\enquote {\bibinfo {title} {\emph{Gravitational Waves
  from a Pati-Salam Phase Transition}},}\ }\href {\doibase
  10.1007/JHEP02(2019)083} {\bibfield  {journal} {\bibinfo  {journal} {JHEP}\
  }\textbf {\bibinfo {volume} {02}},\ \bibinfo {pages} {083} (\bibinfo {year}
  {2019})},\ \Eprint {http://arxiv.org/abs/1812.02747} {arXiv:1812.02747
  [hep-ph]} \BibitemShut {NoStop}%
\bibitem [{\citenamefont {Alves}\ \emph {et~al.}(2019)\citenamefont {Alves},
  \citenamefont {Ghosh}, \citenamefont {Guo}, \citenamefont {Sinha},\ and\
  \citenamefont {Vagie}}]{Alves:2018jsw}%
  \BibitemOpen
  \bibfield  {author} {\bibinfo {author} {\bibfnamefont {A.}~\bibnamefont
  {Alves}}, \bibinfo {author} {\bibfnamefont {T.}~\bibnamefont {Ghosh}},
  \bibinfo {author} {\bibfnamefont {H.-K.}\ \bibnamefont {Guo}}, \bibinfo
  {author} {\bibfnamefont {K.}~\bibnamefont {Sinha}}, \ and\ \bibinfo {author}
  {\bibfnamefont {D.}~\bibnamefont {Vagie}},\ }\bibfield  {title} {\enquote
  {\bibinfo {title} {\emph{Collider and Gravitational Wave Complementarity in
  Exploring the Singlet Extension of the Standard Model}},}\ }\href {\doibase
  10.1007/JHEP04(2019)052} {\bibfield  {journal} {\bibinfo  {journal} {JHEP}\
  }\textbf {\bibinfo {volume} {04}},\ \bibinfo {pages} {052} (\bibinfo {year}
  {2019})},\ \Eprint {http://arxiv.org/abs/1812.09333} {arXiv:1812.09333
  [hep-ph]} \BibitemShut {NoStop}%
\bibitem [{\citenamefont {Breitbach}\ \emph {et~al.}(2019)\citenamefont
  {Breitbach}, \citenamefont {Kopp}, \citenamefont {Madge}, \citenamefont
  {Opferkuch},\ and\ \citenamefont {Schwaller}}]{Breitbach:2018ddu}%
  \BibitemOpen
  \bibfield  {author} {\bibinfo {author} {\bibfnamefont {M.}~\bibnamefont
  {Breitbach}}, \bibinfo {author} {\bibfnamefont {J.}~\bibnamefont {Kopp}},
  \bibinfo {author} {\bibfnamefont {E.}~\bibnamefont {Madge}}, \bibinfo
  {author} {\bibfnamefont {T.}~\bibnamefont {Opferkuch}}, \ and\ \bibinfo
  {author} {\bibfnamefont {P.}~\bibnamefont {Schwaller}},\ }\bibfield  {title}
  {\enquote {\bibinfo {title} {\emph{Dark, Cold, and Noisy: Constraining
  Secluded Hidden Sectors with Gravitational Waves}},}\ }\href {\doibase
  10.1088/1475-7516/2019/07/007} {\bibfield  {journal} {\bibinfo  {journal}
  {JCAP}\ }\textbf {\bibinfo {volume} {07}},\ \bibinfo {pages} {007} (\bibinfo
  {year} {2019})},\ \Eprint {http://arxiv.org/abs/1811.11175} {arXiv:1811.11175
  [hep-ph]} \BibitemShut {NoStop}%
\bibitem [{\citenamefont {Croon}\ \emph {et~al.}(2018)\citenamefont {Croon},
  \citenamefont {Sanz},\ and\ \citenamefont {White}}]{Croon:2018erz}%
  \BibitemOpen
  \bibfield  {author} {\bibinfo {author} {\bibfnamefont {D.}~\bibnamefont
  {Croon}}, \bibinfo {author} {\bibfnamefont {V.}~\bibnamefont {Sanz}}, \ and\
  \bibinfo {author} {\bibfnamefont {G.}~\bibnamefont {White}},\ }\bibfield
  {title} {\enquote {\bibinfo {title} {\emph{Model Discrimination in
  Gravitational Wave spectra from Dark Phase Transitions}},}\ }\href {\doibase
  10.1007/JHEP08(2018)203} {\bibfield  {journal} {\bibinfo  {journal} {JHEP}\
  }\textbf {\bibinfo {volume} {08}},\ \bibinfo {pages} {203} (\bibinfo {year}
  {2018})},\ \Eprint {http://arxiv.org/abs/1806.02332} {arXiv:1806.02332
  [hep-ph]} \BibitemShut {NoStop}%
\bibitem [{\citenamefont {Hall}\ \emph {et~al.}(2020)\citenamefont {Hall},
  \citenamefont {Konstandin}, \citenamefont {McGehee}, \citenamefont
  {Murayama},\ and\ \citenamefont {Servant}}]{Hall:2019ank}%
  \BibitemOpen
  \bibfield  {author} {\bibinfo {author} {\bibfnamefont {E.}~\bibnamefont
  {Hall}}, \bibinfo {author} {\bibfnamefont {T.}~\bibnamefont {Konstandin}},
  \bibinfo {author} {\bibfnamefont {R.}~\bibnamefont {McGehee}}, \bibinfo
  {author} {\bibfnamefont {H.}~\bibnamefont {Murayama}}, \ and\ \bibinfo
  {author} {\bibfnamefont {G.}~\bibnamefont {Servant}},\ }\bibfield  {title}
  {\enquote {\bibinfo {title} {\emph{Baryogenesis From a Dark First-Order Phase
  Transition}},}\ }\href {\doibase 10.1007/JHEP04(2020)042} {\bibfield
  {journal} {\bibinfo  {journal} {JHEP}\ }\textbf {\bibinfo {volume} {04}},\
  \bibinfo {pages} {042} (\bibinfo {year} {2020})},\ \Eprint
  {http://arxiv.org/abs/1910.08068} {arXiv:1910.08068 [hep-ph]} \BibitemShut
  {NoStop}%
\bibitem [{\citenamefont {Ellis}\ \emph {et~al.}(2019)\citenamefont {Ellis},
  \citenamefont {Lewicki}, \citenamefont {No},\ and\ \citenamefont
  {Vaskonen}}]{Ellis:2019oqb}%
  \BibitemOpen
  \bibfield  {author} {\bibinfo {author} {\bibfnamefont {J.}~\bibnamefont
  {Ellis}}, \bibinfo {author} {\bibfnamefont {M.}~\bibnamefont {Lewicki}},
  \bibinfo {author} {\bibfnamefont {J.~M.}\ \bibnamefont {No}}, \ and\ \bibinfo
  {author} {\bibfnamefont {V.}~\bibnamefont {Vaskonen}},\ }\bibfield  {title}
  {\enquote {\bibinfo {title} {\emph{Gravitational Wave Energy Budget in
  Strongly Supercooled Phase Transitions}},}\ }\href {\doibase
  10.1088/1475-7516/2019/06/024} {\bibfield  {journal} {\bibinfo  {journal}
  {JCAP}\ }\textbf {\bibinfo {volume} {06}},\ \bibinfo {pages} {024} (\bibinfo
  {year} {2019})},\ \Eprint {http://arxiv.org/abs/1903.09642} {arXiv:1903.09642
  [hep-ph]} \BibitemShut {NoStop}%
\bibitem [{\citenamefont {Dev}\ \emph {et~al.}(2019)\citenamefont {Dev},
  \citenamefont {Ferrer}, \citenamefont {Zhang},\ and\ \citenamefont
  {Zhang}}]{Dev:2019njv}%
  \BibitemOpen
  \bibfield  {author} {\bibinfo {author} {\bibfnamefont {P.~S.~B.}\
  \bibnamefont {Dev}}, \bibinfo {author} {\bibfnamefont {F.}~\bibnamefont
  {Ferrer}}, \bibinfo {author} {\bibfnamefont {Y.}~\bibnamefont {Zhang}}, \
  and\ \bibinfo {author} {\bibfnamefont {Y.}~\bibnamefont {Zhang}},\ }\bibfield
   {title} {\enquote {\bibinfo {title} {\emph{Gravitational Waves from
  First-Order Phase Transition in a Simple Axion-Like Particle Model}},}\
  }\href {\doibase 10.1088/1475-7516/2019/11/006} {\bibfield  {journal}
  {\bibinfo  {journal} {JCAP}\ }\textbf {\bibinfo {volume} {11}},\ \bibinfo
  {pages} {006} (\bibinfo {year} {2019})},\ \Eprint
  {http://arxiv.org/abs/1905.00891} {arXiv:1905.00891 [hep-ph]} \BibitemShut
  {NoStop}%
\bibitem [{\citenamefont {Hasegawa}\ \emph {et~al.}(2019)\citenamefont
  {Hasegawa}, \citenamefont {Okada},\ and\ \citenamefont
  {Seto}}]{Hasegawa:2019amx}%
  \BibitemOpen
  \bibfield  {author} {\bibinfo {author} {\bibfnamefont {T.}~\bibnamefont
  {Hasegawa}}, \bibinfo {author} {\bibfnamefont {N.}~\bibnamefont {Okada}}, \
  and\ \bibinfo {author} {\bibfnamefont {O.}~\bibnamefont {Seto}},\ }\bibfield
  {title} {\enquote {\bibinfo {title} {\emph{Gravitational Waves from the
  Minimal Gauged $U(1)_{B-L}$ Model}},}\ }\href {\doibase
  10.1103/PhysRevD.99.095039} {\bibfield  {journal} {\bibinfo  {journal} {Phys.
  Rev. D}\ }\textbf {\bibinfo {volume} {99}},\ \bibinfo {pages} {095039}
  (\bibinfo {year} {2019})},\ \Eprint {http://arxiv.org/abs/1904.03020}
  {arXiv:1904.03020 [hep-ph]} \BibitemShut {NoStop}%
\bibitem [{\citenamefont {Von~Harling}\ \emph {et~al.}(2020)\citenamefont
  {Von~Harling}, \citenamefont {Pomarol}, \citenamefont {Pujolas},\ and\
  \citenamefont {Rompineve}}]{VonHarling:2019rgb}%
  \BibitemOpen
  \bibfield  {author} {\bibinfo {author} {\bibfnamefont {B.}~\bibnamefont
  {Von~Harling}}, \bibinfo {author} {\bibfnamefont {A.}~\bibnamefont
  {Pomarol}}, \bibinfo {author} {\bibfnamefont {O.}~\bibnamefont {Pujolas}}, \
  and\ \bibinfo {author} {\bibfnamefont {F.}~\bibnamefont {Rompineve}},\
  }\bibfield  {title} {\enquote {\bibinfo {title} {\emph{Peccei-Quinn Phase
  Transition at LIGO}},}\ }\href {\doibase 10.1007/JHEP04(2020)195} {\bibfield
  {journal} {\bibinfo  {journal} {JHEP}\ }\textbf {\bibinfo {volume} {04}},\
  \bibinfo {pages} {195} (\bibinfo {year} {2020})},\ \Eprint
  {http://arxiv.org/abs/1912.07587} {arXiv:1912.07587 [hep-ph]} \BibitemShut
  {NoStop}%
\bibitem [{\citenamefont {Delle~Rose}\ \emph {et~al.}(2020)\citenamefont
  {Delle~Rose}, \citenamefont {Panico}, \citenamefont {Redi},\ and\
  \citenamefont {Tesi}}]{DelleRose:2019pgi}%
  \BibitemOpen
  \bibfield  {author} {\bibinfo {author} {\bibfnamefont {L.}~\bibnamefont
  {Delle~Rose}}, \bibinfo {author} {\bibfnamefont {G.}~\bibnamefont {Panico}},
  \bibinfo {author} {\bibfnamefont {M.}~\bibnamefont {Redi}}, \ and\ \bibinfo
  {author} {\bibfnamefont {A.}~\bibnamefont {Tesi}},\ }\bibfield  {title}
  {\enquote {\bibinfo {title} {\emph{Gravitational Waves from Supercool
  Axions}},}\ }\href {\doibase 10.1007/JHEP04(2020)025} {\bibfield  {journal}
  {\bibinfo  {journal} {JHEP}\ }\textbf {\bibinfo {volume} {04}},\ \bibinfo
  {pages} {025} (\bibinfo {year} {2020})},\ \Eprint
  {http://arxiv.org/abs/1912.06139} {arXiv:1912.06139 [hep-ph]} \BibitemShut
  {NoStop}%
\bibitem [{\citenamefont {Lewicki}\ and\ \citenamefont
  {Vaskonen}(2020{\natexlab{a}})}]{Lewicki:2019gmv}%
  \BibitemOpen
  \bibfield  {author} {\bibinfo {author} {\bibfnamefont {M.}~\bibnamefont
  {Lewicki}}\ and\ \bibinfo {author} {\bibfnamefont {V.}~\bibnamefont
  {Vaskonen}},\ }\bibfield  {title} {\enquote {\bibinfo {title} {\emph{On
  Bubble Collisions in Strongly Supercooled Phase Transitions}},}\ }\href
  {\doibase 10.1016/j.dark.2020.100672} {\bibfield  {journal} {\bibinfo
  {journal} {Phys. Dark Univ.}\ }\textbf {\bibinfo {volume} {30}},\ \bibinfo
  {pages} {100672} (\bibinfo {year} {2020}{\natexlab{a}})},\ \Eprint
  {http://arxiv.org/abs/1912.00997} {arXiv:1912.00997 [astro-ph.CO]}
  \BibitemShut {NoStop}%
\bibitem [{\citenamefont {Greljo}\ \emph {et~al.}(2020)\citenamefont {Greljo},
  \citenamefont {Opferkuch},\ and\ \citenamefont {Stefanek}}]{Greljo:2019xan}%
  \BibitemOpen
  \bibfield  {author} {\bibinfo {author} {\bibfnamefont {A.}~\bibnamefont
  {Greljo}}, \bibinfo {author} {\bibfnamefont {T.}~\bibnamefont {Opferkuch}}, \
  and\ \bibinfo {author} {\bibfnamefont {B.~A.}\ \bibnamefont {Stefanek}},\
  }\bibfield  {title} {\enquote {\bibinfo {title} {\emph{Gravitational Imprints
  of Flavor Hierarchies}},}\ }\href {\doibase 10.1103/PhysRevLett.124.171802}
  {\bibfield  {journal} {\bibinfo  {journal} {Phys. Rev. Lett.}\ }\textbf
  {\bibinfo {volume} {124}},\ \bibinfo {pages} {171802} (\bibinfo {year}
  {2020})},\ \Eprint {http://arxiv.org/abs/1910.02014} {arXiv:1910.02014
  [hep-ph]} \BibitemShut {NoStop}%
\bibitem [{\citenamefont {Huang}\ \emph {et~al.}(2020)\citenamefont {Huang},
  \citenamefont {Sannino},\ and\ \citenamefont {Wang}}]{Huang:2020bbe}%
  \BibitemOpen
  \bibfield  {author} {\bibinfo {author} {\bibfnamefont {W.-C.}\ \bibnamefont
  {Huang}}, \bibinfo {author} {\bibfnamefont {F.}~\bibnamefont {Sannino}}, \
  and\ \bibinfo {author} {\bibfnamefont {Z.-W.}\ \bibnamefont {Wang}},\
  }\bibfield  {title} {\enquote {\bibinfo {title} {\emph{Gravitational Waves
  from Pati-Salam Dynamics}},}\ }\href {\doibase 10.1103/PhysRevD.102.095025}
  {\bibfield  {journal} {\bibinfo  {journal} {Phys. Rev. D}\ }\textbf {\bibinfo
  {volume} {102}},\ \bibinfo {pages} {095025} (\bibinfo {year} {2020})},\
  \Eprint {http://arxiv.org/abs/2004.02332} {arXiv:2004.02332 [hep-ph]}
  \BibitemShut {NoStop}%
\bibitem [{\citenamefont {Okada}\ \emph {et~al.}(2021)\citenamefont {Okada},
  \citenamefont {Seto},\ and\ \citenamefont {Uchida}}]{Okada:2020vvb}%
  \BibitemOpen
  \bibfield  {author} {\bibinfo {author} {\bibfnamefont {N.}~\bibnamefont
  {Okada}}, \bibinfo {author} {\bibfnamefont {O.}~\bibnamefont {Seto}}, \ and\
  \bibinfo {author} {\bibfnamefont {H.}~\bibnamefont {Uchida}},\ }\bibfield
  {title} {\enquote {\bibinfo {title} {\emph{Gravitational Waves from Breaking
  of an Extra $U(1)$ in $SO(10)$ Grand Unification}},}\ }\href {\doibase
  10.1093/ptep/ptab003} {\bibfield  {journal} {\bibinfo  {journal} {PTEP}\
  }\textbf {\bibinfo {volume} {2021}},\ \bibinfo {pages} {033B01} (\bibinfo
  {year} {2021})},\ \Eprint {http://arxiv.org/abs/2006.01406} {arXiv:2006.01406
  [hep-ph]} \BibitemShut {NoStop}%
\bibitem [{\citenamefont {Lewicki}\ and\ \citenamefont
  {Vaskonen}(2020{\natexlab{b}})}]{Lewicki:2020jiv}%
  \BibitemOpen
  \bibfield  {author} {\bibinfo {author} {\bibfnamefont {M.}~\bibnamefont
  {Lewicki}}\ and\ \bibinfo {author} {\bibfnamefont {V.}~\bibnamefont
  {Vaskonen}},\ }\bibfield  {title} {\enquote {\bibinfo {title}
  {\emph{Gravitational Wave Spectra from Strongly Supercooled Phase
  Transitions}},}\ }\href {\doibase 10.1140/epjc/s10052-020-08589-1} {\bibfield
   {journal} {\bibinfo  {journal} {Eur. Phys. J. C}\ }\textbf {\bibinfo
  {volume} {80}},\ \bibinfo {pages} {1003} (\bibinfo {year}
  {2020}{\natexlab{b}})},\ \Eprint {http://arxiv.org/abs/2007.04967}
  {arXiv:2007.04967 [astro-ph.CO]} \BibitemShut {NoStop}%
\bibitem [{\citenamefont {Ellis}\ \emph
  {et~al.}(2020{\natexlab{a}})\citenamefont {Ellis}, \citenamefont {Lewicki},\
  and\ \citenamefont {Vaskonen}}]{Ellis:2020nnr}%
  \BibitemOpen
  \bibfield  {author} {\bibinfo {author} {\bibfnamefont {J.}~\bibnamefont
  {Ellis}}, \bibinfo {author} {\bibfnamefont {M.}~\bibnamefont {Lewicki}}, \
  and\ \bibinfo {author} {\bibfnamefont {V.}~\bibnamefont {Vaskonen}},\
  }\bibfield  {title} {\enquote {\bibinfo {title} {\emph{Updated Predictions
  for Gravitational Waves Produced in a Strongly Supercooled Phase
  Transition}},}\ }\href {\doibase 10.1088/1475-7516/2020/11/020} {\bibfield
  {journal} {\bibinfo  {journal} {JCAP}\ }\textbf {\bibinfo {volume} {11}},\
  \bibinfo {pages} {020} (\bibinfo {year} {2020}{\natexlab{a}})},\ \Eprint
  {http://arxiv.org/abs/2007.15586} {arXiv:2007.15586 [astro-ph.CO]}
  \BibitemShut {NoStop}%
\bibitem [{\citenamefont {Fornal}\ and\ \citenamefont {Shams
  Es~Haghi}(2020)}]{Fornal:2020esl}%
  \BibitemOpen
  \bibfield  {author} {\bibinfo {author} {\bibfnamefont {B.}~\bibnamefont
  {Fornal}}\ and\ \bibinfo {author} {\bibfnamefont {B.}~\bibnamefont {Shams
  Es~Haghi}},\ }\bibfield  {title} {\enquote {\bibinfo {title} {\emph{Baryon
  and Lepton Number Violation from Gravitational Waves}},}\ }\href {\doibase
  10.1103/PhysRevD.102.115037} {\bibfield  {journal} {\bibinfo  {journal}
  {Phys. Rev. D}\ }\textbf {\bibinfo {volume} {102}},\ \bibinfo {pages}
  {115037} (\bibinfo {year} {2020})},\ \Eprint
  {http://arxiv.org/abs/2008.05111} {arXiv:2008.05111 [hep-ph]} \BibitemShut
  {NoStop}%
\bibitem [{\citenamefont {Han}\ \emph {et~al.}(2021)\citenamefont {Han},
  \citenamefont {Wang},\ and\ \citenamefont {Zhang}}]{Han:2020ekm}%
  \BibitemOpen
  \bibfield  {author} {\bibinfo {author} {\bibfnamefont {X.-F.}\ \bibnamefont
  {Han}}, \bibinfo {author} {\bibfnamefont {L.}~\bibnamefont {Wang}}, \ and\
  \bibinfo {author} {\bibfnamefont {Y.}~\bibnamefont {Zhang}},\ }\bibfield
  {title} {\enquote {\bibinfo {title} {\emph{Dark Matter, Electroweak Phase
  Transition, and Gravitational Waves in the Type II Two-Higgs-Doublet Model
  with a Singlet Scalar Field}},}\ }\href {\doibase
  10.1103/PhysRevD.103.035012} {\bibfield  {journal} {\bibinfo  {journal}
  {Phys. Rev. D}\ }\textbf {\bibinfo {volume} {103}},\ \bibinfo {pages}
  {035012} (\bibinfo {year} {2021})},\ \Eprint
  {http://arxiv.org/abs/2010.03730} {arXiv:2010.03730 [hep-ph]} \BibitemShut
  {NoStop}%
\bibitem [{\citenamefont {Fornal}(2021)}]{Fornal:2020ngq}%
  \BibitemOpen
  \bibfield  {author} {\bibinfo {author} {\bibfnamefont {B.}~\bibnamefont
  {Fornal}},\ }\bibfield  {title} {\enquote {\bibinfo {title}
  {\emph{Gravitational Wave Signatures of Lepton Universality Violation}},}\
  }\href {\doibase 10.1103/PhysRevD.103.015018} {\bibfield  {journal} {\bibinfo
   {journal} {Phys. Rev. D}\ }\textbf {\bibinfo {volume} {103}},\ \bibinfo
  {pages} {015018} (\bibinfo {year} {2021})},\ \Eprint
  {http://arxiv.org/abs/2006.08802} {arXiv:2006.08802 [hep-ph]} \BibitemShut
  {NoStop}%
\bibitem [{\citenamefont {Craig}\ \emph {et~al.}(2020)\citenamefont {Craig},
  \citenamefont {Levi}, \citenamefont {Mariotti},\ and\ \citenamefont
  {Redigolo}}]{Craig:2020jfv}%
  \BibitemOpen
  \bibfield  {author} {\bibinfo {author} {\bibfnamefont {N.}~\bibnamefont
  {Craig}}, \bibinfo {author} {\bibfnamefont {N.}~\bibnamefont {Levi}},
  \bibinfo {author} {\bibfnamefont {A.}~\bibnamefont {Mariotti}}, \ and\
  \bibinfo {author} {\bibfnamefont {D.}~\bibnamefont {Redigolo}},\ }\bibfield
  {title} {\enquote {\bibinfo {title} {\emph{Ripples in Spacetime from Broken
  Supersymmetry}},}\ }\href {\doibase 10.1007/JHEP02(2021)184} {\bibfield
  {journal} {\bibinfo  {journal} {JHEP}\ }\textbf {\bibinfo {volume} {21}},\
  \bibinfo {pages} {184} (\bibinfo {year} {2020})},\ \Eprint
  {http://arxiv.org/abs/2011.13949} {arXiv:2011.13949 [hep-ph]} \BibitemShut
  {NoStop}%
\bibitem [{\citenamefont {Fornal}\ \emph {et~al.}(2021)\citenamefont {Fornal},
  \citenamefont {Shams Es~Haghi}, \citenamefont {Yu},\ and\ \citenamefont
  {Zhao}}]{Fornal:2021ovz}%
  \BibitemOpen
  \bibfield  {author} {\bibinfo {author} {\bibfnamefont {B.}~\bibnamefont
  {Fornal}}, \bibinfo {author} {\bibfnamefont {B.}~\bibnamefont {Shams
  Es~Haghi}}, \bibinfo {author} {\bibfnamefont {J.-H.}\ \bibnamefont {Yu}}, \
  and\ \bibinfo {author} {\bibfnamefont {Y.}~\bibnamefont {Zhao}},\ }\bibfield
  {title} {\enquote {\bibinfo {title} {\emph{Gravitational Waves from
  Mini-Split SUSY}},}\ }\href {\doibase 10.1103/PhysRevD.104.115005} {\bibfield
   {journal} {\bibinfo  {journal} {Phys. Rev. D}\ }\textbf {\bibinfo {volume}
  {104}},\ \bibinfo {pages} {115005} (\bibinfo {year} {2021})},\ \Eprint
  {http://arxiv.org/abs/2104.00747} {arXiv:2104.00747 [hep-ph]} \BibitemShut
  {NoStop}%
\bibitem [{\citenamefont {Di~Bari}\ \emph {et~al.}(2021)\citenamefont
  {Di~Bari}, \citenamefont {Marfatia},\ and\ \citenamefont
  {Zhou}}]{DiBari:2021dri}%
  \BibitemOpen
  \bibfield  {author} {\bibinfo {author} {\bibfnamefont {P.}~\bibnamefont
  {Di~Bari}}, \bibinfo {author} {\bibfnamefont {D.}~\bibnamefont {Marfatia}}, \
  and\ \bibinfo {author} {\bibfnamefont {Y.-L.}\ \bibnamefont {Zhou}},\
  }\bibfield  {title} {\enquote {\bibinfo {title} {\emph{Gravitational Waves
  from First-Order Phase Transitions in Majoron Models of Neutrino Mass}},}\
  }\href {\doibase 10.1007/JHEP10(2021)193} {\bibfield  {journal} {\bibinfo
  {journal} {JHEP}\ }\textbf {\bibinfo {volume} {10}},\ \bibinfo {pages} {193}
  (\bibinfo {year} {2021})},\ \Eprint {http://arxiv.org/abs/2106.00025}
  {arXiv:2106.00025 [hep-ph]} \BibitemShut {NoStop}%
\bibitem [{\citenamefont {Azatov}\ \emph {et~al.}(2021)\citenamefont {Azatov},
  \citenamefont {Vanvlasselaer},\ and\ \citenamefont {Yin}}]{Azatov:2021ifm}%
  \BibitemOpen
  \bibfield  {author} {\bibinfo {author} {\bibfnamefont {A.}~\bibnamefont
  {Azatov}}, \bibinfo {author} {\bibfnamefont {M.}~\bibnamefont
  {Vanvlasselaer}}, \ and\ \bibinfo {author} {\bibfnamefont {W.}~\bibnamefont
  {Yin}},\ }\bibfield  {title} {\enquote {\bibinfo {title} {\emph{Dark Matter
  Production from Relativistic Bubble Walls}},}\ }\href {\doibase
  10.1007/JHEP03(2021)288} {\bibfield  {journal} {\bibinfo  {journal} {JHEP}\
  }\textbf {\bibinfo {volume} {03}},\ \bibinfo {pages} {288} (\bibinfo {year}
  {2021})},\ \Eprint {http://arxiv.org/abs/2101.05721} {arXiv:2101.05721
  [hep-ph]} \BibitemShut {NoStop}%
\bibitem [{\citenamefont {Zhou}\ \emph {et~al.}(2022)\citenamefont {Zhou},
  \citenamefont {Bian},\ and\ \citenamefont {Du}}]{Zhou:2022mlz}%
  \BibitemOpen
  \bibfield  {author} {\bibinfo {author} {\bibfnamefont {R.}~\bibnamefont
  {Zhou}}, \bibinfo {author} {\bibfnamefont {L.}~\bibnamefont {Bian}}, \ and\
  \bibinfo {author} {\bibfnamefont {Y.}~\bibnamefont {Du}},\ }\bibfield
  {title} {\enquote {\bibinfo {title} {\emph{Electroweak Phase Transition and
  Gravitational Waves in the Type-II Seesaw Model}},}\ }\href {\doibase
  10.1007/JHEP08(2022)205} {\bibfield  {journal} {\bibinfo  {journal} {JHEP}\
  }\textbf {\bibinfo {volume} {08}},\ \bibinfo {pages} {205} (\bibinfo {year}
  {2022})},\ \Eprint {http://arxiv.org/abs/2203.01561} {arXiv:2203.01561
  [hep-ph]} \BibitemShut {NoStop}%
\bibitem [{\citenamefont {Benincasa}\ \emph {et~al.}(2022)\citenamefont
  {Benincasa}, \citenamefont {Delle~Rose}, \citenamefont {Kannike},\ and\
  \citenamefont {Marzola}}]{Benincasa:2022elt}%
  \BibitemOpen
  \bibfield  {author} {\bibinfo {author} {\bibfnamefont {N.}~\bibnamefont
  {Benincasa}}, \bibinfo {author} {\bibfnamefont {L.}~\bibnamefont
  {Delle~Rose}}, \bibinfo {author} {\bibfnamefont {K.}~\bibnamefont {Kannike}},
  \ and\ \bibinfo {author} {\bibfnamefont {L.}~\bibnamefont {Marzola}},\
  }\bibfield  {title} {\enquote {\bibinfo {title} {\emph{Multistep Phase
  Transitions and Gravitational Waves in the Inert Doublet Model}},}\ }\href
  {\doibase 10.1088/1475-7516/2022/12/025} {\bibfield  {journal} {\bibinfo
  {journal} {JCAP}\ }\textbf {\bibinfo {volume} {12}},\ \bibinfo {pages} {025}
  (\bibinfo {year} {2022})},\ \Eprint {http://arxiv.org/abs/2205.06669}
  {arXiv:2205.06669 [hep-ph]} \BibitemShut {NoStop}%
\bibitem [{\citenamefont {Kawana}(2022)}]{Kawana:2022fum}%
  \BibitemOpen
  \bibfield  {author} {\bibinfo {author} {\bibfnamefont {K.}~\bibnamefont
  {Kawana}},\ }\bibfield  {title} {\enquote {\bibinfo {title} {\emph{Cosmology
  of a Supercooled Universe}},}\ }\href {\doibase 10.1103/PhysRevD.105.103515}
  {\bibfield  {journal} {\bibinfo  {journal} {Phys. Rev. D}\ }\textbf {\bibinfo
  {volume} {105}},\ \bibinfo {pages} {103515} (\bibinfo {year} {2022})},\
  \Eprint {http://arxiv.org/abs/2201.00560} {arXiv:2201.00560 [hep-ph]}
  \BibitemShut {NoStop}%
\bibitem [{\citenamefont {Costa}\ \emph
  {et~al.}(2022{\natexlab{a}})\citenamefont {Costa}, \citenamefont {Khan},\
  and\ \citenamefont {Kim}}]{Costa:2022oaa}%
  \BibitemOpen
  \bibfield  {author} {\bibinfo {author} {\bibfnamefont {F.}~\bibnamefont
  {Costa}}, \bibinfo {author} {\bibfnamefont {S.}~\bibnamefont {Khan}}, \ and\
  \bibinfo {author} {\bibfnamefont {J.}~\bibnamefont {Kim}},\ }\bibfield
  {title} {\enquote {\bibinfo {title} {\emph{A Two-Component Dark Matter Model
  and its Associated Gravitational Waves}},}\ }\href {\doibase
  10.1007/JHEP06(2022)026} {\bibfield  {journal} {\bibinfo  {journal} {JHEP}\
  }\textbf {\bibinfo {volume} {06}},\ \bibinfo {pages} {026} (\bibinfo {year}
  {2022}{\natexlab{a}})},\ \Eprint {http://arxiv.org/abs/2202.13126}
  {arXiv:2202.13126 [hep-ph]} \BibitemShut {NoStop}%
\bibitem [{\citenamefont {Costa}\ \emph
  {et~al.}(2022{\natexlab{b}})\citenamefont {Costa}, \citenamefont {Khan},\
  and\ \citenamefont {Kim}}]{Costa:2022lpy}%
  \BibitemOpen
  \bibfield  {author} {\bibinfo {author} {\bibfnamefont {F.}~\bibnamefont
  {Costa}}, \bibinfo {author} {\bibfnamefont {S.}~\bibnamefont {Khan}}, \ and\
  \bibinfo {author} {\bibfnamefont {J.}~\bibnamefont {Kim}},\ }\bibfield
  {title} {\enquote {\bibinfo {title} {\emph{A Two-Component Vector WIMP --
  Fermion FIMP Dark Matter Model with an Extended Seesaw Mechanism}},}\ }\href
  {\doibase 10.1007/JHEP12(2022)165} {\bibfield  {journal} {\bibinfo  {journal}
  {JHEP}\ }\textbf {\bibinfo {volume} {12}},\ \bibinfo {pages} {165} (\bibinfo
  {year} {2022}{\natexlab{b}})},\ \Eprint {http://arxiv.org/abs/2209.13653}
  {arXiv:2209.13653 [hep-ph]} \BibitemShut {NoStop}%
\bibitem [{\citenamefont {Fornal}\ and\ \citenamefont
  {Pierre}(2022)}]{Fornal:2022qim}%
  \BibitemOpen
  \bibfield  {author} {\bibinfo {author} {\bibfnamefont {B.}~\bibnamefont
  {Fornal}}\ and\ \bibinfo {author} {\bibfnamefont {E.}~\bibnamefont
  {Pierre}},\ }\bibfield  {title} {\enquote {\bibinfo {title} {\emph{Asymmetric
  Dark Matter from Gravitational Waves}},}\ }\href {\doibase
  10.1103/PhysRevD.106.115040} {\bibfield  {journal} {\bibinfo  {journal}
  {Phys. Rev. D}\ }\textbf {\bibinfo {volume} {106}},\ \bibinfo {pages}
  {115040} (\bibinfo {year} {2022})},\ \Eprint
  {http://arxiv.org/abs/2209.04788} {arXiv:2209.04788 [hep-ph]} \BibitemShut
  {NoStop}%
\bibitem [{\citenamefont {Kierkla}\ \emph {et~al.}(2023)\citenamefont
  {Kierkla}, \citenamefont {Karam},\ and\ \citenamefont
  {Swiezewska}}]{Kierkla:2022odc}%
  \BibitemOpen
  \bibfield  {author} {\bibinfo {author} {\bibfnamefont {M.}~\bibnamefont
  {Kierkla}}, \bibinfo {author} {\bibfnamefont {A.}~\bibnamefont {Karam}}, \
  and\ \bibinfo {author} {\bibfnamefont {B.}~\bibnamefont {Swiezewska}},\
  }\bibfield  {title} {\enquote {\bibinfo {title} {\emph{Conformal Model for
  Gravitational Waves and Dark Matter: A Status Update}},}\ }\href {\doibase
  10.1007/JHEP03(2023)007} {\bibfield  {journal} {\bibinfo  {journal} {JHEP}\
  }\textbf {\bibinfo {volume} {03}},\ \bibinfo {pages} {007} (\bibinfo {year}
  {2023})},\ \Eprint {http://arxiv.org/abs/2210.07075} {arXiv:2210.07075
  [astro-ph.CO]} \BibitemShut {NoStop}%
\bibitem [{\citenamefont {Azatov}\ \emph {et~al.}(2022)\citenamefont {Azatov},
  \citenamefont {Barni}, \citenamefont {Chakraborty}, \citenamefont
  {Vanvlasselaer},\ and\ \citenamefont {Yin}}]{Azatov:2022tii}%
  \BibitemOpen
  \bibfield  {author} {\bibinfo {author} {\bibfnamefont {A.}~\bibnamefont
  {Azatov}}, \bibinfo {author} {\bibfnamefont {G.}~\bibnamefont {Barni}},
  \bibinfo {author} {\bibfnamefont {S.}~\bibnamefont {Chakraborty}}, \bibinfo
  {author} {\bibfnamefont {M.}~\bibnamefont {Vanvlasselaer}}, \ and\ \bibinfo
  {author} {\bibfnamefont {W.}~\bibnamefont {Yin}},\ }\bibfield  {title}
  {\enquote {\bibinfo {title} {\emph{Ultra-Relativistic Bubbles from the
  Simplest Higgs Portal and their Cosmological Consequences}},}\ }\href
  {\doibase 10.1007/JHEP10(2022)017} {\bibfield  {journal} {\bibinfo  {journal}
  {JHEP}\ }\textbf {\bibinfo {volume} {10}},\ \bibinfo {pages} {017} (\bibinfo
  {year} {2022})},\ \Eprint {http://arxiv.org/abs/2207.02230} {arXiv:2207.02230
  [hep-ph]} \BibitemShut {NoStop}%
\bibitem [{\citenamefont {Fornal}\ \emph {et~al.}(2023)\citenamefont {Fornal},
  \citenamefont {Garcia},\ and\ \citenamefont {Pierre}}]{Fornal:2023hri}%
  \BibitemOpen
  \bibfield  {author} {\bibinfo {author} {\bibfnamefont {B.}~\bibnamefont
  {Fornal}}, \bibinfo {author} {\bibfnamefont {K.}~\bibnamefont {Garcia}}, \
  and\ \bibinfo {author} {\bibfnamefont {E.}~\bibnamefont {Pierre}},\
  }\bibfield  {title} {\enquote {\bibinfo {title} {\emph{Testing Unification
  and Dark Matter with Gravitational Waves}},}\ }\href {\doibase
  10.1103/PhysRevD.108.055022} {\bibfield  {journal} {\bibinfo  {journal}
  {Phys. Rev. D}\ }\textbf {\bibinfo {volume} {108}},\ \bibinfo {pages}
  {055022} (\bibinfo {year} {2023})},\ \Eprint
  {http://arxiv.org/abs/2305.12566} {arXiv:2305.12566 [hep-ph]} \BibitemShut
  {NoStop}%
\bibitem [{\citenamefont {Bosch}\ \emph {et~al.}(2023)\citenamefont {Bosch},
  \citenamefont {Delgado}, \citenamefont {Fornal},\ and\ \citenamefont
  {Leon}}]{Bosch:2023spa}%
  \BibitemOpen
  \bibfield  {author} {\bibinfo {author} {\bibfnamefont {J.}~\bibnamefont
  {Bosch}}, \bibinfo {author} {\bibfnamefont {Z.}~\bibnamefont {Delgado}},
  \bibinfo {author} {\bibfnamefont {B.}~\bibnamefont {Fornal}}, \ and\ \bibinfo
  {author} {\bibfnamefont {A.}~\bibnamefont {Leon}},\ }\bibfield  {title}
  {\enquote {\bibinfo {title} {\emph{Gravitational Wave Signatures of Gauged
  Baryon and Lepton Number}},}\ }\href {\doibase 10.1103/PhysRevD.108.095014}
  {\bibfield  {journal} {\bibinfo  {journal} {Phys. Rev. D}\ }\textbf {\bibinfo
  {volume} {108}},\ \bibinfo {pages} {095014} (\bibinfo {year} {2023})},\
  \Eprint {http://arxiv.org/abs/2306.00332} {arXiv:2306.00332 [hep-ph]}
  \BibitemShut {NoStop}%
\bibitem [{\citenamefont {Bunji}\ \emph {et~al.}(2024)\citenamefont {Bunji},
  \citenamefont {Fornal},\ and\ \citenamefont {Garcia}}]{Bunji:2024ovg}%
  \BibitemOpen
  \bibfield  {author} {\bibinfo {author} {\bibfnamefont {N.}~\bibnamefont
  {Bunji}}, \bibinfo {author} {\bibfnamefont {B.}~\bibnamefont {Fornal}}, \
  and\ \bibinfo {author} {\bibfnamefont {K.}~\bibnamefont {Garcia}},\
  }\bibfield  {title} {\enquote {\bibinfo {title} {\emph{Shedding Light on Dark
  Sectors with Gravitational Waves}},}\ }\href@noop {} {\  (\bibinfo {year}
  {2024})},\ \Eprint {http://arxiv.org/abs/2405.17851} {arXiv:2405.17851
  [hep-ph]} \BibitemShut {NoStop}%
\bibitem [{\citenamefont {Caldwell}\ \emph {et~al.}(2022)\citenamefont
  {Caldwell} \emph {et~al.}}]{Caldwell:2022qsj}%
  \BibitemOpen
  \bibfield  {author} {\bibinfo {author} {\bibfnamefont {R.}~\bibnamefont
  {Caldwell}} \emph {et~al.},\ }\bibfield  {title} {\enquote {\bibinfo {title}
  {\emph{Detection of Early-Universe Gravitational-Wave Signatures and
  Fundamental Physics}},}\ }\href {\doibase 10.1007/s10714-022-03027-x}
  {\bibfield  {journal} {\bibinfo  {journal} {Gen. Rel. Grav.}\ }\textbf
  {\bibinfo {volume} {54}},\ \bibinfo {pages} {156} (\bibinfo {year} {2022})},\
  \Eprint {http://arxiv.org/abs/2203.07972} {arXiv:2203.07972 [gr-qc]}
  \BibitemShut {NoStop}%
\bibitem [{\citenamefont {Athron}\ \emph {et~al.}(2024)\citenamefont {Athron},
  \citenamefont {Balazs}, \citenamefont {Fowlie}, \citenamefont {Morris},\ and\
  \citenamefont {Wu}}]{Athron:2023xlk}%
  \BibitemOpen
  \bibfield  {author} {\bibinfo {author} {\bibfnamefont {P.}~\bibnamefont
  {Athron}}, \bibinfo {author} {\bibfnamefont {C.}~\bibnamefont {Balazs}},
  \bibinfo {author} {\bibfnamefont {A.}~\bibnamefont {Fowlie}}, \bibinfo
  {author} {\bibfnamefont {L.}~\bibnamefont {Morris}}, \ and\ \bibinfo {author}
  {\bibfnamefont {L.}~\bibnamefont {Wu}},\ }\bibfield  {title} {\enquote
  {\bibinfo {title} {\emph{Cosmological Phase Transitions: From Perturbative
  Particle Physics to Gravitational Waves}},}\ }\href {\doibase
  10.1016/j.ppnp.2023.104094} {\bibfield  {journal} {\bibinfo  {journal} {Prog.
  Part. Nucl. Phys.}\ }\textbf {\bibinfo {volume} {135}},\ \bibinfo {pages}
  {104094} (\bibinfo {year} {2024})},\ \Eprint
  {http://arxiv.org/abs/2305.02357} {arXiv:2305.02357 [hep-ph]} \BibitemShut
  {NoStop}%
\bibitem [{\citenamefont {Badger}\ \emph {et~al.}(2023)\citenamefont {Badger},
  \citenamefont {Fornal}, \citenamefont {Martinovic}, \citenamefont {Romero},
  \citenamefont {Turbang}, \citenamefont {Guo}, \citenamefont {Mariotti},
  \citenamefont {Sakellariadou}, \citenamefont {Sevrin}, \citenamefont {Yang},\
  and\ \citenamefont {Zhao}}]{LIGO_FOPT}%
  \BibitemOpen
  \bibfield  {author} {\bibinfo {author} {\bibfnamefont {C.}~\bibnamefont
  {Badger}}, \bibinfo {author} {\bibfnamefont {B.}~\bibnamefont {Fornal}},
  \bibinfo {author} {\bibfnamefont {K.}~\bibnamefont {Martinovic}}, \bibinfo
  {author} {\bibfnamefont {A.}~\bibnamefont {Romero}}, \bibinfo {author}
  {\bibfnamefont {K.}~\bibnamefont {Turbang}}, \bibinfo {author} {\bibfnamefont
  {H.}~\bibnamefont {Guo}}, \bibinfo {author} {\bibfnamefont {A.}~\bibnamefont
  {Mariotti}}, \bibinfo {author} {\bibfnamefont {M.}~\bibnamefont
  {Sakellariadou}}, \bibinfo {author} {\bibfnamefont {A.}~\bibnamefont
  {Sevrin}}, \bibinfo {author} {\bibfnamefont {F.-W.}\ \bibnamefont {Yang}}, \
  and\ \bibinfo {author} {\bibfnamefont {Y.}~\bibnamefont {Zhao}},\ }\bibfield
  {title} {\enquote {\bibinfo {title} {\emph{Probing Early Universe Supercooled
  Phase Transitions with Gravitational Wave Data}},}\ }\href {\doibase
  10.1103/PhysRevD.107.023511} {\bibfield  {journal} {\bibinfo  {journal}
  {Phys. Rev. D}\ }\textbf {\bibinfo {volume} {107}},\ \bibinfo {pages}
  {023511} (\bibinfo {year} {2023})},\ \Eprint
  {http://arxiv.org/abs/2209.14707} {arXiv:2209.14707 [hep-ph]} \BibitemShut
  {NoStop}%
\bibitem [{\citenamefont {Kadota}\ \emph {et~al.}(2015)\citenamefont {Kadota},
  \citenamefont {Kawasaki},\ and\ \citenamefont {Saikawa}}]{Kadota:2015dza}%
  \BibitemOpen
  \bibfield  {author} {\bibinfo {author} {\bibfnamefont {K.}~\bibnamefont
  {Kadota}}, \bibinfo {author} {\bibfnamefont {M.}~\bibnamefont {Kawasaki}}, \
  and\ \bibinfo {author} {\bibfnamefont {K.}~\bibnamefont {Saikawa}},\
  }\bibfield  {title} {\enquote {\bibinfo {title} {\emph{Gravitational Waves
  from Domain Walls in the Next-to-Minimal Supersymmetric Standard Model}},}\
  }\href {\doibase 10.1088/1475-7516/2015/10/041} {\bibfield  {journal}
  {\bibinfo  {journal} {JCAP}\ }\textbf {\bibinfo {volume} {10}},\ \bibinfo
  {pages} {041} (\bibinfo {year} {2015})},\ \Eprint
  {http://arxiv.org/abs/1503.06998} {arXiv:1503.06998 [hep-ph]} \BibitemShut
  {NoStop}%
\bibitem [{\citenamefont {Eto}\ \emph {et~al.}(2018{\natexlab{a}})\citenamefont
  {Eto}, \citenamefont {Kurachi},\ and\ \citenamefont {Nitta}}]{Eto:2018hhg}%
  \BibitemOpen
  \bibfield  {author} {\bibinfo {author} {\bibfnamefont {M.}~\bibnamefont
  {Eto}}, \bibinfo {author} {\bibfnamefont {M.}~\bibnamefont {Kurachi}}, \ and\
  \bibinfo {author} {\bibfnamefont {M.}~\bibnamefont {Nitta}},\ }\bibfield
  {title} {\enquote {\bibinfo {title} {\emph{Constraints on Two Higgs Doublet
  Models from Domain Walls}},}\ }\href {\doibase
  10.1016/j.physletb.2018.09.002} {\bibfield  {journal} {\bibinfo  {journal}
  {Phys. Lett. B}\ }\textbf {\bibinfo {volume} {785}},\ \bibinfo {pages}
  {447--453} (\bibinfo {year} {2018}{\natexlab{a}})},\ \Eprint
  {http://arxiv.org/abs/1803.04662} {arXiv:1803.04662 [hep-ph]} \BibitemShut
  {NoStop}%
\bibitem [{\citenamefont {Eto}\ \emph {et~al.}(2018{\natexlab{b}})\citenamefont
  {Eto}, \citenamefont {Kurachi},\ and\ \citenamefont {Nitta}}]{Eto:2018tnk}%
  \BibitemOpen
  \bibfield  {author} {\bibinfo {author} {\bibfnamefont {M.}~\bibnamefont
  {Eto}}, \bibinfo {author} {\bibfnamefont {M.}~\bibnamefont {Kurachi}}, \ and\
  \bibinfo {author} {\bibfnamefont {M.}~\bibnamefont {Nitta}},\ }\bibfield
  {title} {\enquote {\bibinfo {title} {\emph{Non-Abelian Strings and Domain
  Walls in Two Higgs Doublet Models}},}\ }\href {\doibase
  10.1007/JHEP08(2018)195} {\bibfield  {journal} {\bibinfo  {journal} {JHEP}\
  }\textbf {\bibinfo {volume} {08}},\ \bibinfo {pages} {195} (\bibinfo {year}
  {2018}{\natexlab{b}})},\ \Eprint {http://arxiv.org/abs/1805.07015}
  {arXiv:1805.07015 [hep-ph]} \BibitemShut {NoStop}%
\bibitem [{\citenamefont {Chen}\ \emph {et~al.}(2020)\citenamefont {Chen},
  \citenamefont {Li}, \citenamefont {Teng},\ and\ \citenamefont
  {Wu}}]{Chen:2020soj}%
  \BibitemOpen
  \bibfield  {author} {\bibinfo {author} {\bibfnamefont {N.}~\bibnamefont
  {Chen}}, \bibinfo {author} {\bibfnamefont {T.}~\bibnamefont {Li}}, \bibinfo
  {author} {\bibfnamefont {Z.}~\bibnamefont {Teng}}, \ and\ \bibinfo {author}
  {\bibfnamefont {Y.}~\bibnamefont {Wu}},\ }\bibfield  {title} {\enquote
  {\bibinfo {title} {\emph{Collapsing Domain Walls in the Two-Higgs-Doublet
  Model and Deep Insights from the EDM}},}\ }\href {\doibase
  10.1007/JHEP10(2020)081} {\bibfield  {journal} {\bibinfo  {journal} {JHEP}\
  }\textbf {\bibinfo {volume} {10}},\ \bibinfo {pages} {081} (\bibinfo {year}
  {2020})},\ \Eprint {http://arxiv.org/abs/2006.06913} {arXiv:2006.06913
  [hep-ph]} \BibitemShut {NoStop}%
\bibitem [{\citenamefont {Battye}\ \emph {et~al.}(2020)\citenamefont {Battye},
  \citenamefont {Pilaftsis},\ and\ \citenamefont {Viatic}}]{Battye:2020jeu}%
  \BibitemOpen
  \bibfield  {author} {\bibinfo {author} {\bibfnamefont {R.~A.}\ \bibnamefont
  {Battye}}, \bibinfo {author} {\bibfnamefont {A.}~\bibnamefont {Pilaftsis}},  \bibinfo {author} {\bibfnamefont {D.~G.}\ \bibnamefont {Viatic}},\
  }\bibfield  {title} {\enquote {\bibinfo {title} {\emph{Domain Wall
  Constraints on Two-Higgs-Doublet Models with $Z_2$ Symmetry}},}\ }\href
  {\doibase 10.1103/PhysRevD.102.123536} {\bibfield  {journal} {\bibinfo
  {journal} {Phys. Rev. D}\ }\textbf {\bibinfo {volume} {102}},\ \bibinfo
  {pages} {123536} (\bibinfo {year} {2020})},\ \Eprint
  {http://arxiv.org/abs/2010.09840} {arXiv:2010.09840 [hep-ph]} \BibitemShut
  {NoStop}%
\bibitem [{\citenamefont {Craig}\ \emph {et~al.}(2021)\citenamefont {Craig},
  \citenamefont {Garcia~Garcia}, \citenamefont {Koszegi},\ and\ \citenamefont
  {McCune}}]{Craig:2020bnv}%
  \BibitemOpen
  \bibfield  {author} {\bibinfo {author} {\bibfnamefont {N.}~\bibnamefont
  {Craig}}, \bibinfo {author} {\bibfnamefont {I.}~\bibnamefont
  {Garcia~Garcia}}, \bibinfo {author} {\bibfnamefont {G.}~\bibnamefont
  {Koszegi}}, \ and\ \bibinfo {author} {\bibfnamefont {A.}~\bibnamefont
  {McCune}},\ }\bibfield  {title} {\enquote {\bibinfo {title} {\emph{P Not
  PQ}},}\ }\href {\doibase 10.1007/JHEP09(2021)130} {\bibfield  {journal}
  {\bibinfo  {journal} {JHEP}\ }\textbf {\bibinfo {volume} {09}},\ \bibinfo
  {pages} {130} (\bibinfo {year} {2021})},\ \Eprint
  {http://arxiv.org/abs/2012.13416} {arXiv:2012.13416 [hep-ph]} \BibitemShut
  {NoStop}%
\bibitem [{\citenamefont {Dunsky}\ \emph {et~al.}(2022)\citenamefont {Dunsky},
  \citenamefont {Ghoshal}, \citenamefont {Murayama}, \citenamefont
  {Sakakihara},\ and\ \citenamefont {White}}]{Dunsky:2021tih}%
  \BibitemOpen
  \bibfield  {author} {\bibinfo {author} {\bibfnamefont {D.~I.}\ \bibnamefont
  {Dunsky}}, \bibinfo {author} {\bibfnamefont {A.}~\bibnamefont {Ghoshal}},
  \bibinfo {author} {\bibfnamefont {H.}~\bibnamefont {Murayama}}, \bibinfo
  {author} {\bibfnamefont {Y.}~\bibnamefont {Sakakihara}}, \ and\ \bibinfo
  {author} {\bibfnamefont {G.}~\bibnamefont {White}},\ }\bibfield  {title}
  {\enquote {\bibinfo {title} {\emph{GUTs, Hybrid Topological Defects, and
  Gravitational Waves}},}\ }\href {\doibase 10.1103/PhysRevD.106.075030}
  {\bibfield  {journal} {\bibinfo  {journal} {Phys. Rev. D}\ }\textbf {\bibinfo
  {volume} {106}},\ \bibinfo {pages} {075030} (\bibinfo {year} {2022})},\
  \Eprint {http://arxiv.org/abs/2111.08750} {arXiv:2111.08750 [hep-ph]}
  \BibitemShut {NoStop}%
\bibitem [{\citenamefont {Blasi}\ \emph {et~al.}(2023)\citenamefont {Blasi},
  \citenamefont {Mariotti}, \citenamefont {Rase}, \citenamefont {Sevrin},\ and\
  \citenamefont {Turbang}}]{Blasi:2022ayo}%
  \BibitemOpen
  \bibfield  {author} {\bibinfo {author} {\bibfnamefont {S.}~\bibnamefont
  {Blasi}}, \bibinfo {author} {\bibfnamefont {A.}~\bibnamefont {Mariotti}},
  \bibinfo {author} {\bibfnamefont {A.}~\bibnamefont {Rase}}, \bibinfo {author}
  {\bibfnamefont {A.}~\bibnamefont {Sevrin}}, \ and\ \bibinfo {author}
  {\bibfnamefont {K.}~\bibnamefont {Turbang}},\ }\bibfield  {title} {\enquote
  {\bibinfo {title} {\emph{Friction on ALP Domain Walls and Gravitational
  Waves}},}\ }\href {\doibase 10.1088/1475-7516/2023/04/008} {\bibfield
  {journal} {\bibinfo  {journal} {JCAP}\ }\textbf {\bibinfo {volume} {04}},\
  \bibinfo {pages} {008} (\bibinfo {year} {2023})},\ \Eprint
  {http://arxiv.org/abs/2210.14246} {arXiv:2210.14246 [hep-ph]} \BibitemShut
  {NoStop}%
\bibitem [{\citenamefont {Barman}\ \emph {et~al.}(2022)\citenamefont {Barman},
  \citenamefont {Borah}, \citenamefont {Dasgupta},\ and\ \citenamefont
  {Ghoshal}}]{Barman:2022yos}%
  \BibitemOpen
  \bibfield  {author} {\bibinfo {author} {\bibfnamefont {B.}~\bibnamefont
  {Barman}}, \bibinfo {author} {\bibfnamefont {D.}~\bibnamefont {Borah}},
  \bibinfo {author} {\bibfnamefont {A.}~\bibnamefont {Dasgupta}}, \ and\
  \bibinfo {author} {\bibfnamefont {A.}~\bibnamefont {Ghoshal}},\ }\bibfield
  {title} {\enquote {\bibinfo {title} {\emph{Probing High Scale Dirac
  Leptogenesis via Gravitational Waves from Domain Walls}},}\ }\href {\doibase
  10.1103/PhysRevD.106.015007} {\bibfield  {journal} {\bibinfo  {journal}
  {Phys. Rev. D}\ }\textbf {\bibinfo {volume} {106}},\ \bibinfo {pages}
  {015007} (\bibinfo {year} {2022})},\ \Eprint
  {http://arxiv.org/abs/2205.03422} {arXiv:2205.03422 [hep-ph]} \BibitemShut
  {NoStop}%
\bibitem [{\citenamefont {Borah}\ and\ \citenamefont
  {Dasgupta}(2022)}]{Borah:2022wdy}%
  \BibitemOpen
  \bibfield  {author} {\bibinfo {author} {\bibfnamefont {D.}~\bibnamefont
  {Borah}}\ and\ \bibinfo {author} {\bibfnamefont {A.}~\bibnamefont
  {Dasgupta}},\ }\bibfield  {title} {\enquote {\bibinfo {title} {\emph{Probing
  Left-Right Symmetry via Gravitational Waves from Domain Walls}},}\ }\href
  {\doibase 10.1103/PhysRevD.106.035016} {\bibfield  {journal} {\bibinfo
  {journal} {Phys. Rev. D}\ }\textbf {\bibinfo {volume} {106}},\ \bibinfo
  {pages} {035016} (\bibinfo {year} {2022})},\ \Eprint
  {http://arxiv.org/abs/2205.12220} {arXiv:2205.12220 [hep-ph]} \BibitemShut
  {NoStop}%
\bibitem [{\citenamefont {King}\ \emph {et~al.}(2024)\citenamefont {King},
  \citenamefont {Marfatia},\ and\ \citenamefont {Rahat}}]{King:2023cgv}%
  \BibitemOpen
  \bibfield  {author} {\bibinfo {author} {\bibfnamefont {S.~F.}\ \bibnamefont
  {King}}, \bibinfo {author} {\bibfnamefont {D.}~\bibnamefont {Marfatia}}, \
  and\ \bibinfo {author} {\bibfnamefont {M.~H.}\ \bibnamefont {Rahat}},\
  }\bibfield  {title} {\enquote {\bibinfo {title} {\emph{Toward Distinguishing
  Dirac from Majorana Neutrino Mass with Gravitational Waves}},}\ }\href
  {\doibase 10.1103/PhysRevD.109.035014} {\bibfield  {journal} {\bibinfo
  {journal} {Phys. Rev. D}\ }\textbf {\bibinfo {volume} {109}},\ \bibinfo
  {pages} {035014} (\bibinfo {year} {2024})},\ \Eprint
  {http://arxiv.org/abs/2306.05389} {arXiv:2306.05389 [hep-ph]} \BibitemShut
  {NoStop}%
\bibitem [{\citenamefont {Saikawa}(2017)}]{Saikawa:2017hiv}%
  \BibitemOpen
  \bibfield  {author} {\bibinfo {author} {\bibfnamefont {K.}~\bibnamefont
  {Saikawa}},\ }\bibfield  {title} {\enquote {\bibinfo {title} {\emph{A Review
  of Gravitational Waves from Cosmic Domain Walls}},}\ }\href {\doibase
  10.3390/universe3020040} {\bibfield  {journal} {\bibinfo  {journal}
  {Universe}\ }\textbf {\bibinfo {volume} {3}},\ \bibinfo {pages} {40}
  (\bibinfo {year} {2017})},\ \Eprint {http://arxiv.org/abs/1703.02576}
  {arXiv:1703.02576 [hep-ph]} \BibitemShut {NoStop}%
\bibitem [{\citenamefont {Jiang}\ and\ \citenamefont
  {Huang}(2022)}]{Jiang:2022svq}%
  \BibitemOpen
  \bibfield  {author} {\bibinfo {author} {\bibfnamefont {Y.}~\bibnamefont
  {Jiang}}\ and\ \bibinfo {author} {\bibfnamefont {Q.-G.}\ \bibnamefont
  {Huang}},\ }\bibfield  {title} {\enquote {\bibinfo {title}
  {\emph{Implications for Cosmic Domain Walls from the First Three Observing
  Runs of LIGO-Virgo}},}\ }\href {\doibase 10.1103/PhysRevD.106.103036}
  {\bibfield  {journal} {\bibinfo  {journal} {Phys. Rev. D}\ }\textbf {\bibinfo
  {volume} {106}},\ \bibinfo {pages} {103036} (\bibinfo {year} {2022})},\
  \Eprint {http://arxiv.org/abs/2208.00697} {arXiv:2208.00697 [astro-ph.CO]}
  \BibitemShut {NoStop}%
\bibitem [{\citenamefont {Blanco-Pillado}\ and\ \citenamefont
  {Olum}(2017)}]{Blanco-Pillado:2017oxo}%
  \BibitemOpen
  \bibfield  {author} {\bibinfo {author} {\bibfnamefont {J.~J.}\ \bibnamefont
  {Blanco-Pillado}}\ and\ \bibinfo {author} {\bibfnamefont {K.~D.}\
  \bibnamefont {Olum}},\ }\bibfield  {title} {\enquote {\bibinfo {title}
  {\emph{Stochastic Gravitational Wave Background from Smoothed Cosmic String
  Loops}},}\ }\href {\doibase 10.1103/PhysRevD.96.104046} {\bibfield  {journal}
  {\bibinfo  {journal} {Phys. Rev. D}\ }\textbf {\bibinfo {volume} {96}},\
  \bibinfo {pages} {104046} (\bibinfo {year} {2017})},\ \Eprint
  {http://arxiv.org/abs/1709.02693} {arXiv:1709.02693 [astro-ph.CO]}
  \BibitemShut {NoStop}%
\bibitem [{\citenamefont {Ringeval}\ and\ \citenamefont
  {Suyama}(2017)}]{Ringeval:2017eww}%
  \BibitemOpen
  \bibfield  {author} {\bibinfo {author} {\bibfnamefont {C.}~\bibnamefont
  {Ringeval}}\ and\ \bibinfo {author} {\bibfnamefont {T.}~\bibnamefont
  {Suyama}},\ }\bibfield  {title} {\enquote {\bibinfo {title} {\emph{Stochastic
  Gravitational Waves from Cosmic String Loops in Scaling}},}\ }\href {\doibase
  10.1088/1475-7516/2017/12/027} {\bibfield  {journal} {\bibinfo  {journal}
  {JCAP}\ }\textbf {\bibinfo {volume} {12}},\ \bibinfo {pages} {027} (\bibinfo
  {year} {2017})},\ \Eprint {http://arxiv.org/abs/1709.03845} {arXiv:1709.03845
  [astro-ph.CO]} \BibitemShut {NoStop}%
\bibitem [{\citenamefont {Cui}\ \emph {et~al.}(2018)\citenamefont {Cui},
  \citenamefont {Lewicki}, \citenamefont {Morrissey},\ and\ \citenamefont
  {Wells}}]{Cui:2017ufi}%
  \BibitemOpen
  \bibfield  {author} {\bibinfo {author} {\bibfnamefont {Y.}~\bibnamefont
  {Cui}}, \bibinfo {author} {\bibfnamefont {M.}~\bibnamefont {Lewicki}},
  \bibinfo {author} {\bibfnamefont {D.~E.}\ \bibnamefont {Morrissey}}, \ and\
  \bibinfo {author} {\bibfnamefont {J.~D.}\ \bibnamefont {Wells}},\ }\bibfield
  {title} {\enquote {\bibinfo {title} {\emph{Cosmic Archaeology with
  Gravitational Waves from Cosmic Strings}},}\ }\href {\doibase
  10.1103/PhysRevD.97.123505} {\bibfield  {journal} {\bibinfo  {journal} {Phys.
  Rev. D}\ }\textbf {\bibinfo {volume} {97}},\ \bibinfo {pages} {123505}
  (\bibinfo {year} {2018})},\ \Eprint {http://arxiv.org/abs/1711.03104}
  {arXiv:1711.03104 [hep-ph]} \BibitemShut {NoStop}%
\bibitem [{\citenamefont {Cui}\ \emph {et~al.}(2019)\citenamefont {Cui},
  \citenamefont {Lewicki}, \citenamefont {Morrissey},\ and\ \citenamefont
  {Wells}}]{Cui:2018rwi}%
  \BibitemOpen
  \bibfield  {author} {\bibinfo {author} {\bibfnamefont {Y.}~\bibnamefont
  {Cui}}, \bibinfo {author} {\bibfnamefont {M.}~\bibnamefont {Lewicki}},
  \bibinfo {author} {\bibfnamefont {D.~E.}\ \bibnamefont {Morrissey}}, \ and\
  \bibinfo {author} {\bibfnamefont {J.~D.}\ \bibnamefont {Wells}},\ }\bibfield
  {title} {\enquote {\bibinfo {title} {\emph{Probing the Pre-BBN Universe with
  Gravitational Waves from Cosmic Strings}},}\ }\href {\doibase
  10.1007/JHEP01(2019)081} {\bibfield  {journal} {\bibinfo  {journal} {JHEP}\
  }\textbf {\bibinfo {volume} {01}},\ \bibinfo {pages} {081} (\bibinfo {year}
  {2019})},\ \Eprint {http://arxiv.org/abs/1808.08968} {arXiv:1808.08968
  [hep-ph]} \BibitemShut {NoStop}%
\bibitem [{\citenamefont {Guedes}\ \emph {et~al.}(2018)\citenamefont {Guedes},
  \citenamefont {Avelino},\ and\ \citenamefont {Sousa}}]{Guedes:2018afo}%
  \BibitemOpen
  \bibfield  {author} {\bibinfo {author} {\bibfnamefont {G.~S.~F.}\
  \bibnamefont {Guedes}}, \bibinfo {author} {\bibfnamefont {P.~P.}\
  \bibnamefont {Avelino}}, \ and\ \bibinfo {author} {\bibfnamefont
  {L.}~\bibnamefont {Sousa}},\ }\bibfield  {title} {\enquote {\bibinfo {title}
  {\emph{Signature of Inflation in the Stochastic Gravitational Wave Background
  Generated by Cosmic String Networks}},}\ }\href {\doibase
  10.1103/PhysRevD.98.123505} {\bibfield  {journal} {\bibinfo  {journal} {Phys.
  Rev. D}\ }\textbf {\bibinfo {volume} {98}},\ \bibinfo {pages} {123505}
  (\bibinfo {year} {2018})},\ \Eprint {http://arxiv.org/abs/1809.10802}
  {arXiv:1809.10802 [astro-ph.CO]} \BibitemShut {NoStop}%
\bibitem [{\citenamefont {Dror}\ \emph {et~al.}(2020)\citenamefont {Dror},
  \citenamefont {Hiramatsu}, \citenamefont {Kohri}, \citenamefont {Murayama},\
  and\ \citenamefont {White}}]{Dror:2019syi}%
  \BibitemOpen
  \bibfield  {author} {\bibinfo {author} {\bibfnamefont {J.~A.}\ \bibnamefont
  {Dror}}, \bibinfo {author} {\bibfnamefont {T.}~\bibnamefont {Hiramatsu}},
  \bibinfo {author} {\bibfnamefont {K.}~\bibnamefont {Kohri}}, \bibinfo
  {author} {\bibfnamefont {H.}~\bibnamefont {Murayama}}, \ and\ \bibinfo
  {author} {\bibfnamefont {G.}~\bibnamefont {White}},\ }\bibfield  {title}
  {\enquote {\bibinfo {title} {\emph{Testing the Seesaw Mechanism and
  Leptogenesis with Gravitational Waves}},}\ }\href {\doibase
  10.1103/PhysRevLett.124.041804} {\bibfield  {journal} {\bibinfo  {journal}
  {Phys. Rev. Lett.}\ }\textbf {\bibinfo {volume} {124}},\ \bibinfo {pages}
  {041804} (\bibinfo {year} {2020})},\ \Eprint
  {http://arxiv.org/abs/1908.03227} {arXiv:1908.03227 [hep-ph]} \BibitemShut
  {NoStop}%
\bibitem [{\citenamefont {Gouttenoire}\ \emph
  {et~al.}(2020{\natexlab{a}})\citenamefont {Gouttenoire}, \citenamefont
  {Servant},\ and\ \citenamefont {Simakachorn}}]{Gouttenoire:2019rtn}%
  \BibitemOpen
  \bibfield  {author} {\bibinfo {author} {\bibfnamefont {Y.}~\bibnamefont
  {Gouttenoire}}, \bibinfo {author} {\bibfnamefont {G.}~\bibnamefont
  {Servant}}, \ and\ \bibinfo {author} {\bibfnamefont {P.}~\bibnamefont
  {Simakachorn}},\ }\bibfield  {title} {\enquote {\bibinfo {title} {\emph{BSM
  with Cosmic Strings: Heavy, up to EeV Mass, Unstable Particles}},}\ }\href
  {\doibase 10.1088/1475-7516/2020/07/016} {\bibfield  {journal} {\bibinfo
  {journal} {JCAP}\ }\textbf {\bibinfo {volume} {07}},\ \bibinfo {pages} {016}
  (\bibinfo {year} {2020}{\natexlab{a}})},\ \Eprint
  {http://arxiv.org/abs/1912.03245} {arXiv:1912.03245 [hep-ph]} \BibitemShut
  {NoStop}%
\bibitem [{\citenamefont {Buchmuller}\ \emph {et~al.}(2020)\citenamefont
  {Buchmuller}, \citenamefont {Domcke}, \citenamefont {Murayama},\ and\
  \citenamefont {Schmitz}}]{Buchmuller:2019gfy}%
  \BibitemOpen
  \bibfield  {author} {\bibinfo {author} {\bibfnamefont {W.}~\bibnamefont
  {Buchmuller}}, \bibinfo {author} {\bibfnamefont {V.}~\bibnamefont {Domcke}},
  \bibinfo {author} {\bibfnamefont {H.}~\bibnamefont {Murayama}}, \ and\
  \bibinfo {author} {\bibfnamefont {K.}~\bibnamefont {Schmitz}},\ }\bibfield
  {title} {\enquote {\bibinfo {title} {\emph{Probing the Scale of Grand
  Unification with Gravitational Waves}},}\ }\href {\doibase
  10.1016/j.physletb.2020.135764} {\bibfield  {journal} {\bibinfo  {journal}
  {Phys. Lett. B}\ }\textbf {\bibinfo {volume} {809}},\ \bibinfo {pages}
  {135764} (\bibinfo {year} {2020})},\ \Eprint
  {http://arxiv.org/abs/1912.03695} {arXiv:1912.03695 [hep-ph]} \BibitemShut
  {NoStop}%
\bibitem [{\citenamefont {King}\ \emph {et~al.}(2021)\citenamefont {King},
  \citenamefont {Pascoli}, \citenamefont {Turner},\ and\ \citenamefont
  {Zhou}}]{King:2020hyd}%
  \BibitemOpen
  \bibfield  {author} {\bibinfo {author} {\bibfnamefont {Stephen~F.}\
  \bibnamefont {King}}, \bibinfo {author} {\bibfnamefont {Silvia}\ \bibnamefont
  {Pascoli}}, \bibinfo {author} {\bibfnamefont {Jessica}\ \bibnamefont
  {Turner}}, \ and\ \bibinfo {author} {\bibfnamefont {Ye-Ling}\ \bibnamefont
  {Zhou}},\ }\bibfield  {title} {\enquote {\bibinfo {title} {{Gravitational
  Waves and Proton Decay: Complementary Windows into Grand Unified
  Theories}},}\ }\href {\doibase 10.1103/PhysRevLett.126.021802} {\bibfield
  {journal} {\bibinfo  {journal} {Phys. Rev. Lett.}\ }\textbf {\bibinfo
  {volume} {126}},\ \bibinfo {pages} {021802} (\bibinfo {year} {2021})},\
  \Eprint {http://arxiv.org/abs/2005.13549} {arXiv:2005.13549 [hep-ph]}
  \BibitemShut {NoStop}%
\bibitem [{\citenamefont {Gouttenoire}\ \emph
  {et~al.}(2020{\natexlab{b}})\citenamefont {Gouttenoire}, \citenamefont
  {Servant},\ and\ \citenamefont {Simakachorn}}]{Gouttenoire:2019kij}%
  \BibitemOpen
  \bibfield  {author} {\bibinfo {author} {\bibfnamefont {Y.}~\bibnamefont
  {Gouttenoire}}, \bibinfo {author} {\bibfnamefont {G.}~\bibnamefont
  {Servant}}, \ and\ \bibinfo {author} {\bibfnamefont {P.}~\bibnamefont
  {Simakachorn}},\ }\bibfield  {title} {\enquote {\bibinfo {title}
  {\emph{Beyond the Standard Models with Cosmic Strings}},}\ }\href {\doibase
  10.1088/1475-7516/2020/07/032} {\bibfield  {journal} {\bibinfo  {journal}
  {JCAP}\ }\textbf {\bibinfo {volume} {07}},\ \bibinfo {pages} {032} (\bibinfo
  {year} {2020}{\natexlab{b}})},\ \Eprint {http://arxiv.org/abs/1912.02569}
  {arXiv:1912.02569 [hep-ph]} \BibitemShut {NoStop}%
\bibitem [{\citenamefont {Abbott}\ \emph {et~al.}(2021)\citenamefont {Abbott}
  \emph {et~al.}}]{LIGOScientific:2021nrg}%
  \BibitemOpen
  \bibfield  {author} {\bibinfo {author} {\bibfnamefont {R.}~\bibnamefont
  {Abbott}} \emph {et~al.} (\bibinfo {collaboration} {LIGO Scientific, Virgo,
  KAGRA}),\ }\bibfield  {title} {\enquote {\bibinfo {title} {\emph{Constraints
  on Cosmic Strings Using Data from the Third Advanced LIGO\textendash{}Virgo
  Observing Run}},}\ }\href {\doibase 10.1103/PhysRevLett.126.241102}
  {\bibfield  {journal} {\bibinfo  {journal} {Phys. Rev. Lett.}\ }\textbf
  {\bibinfo {volume} {126}},\ \bibinfo {pages} {241102} (\bibinfo {year}
  {2021})},\ \Eprint {http://arxiv.org/abs/2101.12248} {arXiv:2101.12248
  [gr-qc]} \BibitemShut {NoStop}%
\bibitem [{\citenamefont {Debnath}\ and\ \citenamefont
  {Fileviez~Perez}(2023)}]{Debnath:2023akj}%
  \BibitemOpen
  \bibfield  {author} {\bibinfo {author} {\bibfnamefont {H.}~\bibnamefont
  {Debnath}}\ and\ \bibinfo {author} {\bibfnamefont {P.}~\bibnamefont
  {Fileviez~Perez}},\ }\bibfield  {title} {\enquote {\bibinfo {title}
  {\emph{Low Scale Seesaw Mechanism with Local Lepton Number}},}\ }\href
  {\doibase 10.1103/PhysRevD.108.075009} {\bibfield  {journal} {\bibinfo
  {journal} {Phys. Rev. D}\ }\textbf {\bibinfo {volume} {108}},\ \bibinfo
  {pages} {075009} (\bibinfo {year} {2023})},\ \Eprint
  {http://arxiv.org/abs/2307.03646} {arXiv:2307.03646 [hep-ph]} \BibitemShut
  {NoStop}%
\bibitem [{\citenamefont {Reitze}\ \emph {et~al.}(2019)\citenamefont {Reitze}
  \emph {et~al.}}]{Reitze:2019iox}%
  \BibitemOpen
  \bibfield  {author} {\bibinfo {author} {\bibfnamefont {D.}~\bibnamefont
  {Reitze}} \emph {et~al.},\ }\bibfield  {title} {\enquote {\bibinfo {title}
  {\emph{Cosmic Explorer: The U.S. Contribution to Gravitational-Wave Astronomy
  beyond LIGO}},}\ }\href@noop {} {\bibfield  {journal} {\bibinfo  {journal}
  {Bull. Am. Astron. Soc.}\ }\textbf {\bibinfo {volume} {51}},\ \bibinfo
  {pages} {035} (\bibinfo {year} {2019})},\ \Eprint
  {http://arxiv.org/abs/1907.04833} {arXiv:1907.04833 [astro-ph.IM]}
  \BibitemShut {NoStop}%
\bibitem [{\citenamefont {Punturo}\ \emph {et~al.}(2010)\citenamefont {Punturo}
  \emph {et~al.}}]{Punturo:2010zz}%
  \BibitemOpen
  \bibfield  {author} {\bibinfo {author} {\bibfnamefont {M.}~\bibnamefont
  {Punturo}} \emph {et~al.},\ }\bibfield  {title} {\enquote {\bibinfo {title}
  {\emph{The Einstein Telescope: A Third-Generation Gravitational Wave
  Observatory}},}\ }\href {\doibase 10.1088/0264-9381/27/19/194002} {\bibfield
  {journal} {\bibinfo  {journal} {Class. Quant. Grav.}\ }\textbf {\bibinfo
  {volume} {27}},\ \bibinfo {pages} {194002} (\bibinfo {year}
  {2010})}\BibitemShut {NoStop}%
\bibitem [{\citenamefont {Kawamura}\ \emph {et~al.}(2011)\citenamefont
  {Kawamura} \emph {et~al.}}]{Kawamura:2011zz}%
  \BibitemOpen
  \bibfield  {author} {\bibinfo {author} {\bibfnamefont {S.}~\bibnamefont
  {Kawamura}} \emph {et~al.},\ }\bibfield  {title} {\enquote {\bibinfo {title}
  {\emph{The Japanese Space Gravitational Wave Antenna: DECIGO}},}\ }\href
  {\doibase 10.1088/0264-9381/28/9/094011} {\bibfield  {journal} {\bibinfo
  {journal} {Class. Quant. Grav.}\ }\textbf {\bibinfo {volume} {28}},\ \bibinfo
  {pages} {094011} (\bibinfo {year} {2011})}\BibitemShut {NoStop}%
\bibitem [{\citenamefont {Crowder}\ and\ \citenamefont
  {Cornish}(2005)}]{Crowder:2005nr}%
  \BibitemOpen
  \bibfield  {author} {\bibinfo {author} {\bibfnamefont {J.}~\bibnamefont
  {Crowder}}\ and\ \bibinfo {author} {\bibfnamefont {N.~J.}\ \bibnamefont
  {Cornish}},\ }\bibfield  {title} {\enquote {\bibinfo {title} {\emph{Beyond
  LISA: Exploring Future Gravitational Wave Missions}},}\ }\href {\doibase
  10.1103/PhysRevD.72.083005} {\bibfield  {journal} {\bibinfo  {journal} {Phys.
  Rev. D}\ }\textbf {\bibinfo {volume} {72}},\ \bibinfo {pages} {083005}
  (\bibinfo {year} {2005})},\ \Eprint {http://arxiv.org/abs/gr-qc/0506015}
  {arXiv:gr-qc/0506015} \BibitemShut {NoStop}%
\bibitem [{\citenamefont {Amaro-Seoane}\ \emph {et~al.}(2017)\citenamefont
  {Amaro-Seoane} \emph {et~al.}}]{Audley:2017drz}%
  \BibitemOpen
  \bibfield  {author} {\bibinfo {author} {\bibfnamefont {P.}~\bibnamefont
  {Amaro-Seoane}} \emph {et~al.} (\bibinfo {collaboration} {LISA}),\ }\bibfield
   {title} {\enquote {\bibinfo {title} {\emph{Laser Interferometer Space
  Antenna}},}\ }\href@noop {} {\  (\bibinfo {year} {2017})},\ \Eprint
  {http://arxiv.org/abs/1702.00786} {arXiv:1702.00786 [astro-ph.IM]}
  \BibitemShut {NoStop}%
\bibitem [{\citenamefont {Sesana}\ \emph {et~al.}(2021)\citenamefont {Sesana}
  \emph {et~al.}}]{Sesana:2019vho}%
  \BibitemOpen
  \bibfield  {author} {\bibinfo {author} {\bibfnamefont {Alberto}\ \bibnamefont
  {Sesana}} \emph {et~al.},\ }\bibfield  {title} {\enquote {\bibinfo {title}
  {{Unveiling the gravitational universe at $\mu$-Hz frequencies}},}\ }\href
  {\doibase 10.1007/s10686-021-09709-9} {\bibfield  {journal} {\bibinfo
  {journal} {Exper. Astron.}\ }\textbf {\bibinfo {volume} {51}},\ \bibinfo
  {pages} {1333--1383} (\bibinfo {year} {2021})},\ \Eprint
  {http://arxiv.org/abs/1908.11391} {arXiv:1908.11391 [astro-ph.IM]}
  \BibitemShut {NoStop}%
\bibitem [{\citenamefont {Arzoumanian}\ \emph {et~al.}(2018)\citenamefont
  {Arzoumanian} \emph {et~al.}}]{NANOGRAV:2018hou}%
  \BibitemOpen
  \bibfield  {author} {\bibinfo {author} {\bibfnamefont {Z.}~\bibnamefont
  {Arzoumanian}} \emph {et~al.} (\bibinfo {collaboration} {NANOGRAV}),\
  }\bibfield  {title} {\enquote {\bibinfo {title} {\emph{The NANOGrav 11-Year
  Data Set: Pulsar-Timing Constraints on the Stochastic Gravitational-Wave
  Background}},}\ }\href {\doibase 10.3847/1538-4357/aabd3b} {\bibfield
  {journal} {\bibinfo  {journal} {Astrophys. J.}\ }\textbf {\bibinfo {volume}
  {859}},\ \bibinfo {pages} {47} (\bibinfo {year} {2018})},\ \Eprint
  {http://arxiv.org/abs/1801.02617} {arXiv:1801.02617 [astro-ph.HE]}
  \BibitemShut {NoStop}%
\bibitem [{\citenamefont {Weltman}\ \emph {et~al.}(2020)\citenamefont {Weltman}
  \emph {et~al.}}]{Weltman:2018zrl}%
  \BibitemOpen
  \bibfield  {author} {\bibinfo {author} {\bibfnamefont {A.}~\bibnamefont
  {Weltman}} \emph {et~al.},\ }\bibfield  {title} {\enquote {\bibinfo {title}
  {\emph{Fundamental Physics with the Square Kilometre Array}},}\ }\href
  {\doibase 10.1017/pasa.2019.42} {\bibfield  {journal} {\bibinfo  {journal}
  {Publ. Astron. Soc. Austral.}\ }\textbf {\bibinfo {volume} {37}},\ \bibinfo
  {pages} {e002} (\bibinfo {year} {2020})},\ \Eprint
  {http://arxiv.org/abs/1810.02680} {arXiv:1810.02680 [astro-ph.CO]}
  \BibitemShut {NoStop}%
\bibitem [{\citenamefont {Garcia-Bellido}\ \emph {et~al.}(2021)\citenamefont
  {Garcia-Bellido}, \citenamefont {Murayama},\ and\ \citenamefont
  {White}}]{Garcia-Bellido:2021zgu}%
  \BibitemOpen
  \bibfield  {author} {\bibinfo {author} {\bibfnamefont {Juan}\ \bibnamefont
  {Garcia-Bellido}}, \bibinfo {author} {\bibfnamefont {Hitoshi}\ \bibnamefont
  {Murayama}}, \ and\ \bibinfo {author} {\bibfnamefont {Graham}\ \bibnamefont
  {White}},\ }\bibfield  {title} {\enquote {\bibinfo {title} {{Exploring the
  early Universe with Gaia and Theia}},}\ }\href {\doibase
  10.1088/1475-7516/2021/12/023} {\bibfield  {journal} {\bibinfo  {journal}
  {JCAP}\ }\textbf {\bibinfo {volume} {12}},\ \bibinfo {pages} {023} (\bibinfo
  {year} {2021})},\ \Eprint {http://arxiv.org/abs/2104.04778} {arXiv:2104.04778
  [hep-ph]} \BibitemShut {NoStop}%
\bibitem [{\citenamefont {{Brown}}\ \emph {et~al.}(2018)\citenamefont {{Brown}}
  \emph {et~al.}}]{Gaia1}%
  \BibitemOpen
  \bibfield  {author} {\bibinfo {author} {\bibfnamefont {A.~G.~A.}\
  \bibnamefont {{Brown}}} \emph {et~al.} (\bibinfo {collaboration} {Gaia}),\
  }\bibfield  {title} {\enquote {\bibinfo {title} {\emph{Gaia Data Release 2.
  Summary of the Contents and Survey Properties}},}\ }\href {\doibase
  10.1051/0004-6361/201833051} {\bibfield  {journal} {\bibinfo  {journal}
  {Astron. Astrophys.}\ }\textbf {\bibinfo {volume} {616}},\ \bibinfo {eid}
  {A1} (\bibinfo {year} {2018})},\ \Eprint {http://arxiv.org/abs/1804.09365}
  {arXiv:1804.09365 [astro-ph.GA]} \BibitemShut {NoStop}%
\bibitem [{\citenamefont {Moore}\ \emph {et~al.}(2017)\citenamefont {Moore},
  \citenamefont {Mihaylov}, \citenamefont {Lasenby},\ and\ \citenamefont
  {Gilmore}}]{Moore:2017ity}%
  \BibitemOpen
  \bibfield  {author} {\bibinfo {author} {\bibfnamefont {C.~J.}\ \bibnamefont
  {Moore}}, \bibinfo {author} {\bibfnamefont {D.~P.}\ \bibnamefont {Mihaylov}},
  \bibinfo {author} {\bibfnamefont {A.}~\bibnamefont {Lasenby}}, \ and\
  \bibinfo {author} {\bibfnamefont {G.}~\bibnamefont {Gilmore}},\ }\bibfield
  {title} {\enquote {\bibinfo {title} {\emph{Astrometric Search Method for
  Individually Resolvable Gravitational Wave Sources with Gaia}},}\ }\href
  {\doibase 10.1103/PhysRevLett.119.261102} {\bibfield  {journal} {\bibinfo
  {journal} {Phys. Rev. Lett.}\ }\textbf {\bibinfo {volume} {119}},\ \bibinfo
  {pages} {261102} (\bibinfo {year} {2017})},\ \Eprint
  {http://arxiv.org/abs/1707.06239} {arXiv:1707.06239 [astro-ph.IM]}
  \BibitemShut {NoStop}%
\bibitem [{\citenamefont {Heeck}\ and\ \citenamefont
  {Patel}(2019)}]{Heeck:2019guh}%
  \BibitemOpen
  \bibfield  {author} {\bibinfo {author} {\bibfnamefont {J.}~\bibnamefont
  {Heeck}}\ and\ \bibinfo {author} {\bibfnamefont {H.~H.}\ \bibnamefont
  {Patel}},\ }\bibfield  {title} {\enquote {\bibinfo {title} {\emph{Majoron at
  Two Loops}},}\ }\href {\doibase 10.1103/PhysRevD.100.095015} {\bibfield
  {journal} {\bibinfo  {journal} {Phys. Rev. D}\ }\textbf {\bibinfo {volume}
  {100}},\ \bibinfo {pages} {095015} (\bibinfo {year} {2019})},\ \Eprint
  {http://arxiv.org/abs/1909.02029} {arXiv:1909.02029 [hep-ph]} \BibitemShut
  {NoStop}%
\bibitem [{\citenamefont {Aghanim}\ \emph {et~al.}(2020)\citenamefont {Aghanim}
  \emph {et~al.}}]{Planck:2018vyg}%
  \BibitemOpen
  \bibfield  {author} {\bibinfo {author} {\bibfnamefont {N.}~\bibnamefont
  {Aghanim}} \emph {et~al.} (\bibinfo {collaboration} {Planck}),\ }\bibfield
  {title} {\enquote {\bibinfo {title} {\emph{Planck 2018 Results. VI.
  Cosmological Parameters}},}\ }\href {\doibase 10.1051/0004-6361/201833910}
  {\bibfield  {journal} {\bibinfo  {journal} {Astron. Astrophys.}\ }\textbf
  {\bibinfo {volume} {641}},\ \bibinfo {pages} {A6} (\bibinfo {year} {2020})},\
  \bibinfo {note} {[Erratum: Astron.Astrophys. 652, C4 (2021)]},\ \Eprint
  {http://arxiv.org/abs/1807.06209} {arXiv:1807.06209 [astro-ph.CO]}
  \BibitemShut {NoStop}%
\bibitem [{\citenamefont {Quiros}(1999)}]{Quiros:1999jp}%
  \BibitemOpen
  \bibfield  {author} {\bibinfo {author} {\bibfnamefont {M.}~\bibnamefont
  {Quiros}},\ }\bibfield  {title} {\enquote {\bibinfo {title} {\emph{Finite
  Temperature Field Theory and Phase Transitions}},}\ }in\ \href@noop {} {\emph
  {\bibinfo {booktitle} {{ICTP Summer School in High-Energy Physics and
  Cosmology}}}}\ (\bibinfo {year} {1999})\ pp.\ \bibinfo {pages} {187--259},\
  \Eprint {http://arxiv.org/abs/hep-ph/9901312} {arXiv:hep-ph/9901312}
  \BibitemShut {NoStop}%
\bibitem [{\citenamefont {Ginzburg}\ and\ \citenamefont
  {Krawczyk}(2005)}]{Ginzburg:2004vp}%
  \BibitemOpen
  \bibfield  {author} {\bibinfo {author} {\bibfnamefont {I.~F.}\ \bibnamefont
  {Ginzburg}}\ and\ \bibinfo {author} {\bibfnamefont {M.}~\bibnamefont
  {Krawczyk}},\ }\bibfield  {title} {\enquote {\bibinfo {title}
  {\emph{Symmetries of Two Higgs Doublet Model and CP Violation}},}\ }\href
  {\doibase 10.1103/PhysRevD.72.115013} {\bibfield  {journal} {\bibinfo
  {journal} {Phys. Rev. D}\ }\textbf {\bibinfo {volume} {72}},\ \bibinfo
  {pages} {115013} (\bibinfo {year} {2005})},\ \Eprint
  {http://arxiv.org/abs/hep-ph/0408011} {arXiv:hep-ph/0408011} \BibitemShut
  {NoStop}%
\bibitem [{\citenamefont {Battye}\ \emph {et~al.}(2011)\citenamefont {Battye},
  \citenamefont {Brawn},\ and\ \citenamefont {Pilaftsis}}]{Battye:2011jj}%
  \BibitemOpen
  \bibfield  {author} {\bibinfo {author} {\bibfnamefont {R.~A.}\ \bibnamefont
  {Battye}}, \bibinfo {author} {\bibfnamefont {G.~D.}\ \bibnamefont {Brawn}}, \
  and\ \bibinfo {author} {\bibfnamefont {A.}~\bibnamefont {Pilaftsis}},\
  }\bibfield  {title} {\enquote {\bibinfo {title} {\emph{Vacuum Topology of the
  Two Higgs Doublet Model}},}\ }\href {\doibase 10.1007/JHEP08(2011)020}
  {\bibfield  {journal} {\bibinfo  {journal} {JHEP}\ }\textbf {\bibinfo
  {volume} {08}},\ \bibinfo {pages} {020} (\bibinfo {year} {2011})},\ \Eprint
  {http://arxiv.org/abs/1106.3482} {arXiv:1106.3482 [hep-ph]} \BibitemShut
  {NoStop}%
\bibitem [{\citenamefont {{Wolfram Research{,} Inc.}}()}]{Wolfram}%
  \BibitemOpen
  \bibfield  {author} {\bibinfo {author} {\bibnamefont {{Wolfram Research{,}
  Inc.}}},\ }\bibfield  {title} {\enquote {\bibinfo {title} {\emph{Wolfram
  Programming Lab, {V}ersion 13.2}},}\ }\href@noop {} {\bibinfo  {journal}
  {{Champaign, IL, 2023}}\ }\BibitemShut {NoStop}%
\bibitem [{\citenamefont {Linde}(1983)}]{LINDE1983421}%
  \BibitemOpen
\bibfield  {journal} {  }\bibfield  {author} {\bibinfo {author} {\bibfnamefont
  {A.~D.}\ \bibnamefont {Linde}},\ }\bibfield  {title} {\enquote {\bibinfo
  {title} {\emph{Decay of the False Vacuum at Finite Temperature}},}\ }\href
  {\doibase https://doi.org/10.1016/0550-3213(83)90293-6} {\bibfield  {journal}
  {\bibinfo  {journal} {Nuclear Physics B}\ }\textbf {\bibinfo {volume}
  {216}},\ \bibinfo {pages} {421 -- 445} (\bibinfo {year} {1983})}\BibitemShut
  {NoStop}%
\bibitem [{\citenamefont {Espinosa}\ \emph {et~al.}(2010)\citenamefont
  {Espinosa}, \citenamefont {Konstandin}, \citenamefont {No},\ and\
  \citenamefont {Servant}}]{Espinosa:2010hh}%
  \BibitemOpen
  \bibfield  {author} {\bibinfo {author} {\bibfnamefont {J.~R.}\ \bibnamefont
  {Espinosa}}, \bibinfo {author} {\bibfnamefont {T.}~\bibnamefont
  {Konstandin}}, \bibinfo {author} {\bibfnamefont {J.~M.}\ \bibnamefont {No}},
  \ and\ \bibinfo {author} {\bibfnamefont {G.}~\bibnamefont {Servant}},\
  }\bibfield  {title} {\enquote {\bibinfo {title} {\emph{Energy Budget of
  Cosmological First-Order Phase Transitions}},}\ }\href {\doibase
  10.1088/1475-7516/2010/06/028} {\bibfield  {journal} {\bibinfo  {journal}
  {JCAP}\ }\textbf {\bibinfo {volume} {06}},\ \bibinfo {pages} {028} (\bibinfo
  {year} {2010})},\ \Eprint {http://arxiv.org/abs/1004.4187} {arXiv:1004.4187
  [hep-ph]} \BibitemShut {NoStop}%
\bibitem [{\citenamefont {Caprini}\ \emph {et~al.}(2016)\citenamefont {Caprini}
  \emph {et~al.}}]{Caprini:2015zlo}%
  \BibitemOpen
  \bibfield  {author} {\bibinfo {author} {\bibfnamefont {C.}~\bibnamefont
  {Caprini}} \emph {et~al.},\ }\bibfield  {title} {\enquote {\bibinfo {title}
  {\emph{Science with the Space-Based Interferometer eLISA. II: Gravitational
  Waves from Cosmological Phase Transitions}},}\ }\href {\doibase
  10.1088/1475-7516/2016/04/001} {\bibfield  {journal} {\bibinfo  {journal}
  {JCAP}\ }\textbf {\bibinfo {volume} {04}},\ \bibinfo {pages} {001} (\bibinfo
  {year} {2016})},\ \Eprint {http://arxiv.org/abs/1512.06239} {arXiv:1512.06239
  [astro-ph.CO]} \BibitemShut {NoStop}%
\bibitem [{\citenamefont {Huber}\ and\ \citenamefont
  {Konstandin}(2008)}]{Huber:2008hg}%
  \BibitemOpen
  \bibfield  {author} {\bibinfo {author} {\bibfnamefont {S.~J.}\ \bibnamefont
  {Huber}}\ and\ \bibinfo {author} {\bibfnamefont {T.}~\bibnamefont
  {Konstandin}},\ }\bibfield  {title} {\enquote {\bibinfo {title}
  {\emph{Gravitational Wave Production by Collisions: More Bubbles}},}\ }\href
  {\doibase 10.1088/1475-7516/2008/09/022} {\bibfield  {journal} {\bibinfo
  {journal} {JCAP}\ }\textbf {\bibinfo {volume} {09}},\ \bibinfo {pages} {022}
  (\bibinfo {year} {2008})},\ \Eprint {http://arxiv.org/abs/0806.1828}
  {arXiv:0806.1828 [hep-ph]} \BibitemShut {NoStop}%
\bibitem [{\citenamefont {Lewicki}\ and\ \citenamefont
  {Vaskonen}(2021)}]{Lewicki:2020azd}%
  \BibitemOpen
  \bibfield  {author} {\bibinfo {author} {\bibfnamefont {M.}~\bibnamefont
  {Lewicki}}\ and\ \bibinfo {author} {\bibfnamefont {V.}~\bibnamefont
  {Vaskonen}},\ }\bibfield  {title} {\enquote {\bibinfo {title}
  {\emph{Gravitational Waves from Colliding Vacuum Bubbles in Gauge
  Theories}},}\ }\href {\doibase 10.1140/epjc/s10052-021-09232-3} {\bibfield
  {journal} {\bibinfo  {journal} {Eur. Phys. J. C}\ }\textbf {\bibinfo {volume}
  {81}},\ \bibinfo {pages} {437} (\bibinfo {year} {2021})},\ \Eprint
  {http://arxiv.org/abs/2012.07826} {arXiv:2012.07826 [astro-ph.CO]}
  \BibitemShut {NoStop}%
\bibitem [{\citenamefont {Kamionkowski}\ \emph {et~al.}(1994)\citenamefont
  {Kamionkowski}, \citenamefont {Kosowsky},\ and\ \citenamefont
  {Turner}}]{Kamionkowski:1993fg}%
  \BibitemOpen
  \bibfield  {author} {\bibinfo {author} {\bibfnamefont {M.}~\bibnamefont
  {Kamionkowski}}, \bibinfo {author} {\bibfnamefont {A.}~\bibnamefont
  {Kosowsky}}, \ and\ \bibinfo {author} {\bibfnamefont {M.~S.}\ \bibnamefont
  {Turner}},\ }\bibfield  {title} {\enquote {\bibinfo {title}
  {\emph{Gravitational Radiation from First Order Phase Transitions}},}\ }\href
  {\doibase 10.1103/PhysRevD.49.2837} {\bibfield  {journal} {\bibinfo
  {journal} {Phys. Rev. D}\ }\textbf {\bibinfo {volume} {49}},\ \bibinfo
  {pages} {2837--2851} (\bibinfo {year} {1994})},\ \Eprint
  {http://arxiv.org/abs/astro-ph/9310044} {arXiv:astro-ph/9310044} \BibitemShut
  {NoStop}%
\bibitem [{\citenamefont {Hindmarsh}\ \emph {et~al.}(2014)\citenamefont
  {Hindmarsh}, \citenamefont {Huber}, \citenamefont {Rummukainen},\ and\
  \citenamefont {Weir}}]{Hindmarsh:2013xza}%
  \BibitemOpen
  \bibfield  {author} {\bibinfo {author} {\bibfnamefont {M.}~\bibnamefont
  {Hindmarsh}}, \bibinfo {author} {\bibfnamefont {S.~J.}\ \bibnamefont
  {Huber}}, \bibinfo {author} {\bibfnamefont {K.}~\bibnamefont {Rummukainen}},
  \ and\ \bibinfo {author} {\bibfnamefont {D.~J.}\ \bibnamefont {Weir}},\
  }\bibfield  {title} {\enquote {\bibinfo {title} {\emph{Gravitational Waves
  from the Sound of a First Order Phase Transition}},}\ }\href {\doibase
  10.1103/PhysRevLett.112.041301} {\bibfield  {journal} {\bibinfo  {journal}
  {Phys. Rev. Lett.}\ }\textbf {\bibinfo {volume} {112}},\ \bibinfo {pages}
  {041301} (\bibinfo {year} {2014})},\ \Eprint {http://arxiv.org/abs/1304.2433}
  {arXiv:1304.2433 [hep-ph]} \BibitemShut {NoStop}%
\bibitem [{\citenamefont {Ellis}\ \emph
  {et~al.}(2020{\natexlab{b}})\citenamefont {Ellis}, \citenamefont {Lewicki},\
  and\ \citenamefont {No}}]{Ellis:2020awk}%
  \BibitemOpen
  \bibfield  {author} {\bibinfo {author} {\bibfnamefont {J.}~\bibnamefont
  {Ellis}}, \bibinfo {author} {\bibfnamefont {M.}~\bibnamefont {Lewicki}}, \
  and\ \bibinfo {author} {\bibfnamefont {J.~M.}\ \bibnamefont {No}},\
  }\bibfield  {title} {\enquote {\bibinfo {title} {\emph{Gravitational Waves
  from First-Order Cosmological Phase Transitions: Lifetime of the Sound Wave
  Source}},}\ }\href {\doibase 10.1088/1475-7516/2020/07/050} {\bibfield
  {journal} {\bibinfo  {journal} {JCAP}\ }\textbf {\bibinfo {volume} {07}},\
  \bibinfo {pages} {050} (\bibinfo {year} {2020}{\natexlab{b}})},\ \Eprint
  {http://arxiv.org/abs/2003.07360} {arXiv:2003.07360 [hep-ph]} \BibitemShut
  {NoStop}%
\bibitem [{\citenamefont {Guo}\ \emph {et~al.}(2021)\citenamefont {Guo},
  \citenamefont {Sinha}, \citenamefont {Vagie},\ and\ \citenamefont
  {White}}]{Guo:2020grp}%
  \BibitemOpen
  \bibfield  {author} {\bibinfo {author} {\bibfnamefont {H.-K.}\ \bibnamefont
  {Guo}}, \bibinfo {author} {\bibfnamefont {K.}~\bibnamefont {Sinha}}, \bibinfo
  {author} {\bibfnamefont {D.}~\bibnamefont {Vagie}}, \ and\ \bibinfo {author}
  {\bibfnamefont {G.}~\bibnamefont {White}},\ }\bibfield  {title} {\enquote
  {\bibinfo {title} {\emph{Phase Transitions in an Expanding Universe:
  Stochastic Gravitational Waves in Standard and Non-Standard Histories}},}\
  }\href {\doibase 10.1088/1475-7516/2021/01/001} {\bibfield  {journal}
  {\bibinfo  {journal} {JCAP}\ }\textbf {\bibinfo {volume} {01}},\ \bibinfo
  {pages} {001} (\bibinfo {year} {2021})},\ \Eprint
  {http://arxiv.org/abs/2007.08537} {arXiv:2007.08537 [hep-ph]} \BibitemShut
  {NoStop}%
\bibitem [{\citenamefont {Caprini}\ and\ \citenamefont
  {Durrer}(2006)}]{Caprini:2006jb}%
  \BibitemOpen
  \bibfield  {author} {\bibinfo {author} {\bibfnamefont {C.}~\bibnamefont
  {Caprini}}\ and\ \bibinfo {author} {\bibfnamefont {R.}~\bibnamefont
  {Durrer}},\ }\bibfield  {title} {\enquote {\bibinfo {title}
  {\emph{Gravitational Waves from Stochastic Relativistic Sources: Primordial
  Turbulence and Magnetic Fields}},}\ }\href {\doibase
  10.1103/PhysRevD.74.063521} {\bibfield  {journal} {\bibinfo  {journal} {Phys.
  Rev. D}\ }\textbf {\bibinfo {volume} {74}},\ \bibinfo {pages} {063521}
  (\bibinfo {year} {2006})},\ \Eprint {http://arxiv.org/abs/astro-ph/0603476}
  {arXiv:astro-ph/0603476} \BibitemShut {NoStop}%
\bibitem [{\citenamefont {Caprini}\ \emph {et~al.}(2009)\citenamefont
  {Caprini}, \citenamefont {Durrer},\ and\ \citenamefont
  {Servant}}]{Caprini:2009yp}%
  \BibitemOpen
  \bibfield  {author} {\bibinfo {author} {\bibfnamefont {C.}~\bibnamefont
  {Caprini}}, \bibinfo {author} {\bibfnamefont {R.}~\bibnamefont {Durrer}}, \
  and\ \bibinfo {author} {\bibfnamefont {G.}~\bibnamefont {Servant}},\
  }\bibfield  {title} {\enquote {\bibinfo {title} {\emph{The Stochastic
  Gravitational Wave Background from Turbulence and Magnetic Fields Generated
  by a First-Order Phase Transition}},}\ }\href {\doibase
  10.1088/1475-7516/2009/12/024} {\bibfield  {journal} {\bibinfo  {journal}
  {JCAP}\ }\textbf {\bibinfo {volume} {12}},\ \bibinfo {pages} {024} (\bibinfo
  {year} {2009})},\ \Eprint {http://arxiv.org/abs/0909.0622} {arXiv:0909.0622
  [astro-ph.CO]} \BibitemShut {NoStop}%
\bibitem [{\citenamefont {Aasi}\ \emph {et~al.}(2015)\citenamefont {Aasi} \emph
  {et~al.}}]{LIGOScientific:2014pky}%
  \BibitemOpen
  \bibfield  {author} {\bibinfo {author} {\bibfnamefont {J.}~\bibnamefont
  {Aasi}} \emph {et~al.} (\bibinfo {collaboration} {LIGO Scientific}),\
  }\bibfield  {title} {\enquote {\bibinfo {title} {\emph{Advanced LIGO}},}\
  }\href {\doibase 10.1088/0264-9381/32/7/074001} {\bibfield  {journal}
  {\bibinfo  {journal} {Class. Quant. Grav.}\ }\textbf {\bibinfo {volume}
  {32}},\ \bibinfo {pages} {074001} (\bibinfo {year} {2015})},\ \Eprint
  {http://arxiv.org/abs/1411.4547} {arXiv:1411.4547 [gr-qc]} \BibitemShut
  {NoStop}%
\bibitem [{\citenamefont {Hiramatsu}\ \emph {et~al.}(2014)\citenamefont
  {Hiramatsu}, \citenamefont {Kawasaki},\ and\ \citenamefont
  {Saikawa}}]{Hiramatsu:2013qaa}%
  \BibitemOpen
  \bibfield  {author} {\bibinfo {author} {\bibfnamefont {T.}~\bibnamefont
  {Hiramatsu}}, \bibinfo {author} {\bibfnamefont {M.}~\bibnamefont {Kawasaki}},
  \ and\ \bibinfo {author} {\bibfnamefont {K.}~\bibnamefont {Saikawa}},\
  }\bibfield  {title} {\enquote {\bibinfo {title} {\emph{On the Estimation of
  Gravitational Wave Spectrum from Cosmic Domain Walls}},}\ }\href {\doibase
  10.1088/1475-7516/2014/02/031} {\bibfield  {journal} {\bibinfo  {journal}
  {JCAP}\ }\textbf {\bibinfo {volume} {02}},\ \bibinfo {pages} {031} (\bibinfo
  {year} {2014})},\ \Eprint {http://arxiv.org/abs/1309.5001} {arXiv:1309.5001
  [astro-ph.CO]} \BibitemShut {NoStop}%
\bibitem [{\citenamefont {Clarke}\ \emph {et~al.}(2020)\citenamefont {Clarke},
  \citenamefont {Copeland},\ and\ \citenamefont {Moss}}]{Clarke:2020bil}%
  \BibitemOpen
  \bibfield  {author} {\bibinfo {author} {\bibfnamefont {T.~J.}\ \bibnamefont
  {Clarke}}, \bibinfo {author} {\bibfnamefont {E.~J.}\ \bibnamefont
  {Copeland}}, \ and\ \bibinfo {author} {\bibfnamefont {A.}~\bibnamefont
  {Moss}},\ }\bibfield  {title} {\enquote {\bibinfo {title} {\emph{Constraints
  on Primordial Gravitational Waves from the Cosmic Microwave Background}},}\
  }\href {\doibase 10.1088/1475-7516/2020/10/002} {\bibfield  {journal}
  {\bibinfo  {journal} {JCAP}\ }\textbf {\bibinfo {volume} {10}},\ \bibinfo
  {pages} {002} (\bibinfo {year} {2020})},\ \Eprint
  {http://arxiv.org/abs/2004.11396} {arXiv:2004.11396 [astro-ph.CO]}
  \BibitemShut {NoStop}%
\end{thebibliography}%

\end{document}